\documentclass[english, 11pt]{article}
\usepackage[T1]{fontenc}
\usepackage[utf8]{inputenc}
\usepackage{geometry}
\usepackage{float}
\usepackage{graphicx}
\usepackage{babel}
\usepackage{bm}
\usepackage{amsmath}
\usepackage{amssymb} 
\usepackage{mathtools}
\usepackage{bm}
\usepackage{fixmath} 
\usepackage{subcaption}
\usepackage[cal=cm]{mathalfa}
\usepackage{caption}
\usepackage{soul}
\usepackage{subcaption}
\usepackage{xcolor}
\usepackage[round, sort&compress,comma, numbers]{natbib}
\usepackage{authblk}
\usepackage[normalem]{ulem}
\usepackage{dsfont}
\usepackage{pifont} 

\usepackage{setspace}

\useunder{\uline}{\ul}{}
\usepackage{multirow}
\usepackage{tikz}
\usepackage{adjustbox}
\usepackage{calc}

\usetikzlibrary{positioning, shapes, calc, fit}
\usepackage{array}
\newcolumntype{P}[1]{>{\centering\arraybackslash}p{#1}}
\newcolumntype{M}[1]{>{\centering\arraybackslash}m{#1}}

\usepackage{hyperref} 
\hypersetup{colorlinks=true,linkcolor=blue,filecolor=blue,citecolor=blue}
\newcommand{\argmin}{\mathop{\mathrm{argmin}}}
\newcommand{\indep}{\perp \!\!\! \perp}

\newtheorem{theorem}{Theorem}

\newtheorem{lemma}{Lemma}
\newtheorem{assumption}{A\hspace{-0.3em}}
\newtheorem{algorithm}{Algorithm}

\newcommand{\blind}{0}

\begin{document}

\title{A Statistical Framework for Understanding Causal Effects that Vary by Treatment Initiation Time in EHR-based Studies}

\if0\blind {
\author[1]{Luke Benz}
\author[2]{Rajarshi Mukherjee}
\author[2,3,4]{Rui Wang}
\author[5]{David Arterburn}
\author[6]{Heidi Fischer}
\author[7]{Catherine Lee}
\author[8,9]{Susan M. Shortreed}
\author[10]{Alexander W. Levis*}
\author[2]{Sebastien Haneuse*}
\affil[1]{Global Statistical Sciences,
Eli Lilly \& Company, Indianapolis, IN, USA}
\affil[2]{Department of Biostatistics,
Harvard T.H. Chan School of Public Health, Boston, MA, USA}
\affil[3]{Department of Population Medicine, Harvard Pilgrim Health Care Institute, Boston, MA, USA}
\affil[4]{Department of Population Medicine, Harvard Medical School, Boston, MA, USA}
\affil[5]{Kaiser Permanente Washington Health Research Institute, Seattle, WA, USA}
\affil[6]{Department of Research \& Evaluation, Kaiser Permanente Southern California, Pasadena, CA, USA}
\affil[7]{Department of Epidemiology and Biostatistics, University of California San Francisco, San Francisco, CA, USA}
\affil[8]{Biostatistics Division, Kaiser Permanente Washington Health Research Institute, Seattle, WA, USA}
\affil[9]{Department of Biostatistics, University of Washington School of Public Health, Seattle, WA, USA}
\affil[10]{Department of Biostatistics, Epidemiology and Informatics, University of Pennsylvania, Philadelphia, PA, USA }

\begingroup
\renewcommand{\thefootnote}{}%
\renewcommand{\footnoterule}{}%
\footnotetext{\noindent $^*$ denotes co-last author (AWL and SH)}
\footnotetext{Corresponding Author: Luke Benz (\url{lukesbenz@gmail.com})}
\endgroup

\date{
    \today 
}

\maketitle

}\fi

\if1\blind{

\author{}

\date{
    \today 
}

\vspace{-3in}

\maketitle

}\fi

\begin{abstract}

\noindent Standard practice in electronic health record (EHR)-based studies evaluating the comparative effectiveness of bariatric surgery relative to no surgery is to estimate and report a constant treatment effect across calendar time. However, real-world treatment strategies can evolve, particularly when comparators include standard of care or surgical procedures where techniques may improve, making it clinically important to ascertain whether efficacy of bariatric surgery has changed over time. Efforts to determine whether treatment efficacy itself is evolving are complicated by changing patient populations, with potential covariate shift in key effect modifiers. Through a comprehensive analysis of EHR data from Kaiser Permanente comparing two bariatric surgical procedures to standard of care, we develop a statistical framework to estimate calendar time-specific average treatment effects and describe both how and why effects vary across treatment initiation time in EHR-based studies. Our approach projects doubly robust, time-specific treatment effect estimates onto candidate marginal structural models and uses model selection to best describe how effects vary by calendar time. We further introduce a novel summary metric, based on standardization, to quantify the degree to which observed effect changes may be attributed to covariate shift in the population receiving treatment, as opposed to changes in underlying treatment efficacy.

\end{abstract}

{\footnotesize {\textbf{Keywords}: causal inference, treatment effect heterogeneity,  bariatric surgery, electronic health records} }

\if0\blind
\newpage{}
\fi

\doublespacing 

\section{Introduction}
\label{sec:introduction}

Electronic health record (EHR) databases provide a useful data source for comparative effectiveness studies, allowing for larger sample sizes from patient populations which are more diverse than those typical of clinical trials. In EHR-based studies, study entry periods frequently extend over a decade or more \citep{arterburn2020, fisher2018association, coleman2016longterm, Harrington2024, haneuse2025}, perhaps because (a) treatment initiation is uncommon, (b) long-term outcomes are of interest, and/or (c) extending data collection further back in time requires little additional effort. For example, in a large EHR-based study, Arterburn et al.~\cite{arterburn2020} examined long-term weight loss outcomes of patients undergoing bariatric surgery between 2005-2015 in comparison to matched non-surgical controls. In contrast to treatment protocols in tightly controlled randomized clinical trials, real-world treatments, such as those examined by Arterburn et al.~\cite{arterburn2020}, can evolve over time, especially if comparators include a somewhat loosely defined standard of care (e.g., for obesity management), or bariatric surgical procedures where techniques or best practices may change \citep{Sheetz2020, Faria2017, Felsenreich2021, Guldogan2020Sleeve}. As such, causal effects of interest in EHR-based studies with long study entry periods may vary over the course of treatment initiation calendar time (i.e., the date at which a treatment is initiated), which we also refer to as ``calendar time'' throughout.

Heterogeneity in treatment initiation calendar time has been discussed previously in the context of studies in pharmacoepidemiology. Common across such studies, however, is that variation in treatment initiation time is most often viewed as a potential source of bias towards estimating an implicitly assumed \textit{constant} treatment effect across calendar time (though for longitudinal or time-to-event outcomes, the effect may vary across time since initiation). For example, stratification of propensity score models by calendar year has been suggested as a solution to address changes in physician prescription patterns or characteristics of treated patients over time \citep{Dusetzina2013, Mack2013, Dilokthornsakul2014CalendarTime}. When comparisons of interest involve, say, two drugs approved at different points in time, analysis is often restricted to a period where both medications are available to reduce differences attributable to calendar time \citep{Gagne2013TreatmentDynamics, Suissa2017PrevalentNewUser}. Strategies based on instrumental variables \citep{Gokhale2018CalendarTime} and matching \citep{HelinSalmivaara2014SequentialCohort} have also been proposed. In all, the predominant viewpoint has been that variation in causal contrasts by calendar time is something that must be adjusted for, rather than adoption of causal estimands which explicitly use calendar time in their definition. That this is the case likely owes to the fact that in pharmacoepidemiologic studies, typical treatments (e.g., drugs or medications) are not changing over time. In such cases, there may be strong reason to believe there exists a common effect over calendar time in a static population, though the underlying populations may be shifting.

Causal estimands which depend on calendar time have recently been explored in the context of stepped-wedge cluster randomized trials \citep{wang2024model, Lee2025SteppedWedge} and staggered-adoption difference-in-difference settings \citep{callaway2021did, sun2021dynamic}, the latter of which often considers analysis of multi-year periods during which a policy is rolled out. Because intervention timing is planned or quasi-experimental across clusters, both study designs provide a natural way to define calendar time-specific effects for groups of clusters initiating treatment at the same time. By contrast, EHR-based studies analyze patient-level outcomes under naturally occurring treatment patterns, which do not follow any controlled schedule, and thus strategies from these literatures do not directly apply. 

Determining whether causal effects vary over calendar time in EHR-based studies is of clinical interest as such efforts can help characterize the real-world history of a treatment and inform whether some fundamental aspect of its efficacy is changing. Moreover, describing the structure of causal effects over calendar time yields an estimate of the treatment effect at the end of the study period (i.e., that which most closely reflects current practice) and informs how historical data might be leveraged to guide present-day clinical decision-making. However, such efforts are complicated by changing patient populations, and thus, potentially, covariate shift in key effect modifiers. As such, it is equally important to understand why causal effect estimates may vary over time, and the degree to which effect modification and changing patient populations plays a role. 

In this work, motivated by our experience in a series of studies of long-term outcomes following bariatric surgery \citep{arterburn2020, coleman2016longterm, obrien2018microvascular, coleman2022bariatric, fisher2018association, Harrington2024, haneuse2025}, we propose a statistical framework to estimate and describe how and why causal effects vary over calendar time in EHR-based studies. First, doubly-robust, calendar time-specific average treatment effects (ATE) are estimated on the basis of pooled nonparametric nuisance models. In an effort to summarize temporal trends, estimates are projected onto candidate marginal structural models (MSM), each representing a different pattern in how effect estimates vary across calendar time (e.g., constant, linear, etc.). An MSM selection procedure is then applied to choose which structure best describes temporal variation in effect estimates. Next, in an attempt to characterize the degree to which observed changes in treatment effects may be attributed to shifts in the patient population receiving treatment, as opposed to changes in underlying treatment efficacy, we define a causal estimand that corresponds to the treatment effect which would have been observed if the covariate distribution among eligible patients at a given time instead reflected that at some alternative time. Finally, we introduce a metric summarizing variability across these ``cross-temporal'' contrasts to quantify the role of covariate shift in explaining observed effect changes across calendar time. To our knowledge, this work is the first to apply tools from semiparametric theory \citep{tsiatis_semiparametric_2006, kennedy2023semiparametric} to study calendar time-varying treatment effects in EHR-based studies.


\section{An EHR-based Study of Weight Loss Following Bariatric Surgery}
\label{sec:bariatric_surgery}

\subsection{Bariatric Surgery}
\label{sec:bs1}

Bariatric surgery is an intervention to mitigate obesity and related comorbidities, with typical candidates having a body mass index (BMI) exceeding 35 kg/m$^2$. Due to the cost and complexity of conducting large-scale trials for bariatric surgery \citep{Courcoulas2024}, EHR-based studies are a critical tool to study long-term outcomes following bariatric surgery, particularly in relation to similarly obese patients not undergoing surgery \citep{arterburn2020, obrien2018microvascular, fisher2018association, Harrington2024,  coleman2022bariatric}. Despite its success as a weight-loss intervention, only 1-2\% of eligible candidates elect to undergo bariatric surgery annually \citep{arterburn2014bariatric}. Therefore, EHR-based studies of bariatric surgery must consider a multi-year study entry period to obtain a meaningful number of patients undergoing surgery. 

The two most common bariatric surgical procedures are Roux-en-Y gastric bypass (RYGB) and sleeve gastrectomy (SG), with SG having surpassed RYGB in frequency in the past 15 years \citep{arterburn2014bariatric}. SG is a less invasive and technically easier procedure than RYGB, and as such is offered by more surgeons. Relevant to this work is that the surgical technique for SG has historically been somewhat less standardized than for RYGB, particularly during the time-frame during which SG became the more commonly performed procedure \citep{Guldogan2020Sleeve}. Given the rapid shifts in the relative popularity of RYGB and SG in recent years, along with possibly evolving surgical techniques and standards of care for obesity management, understanding potential temporal changes in the effectiveness of bariatric surgery is of particular interest. Not only does characterization of this temporal structure offer insight into these procedures during an important change in their utilization history, but it also includes assessments of treatment effects at the end of the study period that are more relevant to current clinical practice when the data are proximal to present day.

\subsection{The DURABLE EHR Database}
\label{sec:bs2}

With this backdrop, we examined relative weight change at four pre-specified times post-surgery (6 months, 1 year, 2 years, 3 years), comparing 17,905 patients undergoing bariatric surgery between 2005-2011 at one of three Kaiser Permanente (KP) sites (KP Washington, KP Northern California, KP Southern California) with 933,044 non-surgical patients. In particular, we compared outcomes between surgery and no surgery, each of RYGB/SG versus no surgery separately, and RYGB versus SG. We leveraged EHR data from DURABLE, an NIH-funded study examining long-term outcomes of bariatric surgery across Kaiser Permanente sites \citep{arterburn2020, coleman2016longterm, obrien2018microvascular, coleman2022bariatric, fisher2018association, Harrington2024, haneuse2025}.

Using this data source, Arterburn et al.~\cite{arterburn2020} previously characterized longitudinal weight trajectories following RYGB and SG in comparison to matched non-surgical patients, though calendar time was considered only as a potential confounder. As an alternative to the matched cohort design employed by Arterburn et al.~\cite{arterburn2020}, we used a sequential target trial emulation approach \citep{benz2024, madenci2024, haneuse2025, danaei2013observational}. Specifically, we considered a sequence of 84 target trials (formalized in Section \ref{sec:setting}), initiating at monthly intervals between January 2005 and December 2011. Following Arterburn et al., our study population included adults aged 19-79 years with BMI exceeding 35 kg/m$^2$. We excluded patients without a year of continuous enrollment in KP insurance and those pregnant in the year prior to the start of a particular target trial. In total, 17,596 surgical patients (98.3\%) and 763,016 non-surgical patients (81.8\%) met the study inclusion criteria at baseline of at least one target trial.

\begin{figure}
    \centering
    \includegraphics[width=\textwidth]{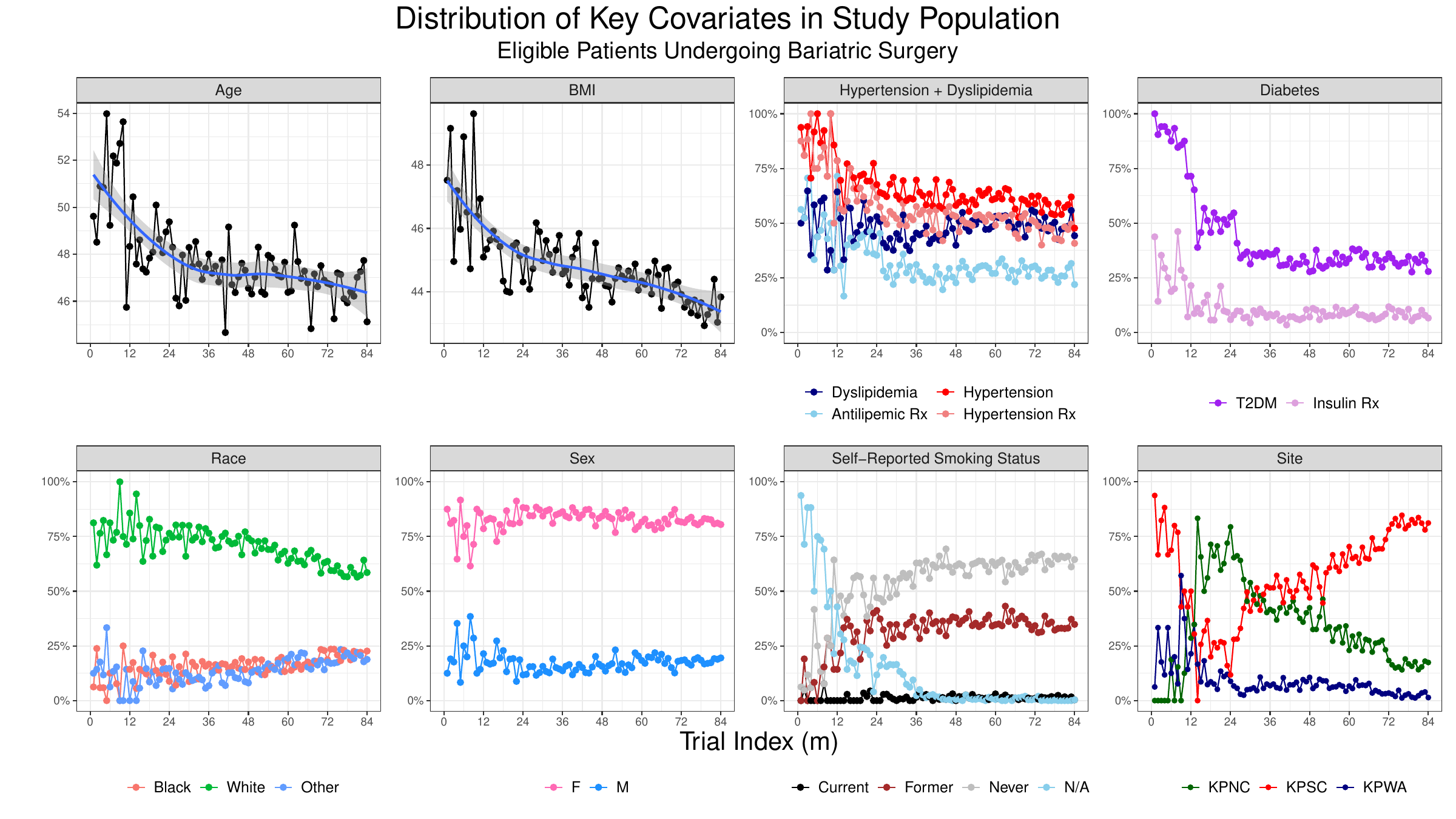}
    \caption{Distribution of patient characteristics in DURABLE electronic health record database over calendar time for eligible patients undergoing bariatric surgery between 2005-2011. Covariate mean values (continuous covariates) or frequency (binary/categorical covariates) are plotted by month of surgery. $m = 1$ corresponds to January 2005 and $m = 84$ corresponds to December 2011.}
    \label{fig:surg_dist}
\end{figure}

\subsection{Calendar Time-Varying Effects in Bariatric Surgery}
\label{sec:bs3}

Towards quantifying the relative benefits of surgery versus no surgery, the standard way forward would be to estimate and report a common effect across calendar time, for example, on 3-year relative weight change. Though formal details are deferred until Section \ref{sec:application}, such an approach produces estimates of -19.8\% (-19.9\%, -19.6\%) for surgery overall, with corresponding estimates of -23.1\% (-23.3\%, -22.9\%) for RYGB and -16.3\% (-16.6\%, -16.1\%) for SG. Note, these are in close agreement with estimates of -23.5\% and -16.7\% reported by Arterburn et al.~\cite{arterburn2020}, respectively. While such summaries have appeal, particularly for communication to broad audiences, they may conceal meaningful changes in the effectiveness of bariatric surgery over calendar time, whether due to evolving surgical techniques, shifting patient populations, or perhaps both.

Towards gaining preliminary insight, Figure \ref{fig:surg_dist} summarizes the distribution of several key covariates among eligible bariatric patients across the 84 target trials (i.e., across calendar time). Notably, the distribution of several effect modifiers highlighted by Arterburn et al.~\cite{arterburn2020} (age, BMI, race, and diabetes status) varies meaningfully across calendar time. For example, mean BMI of patients undergoing bariatric surgery in a given calendar month declines from 47.5 kg/m$^2$ in January 2005 to 43.8 kg/m$^2$ in December 2011, with monthly averages ranging between 42.9 kg/m$^2$ and 49.6 kg/m$^2$. Accordingly, any attempt to study how the effect of bariatric surgery may have changed over the course of the study period should acknowledge and account for this covariate shift. A similar figure displaying covariate shift for non-surgical patients is available in the Supplementary Materials (Figure S1).  

\section{Problem Setting}
\label{sec:setting}

In a hypothetical clinical trial, each patient would be randomized to undergo or not undergo surgery and be followed from the point of treatment assignment. By contrast, patients in our EHR-based study are not randomized to their respective treatment assignment. A key challenge in addressing the primary questions laid out in Section \ref{sec:bariatric_surgery} is that, in contrast to patients who undergo surgery, there is no obvious way to assign a ``time zero'' to patients who are in the ``no surgery'' arm \citep{hernan2016immortal}. 

Many prior studies of bariatric surgery, \cite{coleman2016longterm, obrien2018microvascular, coleman2022bariatric, fisher2018association, Harrington2024}, including Arterburn et al.~\citep{arterburn2020}, have dealt with this problem with a matched cohort design, matching non-surgical controls to surgical patients at the index date of surgery. More recent studies of bariatric surgery \citep{benz2024, madenci2024, haneuse2025} have utilized sequential target trial emulation \citep{hernan2008, danaei2013observational}, which partitions the study entry period into a sequence of $M$ target trials, indexed by $m \in \{1, \dots, M\}$. At baseline of each target trial, study eligibility is queried for all patients in the EHR, and patients are included in each trial for which they meet the eligibility criteria. We employ target trial emulation in our study, as the sequential nature of the design gives rise to a very natural definition of initiator cohorts through which to define calendar time-specific average treatment effects. In our study, $M = 84$ and component target trials are conducted at monthly intervals between January 2005 ($m = 1$) and December 2011 ($m = M = 84$). 

\subsection{Notation}
\label{sec:notation}
 To the best of our knowledge, the sequential target trial emulation design has been applied exclusively in applications examining time-to-event outcomes \citep{scola2023implementation}. Since relative weight change is not a time-to-event endpoint, we describe notation to formalize this design for univariate continuous outcomes.  Let $A_m \in \{0, 1\}$ denote a binary point exposure (e.g., surgery vs. no surgery) received at baseline of trial $m$, $Y_m \in \mathbb{R}$ denote an outcome of interest (e.g., relative weight change since baseline of trial $m$), and $\bm L_m \in \mathbb{R}^p$ denote a set of covariates at baseline of trial $m$ assumed sufficient for control of confounding, such as those in Figure \ref{fig:surg_dist}. Let $E_m = g(\bm L_m)$ be a binary eligibility indicator for inclusion in target trial $m$, where $g(\cdot)$ represents a fixed and known eligibility rule \citep{benz2024, benz2025}. Practically, because not all patients are in the EHR for the entire duration of the study entry period, we treat subjects as ineligible $(E_m = 0)$ for periods before they appear and/or after they disappear from the EHR. Thus, at each of $M$ regularly occurring intervals, we ascertain for each patient $i$ their eligibility $E_{m,i}$ for trial $m$, and, if eligible, $\bm Z_{m,i} \coloneqq (\bm L_{m,i}, A_{m,i}, Y_{m,i})$. That is, patients contribute $\bm Z_{m,i}$ for all target trials where they meet study eligibility. Formally, we work with coarsened observed data units

$$
O_i =  (E_{1,i}, E_{1,i}\bm Z_{1,i}, \dots E_{m,i}, E_{m,i}\bm Z_{m,i} , \dots E_{M,i}, E_{M,i}\bm Z_{M,i}) 
$$

We observe a sample $(O_1, \ldots, O_n)$ of $n$ independent and identically distributed observations from an underlying and unknown data distribution, $P$, i.e., the joint distribution of information across both patients and trials. For succinctness, we often drop the patient index $i$ and work with a generic observation $O$. Implicitly throughout, we assume covariates across trials share a common support contained in $\mathcal{L}$. As will become clear in Sections \ref{sec:ctate} and \ref{sec:estimation}, covariate, treatment and outcome values of ineligible patients are not required for the definition of our calendar time-specific causal estimands, nor for estimation.



For any $P$-integrable function $f$ we define $Pf = \mathbb{E}_P[f] = \int f \,dP$ to denote the expectation of $f$ under $P$, and analogously define $\mathbb{P}_nf = \frac{1}{n}\sum_{i = 1}^n f(O_i)$ to express the sample mean of $f$ on data units $(O_1, ..., O_n)$. In particular, we use $\mathbb{P}_n[E_m] = n^{-1}\sum_{i=1}^n \mathds{1}(E_{m,i} = 1)$ denotes the empirical proportion of patients eligible for trial $m$, while $P(E_m = 1)$ denotes its population analogue. Lastly, we let $Y_m(a_m)$ denote counterfactual outcomes for the $m^\text{th}$ trial, that is the outcome that would have been observed under treatment $A_m = a_m$ \citep{hernan2024}. 

\subsection{Calendar Time-Specific Average Treatment Effects}
\label{sec:ctate}

Towards understanding how and why treatment efficacy may vary over treatment initiation calendar time, we define the following causal estimand:

$$
\chi_m(P) = \mathbb{E}[Y_m(1) - Y_m(0)~|~E_m = 1].
$$

\noindent Notice that $\chi_m(P)$ is simply the calendar time-specific ATE among patients eligible for trial $m$. Implicit in the definition of $\chi_m(P)$ is that comparisons are being made between the specific versions of treatment (i.e., surgical technique, standard of care) in place at time $m$. We are not interested in defining nor estimating effects of dynamic treatment regimes and note that consistent with the definition of $A_m$ as a point exposure, $\chi_m(P)$ captures the effect of initiating treatment at time $m$ $(a_m = 1)$ versus not initiating $(a_m = 0)$ without regard to what happens at future time points. In particular, the definition of $\chi_m(P)$ is not concerned that treatment definitions may evolve over time (e.g., $a_j = 0$ and $a_m = 0$ may mean slightly different things for $j > m$), nor for the possibility of treatment switching; see Section~\ref{sec:discussion} for a discussion of how our framing relates to the consistency assumption and ``multiple versions of treatment''. Due to the rarity of initiating surgery relative to standard of care and since bariatric surgery is effectively an irreversible procedure (at least in so far as it relates to a patient adhering to surgery or standard of care), non-adherence is extremely uncommon in our study.

Just as a constant hazard ratio represents a weighted average of time-varying hazard ratios (across time \textit{since} treatment initiation) \citep{hernan2010hazards}, a sequential target trial emulation estimating a constant ATE across calendar time implicitly targets some weighted average of trial-specific effects $\chi_m(P)$. However, the precise definition and causal interpretation of such contrast is rarely formalized or made transparent.

\section{Estimation of Calendar Time-Specific Average Treatment Effects}
\label{sec:estimation}

In order to answer the clinical questions raised in Section \ref{sec:bariatric_surgery}, one might initially target the nonparametric curve composed of the collection $\bigr\{\chi_m(P)\}_{m=1}^M$. As we elaborate upon, however, a small number of surgical patients in any particular target trial presents a challenge towards efficient estimation of any individual $\chi_m(P)$. MSMs have proven useful for analysis of continuous exposures by parameterizing a dose-response curve to make causal inference tractable \cite{robins2000marginal, naimi2014constructing, hernan2024}. In that same vein, pragmatic adoption of MSMs in the present context provides structure on top of $\bigr\{\chi_m(P)\}_{m=1}^M$, allowing information to be borrowed across target trials to describe trends over calendar time.

\subsection{Assumptions}
\label{sec:assumptions}

Given that treatment decisions in EHR-based settings are not randomized, we begin by articulating standard causal inference assumptions to address the potential for confounding. Note that these assumptions are only required to hold among the subset of patients eligible for each trial. This is particularly salient for comparisons of bariatric surgery to no surgery, where study eligibility criteria are often designed to exclude patients who are not realistic candidates for surgery.

\begin{assumption}[Consistency]\label{as:consistency}
  $Y_m(A_m) = Y_m$ when $E_m = 1$, for all $m$.
\end{assumption}

\begin{assumption}[Positivity]\label{as:positivity}
  $\exists ~\epsilon > 0$ such that $ \epsilon \leq P(A_m = 1 ~|~\bm  L_m, E_m = 1) \leq 1 - \epsilon $, almost surely, for all $m$. 
\end{assumption}

\begin{assumption}[No Unmeasured Confounding]\label{as:NUC}
  $Y_m(a_m) \indep A_m~|~\bm L_m, E_m = 1$ for $a_m \in \{0,1\}$ and all $m$.
\end{assumption}

\noindent Following closely to Arterburn et al.~\citep{arterburn2020}, we take $\bm L_m$ to be the most recent values of covariates outlined in Figure \ref{fig:surg_dist}, although we note that analysts can also include in this vector appropriate summaries of covariate and/or treatment history prior to baseline of trial $m$, if warranted in their application. A careful discussion of Assumption \ref{as:consistency}, particularly in the context of multiple versions of treatment and changing treatment efficacy, is deferred until Section \ref{sec:discussion}.

\subsection{Motivating A Projection Parameter Approach}
\label{sec:projection_motivation}

Let $\mu_m(a_m, \bm L_m) = \mathbb{E}[Y_m~|~A_m = a_m , \bm L_m, E_m = 1]$ and $\pi_m(L_m) = P(A_m = 1~|~\bm L_m, E_m = 1)$ denote the trial $m$-specific outcome regression and propensity score models, respectively. Under Assumptions \ref{as:consistency}-\ref{as:NUC}, $\chi_m(P)$ is identified by the $g$-formula, $\mathbb{E}[\mu_m(1,\bm L_m)-\mu_m(0,\bm L_m)~|~E_m = 1]$ or via inverse probability weighting, $\mathbb{E}\big[\big(\frac{A_m}{\pi_m(\bm L_m)}- \frac{1-A_m}{1-\pi_m(\bm L_m)}\big)Y_m~|~E_m = 1\big]$ \citep{hernan2024}. One might then proceed to estimate $\chi_m(P)$ on the basis of its nonparametric influence function \citep{tsiatis_semiparametric_2006, kennedy2023semiparametric},

$$
\begin{aligned}
\dot\chi^*_m(O; P) = \frac{\mathds{1}(E_m = 1)}{P(E_m = 1)}&\biggr\{\mu_m(1, \bm L_m) - \mu_m(0, \bm L_m) - \chi_m(P) \\
&~~~ +\biggr(\frac{A_m}{\pi_m(\bm L_m)} - \frac{1-A_m}{1-\pi_m(\bm L_m)}\biggr)\Bigr(Y_m - \mu_m(A_m, \bm L_m)\Bigr)\biggr\}.
\end{aligned}
$$

\noindent Unfortunately, estimating $\mu_m$ or $\pi_m$ by fitting a separate model for each trial would be difficult in practice, especially in settings where a relatively small number of patients initiate a particular treatment during trial $m$. For example, 24 of 84 target trials in our study have fewer than 100 patients undergoing bariatric surgery at the corresponding baseline month, and 12 of 84 have fewer than 25 surgical patients. In fact, the lack of power from small sample sizes inherent to emulating a target trial from a single enrollment window was one of the primary motivations behind development of the sequential target trial design \citep{hernan2008}.

The predominant analytic strategy in sequential target trial emulations for time-to-event outcomes targets (discrete time) hazard ratios by fitting a single parametric outcome and/or propensity score model pooling information across trials \citep{scola2023implementation} (e.g., pooled logistic regression \citep{dagostino1990}), perhaps with spline terms to enable non-proportional hazards and flexibly adjust for calendar time as a confounder \citep{dickerman2019, madenci2024}. Given that trends in $\chi_m(P)$ over calendar time may reflect a complex interplay between changing treatment efficacy and shifts in effect modifier distributions, applying a pooled parametric model in this setting would require correctly specifying a substantial number of flexible interactions between treatment, confounders, and calendar time. Such a task may be unrealistic in many instances---arguably including our study of bariatric surgery given the heterogeneity reported in Figure \ref{fig:surg_dist}---and might instead motivate the use of pooled models based on flexible machine learning methods. 

Ultimately, while we are interested in trial-specific ATEs, we are most interested in characterization of the underlying trend in $\chi_m(P)$ across calendar time. To formalize estimation of this trend across calendar time, we follow the strategy of Kennedy et al.~\cite{kennedy2019} and others \citep{kennedy2023density, neugebauer2007nonparametric, rosenblum2010targeted}, and formulate our estimand as a projection of all $M$ trial-specific effects onto a candidate parametric marginal structural model. Specifically, following the pragmatic use of MSMs for analysis of continuous exposures \citep{kennedy2019, naimi2014constructing, hernan2024, robins2000marginal}, we define a MSM for mean trial-specific counterfactual outcomes.

\begin{equation}\label{eqn:msm}
    \mathbb{E}[Y_m(a_m)~|~E_m = 1] = \psi_0(m) + \psi(m)a_m.
\end{equation}

\noindent When the MSM in Equation \eqref{eqn:msm} is saturated, $\psi(m) = \chi_m(P)$, that is, we are targeting the true nonparametric effect curve across calendar time. Because such contrast is not easily estimable due to challenges described above, we adopt a working parametric model for $\psi(m)$, denoted $\psi(m; \bm \beta)$, and take our estimand to be the projection of $\chi_m(P)$ onto $\psi(m; \bm \beta)$. Despite being a working model, $\psi(m; \bm \beta)$ remains a useful summary of how treatment efficacy has evolved over calendar time. Moreover, when study data are proximal to the present, the more recent end of this curve, $\psi(M; \bm \beta)$, offers direct insight into treatment efficacy in current clinical  practice.

Formally, $\bm\beta(P)$ is the projection parameter defined by 

$$
\bm\beta(P) = \argmin_{\bm\beta \in \mathbb{R}^k} \sum_{m = 1}^M w(m)P(E_m = 1)\Bigr(\chi_m(P) - \psi(m;\bm\beta)\Bigr)^2,
$$

\noindent where $P(E_m = 1)$ weights trial contributions according to their effective sample size and $w(m)$ allows for optional weighting by other additional factors. We emphasize that $\psi(m; \bm \beta)$ is not believed to be the true effect curve but rather a tractable summary of trends across calendar time. Because the form of $\psi(m; \bm \beta)$ that best approximates the nonparametric curve $\bigr\{\chi_m(P)\}_{m=1}^M$ is unknown, we outline a model selection procedure (described in Section \ref{sec:model_selection}) for comparing candidate MSMs to identify the functional form that best describes how treatment effects vary over calendar time, including the possibility of a single, constant effect.


\subsection{Efficient Estimation of Projection Parameter}
\label{sec:projection_estimation}

We now introduce preliminary efficiency theory for the projection parameters $\bm \beta(P)$, which will motivate our proposed estimator. Since influence functions are by definition mean zero, it is helpful when constructing estimators to work with uncentered versions, $\dot\chi_m(O; P) \coloneqq  \frac{\mathds{1}(E_m = 1)}{P(E_m = 1)}\dot\chi^\dagger_m(O; P)$, which satisfy $\mathbb{E}_P[\dot\chi_m(O; P)] = \chi_m(P)$, where 

$$
\dot\chi^\dagger_m(O; P) \coloneqq \mu_m(1, \bm L_m) - \mu_m(0, \bm L_m) + \biggr(\frac{A_m}{\pi_m(\bm L_m)} - \frac{1-A_m}{1-\pi_m(\bm L_m)}\biggr)\Bigr(Y_m - \mu_m(A_m, \bm L_m)\Bigr).
$$

\begin{theorem}\label{thm:beta_IF}
Suppose that the candidate MSM, $\psi(m;\bm\beta)$ is twice differentiable in $\bm \beta$. Under Assumptions \ref{as:consistency}-\ref{as:NUC}, $\bm\beta(P)$ is identified as the solution to the estimating equation,
$$
0 = \sum_{m = 1}^M w(m) P(E_m = 1)\nabla_{\bm\beta}\psi(m;\bm\beta)[\chi_m(P) - \psi(m;\bm\beta)].
$$
Furthermore, $\dot{\bm \beta}^*(O;P)$, the influence function of $\bm\beta(P)$ (up to proportionality), is given by 
\begin{equation}\label{eqn:beta_IF}
    \dot{\bm\beta}^*(O; P) \propto \dot{\bm \beta}^\dagger(O;P) = \sum_{m = 1}^M \mathds{1}(E_m  =1) w(m)\nabla_{\bm \beta}  \psi(m; \bm \beta)|_{\bm \beta = \bm \beta(P)}\Bigr(\dot \chi^\dagger_m(O;P) - \psi\bigr(m;\bm\beta(P)\bigr)\Bigr).
\end{equation}
\end{theorem}

There are several observations from Theorem \ref{thm:beta_IF} worth pointing out. First, Equation \eqref{eqn:beta_IF} motivates an estimator $\widehat{\bm\beta}$ which solves $\mathbb{P}_n[\dot{\bm \beta}^\dagger(O;\widehat P)]$ = 0, that is, 

\begin{equation}\label{eqn:estimating_equation}
0 = \sum_{m = 1}^M w(m) \mathbb{P}_n(E_m = 1)\nabla_{\bm\beta}\psi(m;\bm \beta)|_{\bm \beta = \widehat{\bm\beta}}\Bigr\{\widehat \chi_m - \psi(m; \widehat{\bm\beta})  \Bigr\},
\end{equation}

\noindent where $\widehat{\chi}_m$ is the influence function-based estimator of $\chi_m(P)$ mentioned in Section~\ref{sec:projection_motivation}, and $\widehat P$ indicates that any necessary nuisance functions are replaced by estimated quantities. Equations \eqref{eqn:beta_IF} and \eqref{eqn:estimating_equation} make clear, however, that to even estimate $\bm\beta(P)$, estimates of $\chi_m(P)$, and thus estimates of nuisance functions $\mu_m$ and $\pi_m$, are required. 

A subtle but important point is that the projection-based estimate of the treatment effect trend, $\psi(m; \widehat{\bm\beta})$, will reflect patterns in our estimates of $\chi_m(P)$, and thus, projections based on estimates $\widehat \chi_m$ from pooled parametric models will mirror any structure implied by the adopted models. As such, it is important to maintain a sufficient degree of flexibility in estimating $\chi_m(P)$ and only adopt any structure (e.g., to characterize the underlying trend) after trial-specific ATEs have been estimated. We revisit the topic of selecting how much structure to adopt on top of $\widehat\chi_m$ in Section \ref{sec:model_selection}.

To estimate $\chi_m(P)$, we first create a pooled dataset $\mathcal{D}$ of patient information across all trials in which they are eligible, that is,

\begin{equation}\label{eqn:pooled_data}
\mathcal{D} = \bigcup_{m = 1}^M \Bigr\{(\bm L_{m,i}, A_{m,i}, Y_{m,i}, m)\Bigr\}_{i: E_{m,i} = 1}.
\end{equation}

\noindent We then use $\mathcal{D}$ to estimate a pooled outcome regression $\widetilde \mu(A_m, \bm L_m, m)$ and propensity score model $\widetilde \pi(\bm L_m, m)$. Notice that the trial index, $m$, is included as a covariate in the estimation of these pooled models. Not only does this ensure that fitted nuisance functions $\widehat\mu_m(A_m, \bm L_m) \equiv \widetilde \mu(A_m, \bm L_m, m)$ and $\widehat \pi_m(\bm L_m) \equiv \widetilde\pi(\bm L_m, m)$ maintain a dependence on calendar time, but also when flexible modeling strategies are employed, it allows the strength of association between covariates in $\bm L$ and treatment or outcome to vary over calendar time as well. Following common practice in target trial emulations \citep{hernan2008, danaei2013observational, dickerman2019}, we recommend stratification of the outcome model by treatment status, that is estimating separate pooled models $\widetilde \mu_1(\bm L_m, m)$ for $A_m = 1$ and $\widetilde \mu_0(\bm L_m, m)$ for $A_m = 0$. Using nuisance function estimates from pooled models, we then estimate $\chi_m(P)$ with the influence function-based estimator $\mathbb{P}_n\big[\frac{\mathds{1}(E_m = 1)}{\mathbb{P}_n[E_m]}\dot\chi_m^{\dagger}(O;\widehat P)\big]$, i.e., which solves $\mathbb{P}_n[\dot\chi^*_m(O;\widehat P)] = 0$.

Routines for estimation of influence function-based estimators like $\widehat\chi_m$ or $\widehat{\bm\beta}$ typically involve sample splitting or cross-fitting, particularly when using nonparametric and/or machine learning techniques to estimate component nuisance functions \citep{chernozhukov2018}. We describe the complete procedure for estimating $\widehat\chi_m$ and $\widehat{\bm\beta}$ in Algorithm \ref{alg:estimation} below.

\singlespacing
\begin{algorithm}[Computation of $\widehat\chi_m$ and $\widehat{\bm\beta}$]\label{alg:estimation}
Let $S_1$ and $S_2$ be two disjoint, independent splits of $(O_1,\ldots,O_n)$ of sizes $n_1$ and $n_2$, respectively, with $n = n_1 + n_2$. The analytical procedure to compute the estimator $\widehat\chi_m$ and $\widehat{\bm\beta}$ is as follows:

\begin{enumerate}
    \item Create pooled dataset of eligible subject-trials $\mathcal{D}_1$ among the $n_1$ subjects in split $S_1$, as in Equation \eqref{eqn:pooled_data}.
    \item Fit pooled models $(\widetilde \mu_1, \widetilde\mu_0, \widetilde\pi)$ on $\mathcal{D}_1$ to obtain nuisance estimators $\widehat{P}_1 \coloneqq\{(\widehat \mu_m, \widehat \pi_m)\}_{m = 1}^M$.
     \item For $O_{n_1 + 1}, O_{n_1 + 2}, ..., O_{n} \in S_2$, construct plug-in estimates of uncentered influence functions $\{\dot\chi_m(O_i;\widehat P_1)\}_{i = n_1+1}^n$ for each $m \in \{1, \dots, M\}$, where $\dot\chi_m(O_i;\widehat P_1) \equiv \frac{\mathds{1}(E_m = 1)}{\mathbb{P}_n[E_m]}\dot\chi_m^{\dagger}(O_i;\widehat P_1)$.
    \item Repeat steps 1-3, swapping the roles of $S_1$ and $S_2$ to create, for each $m \in \{1, \dots, M\}$, complete sets of influence function contributions from corresponding out-of-sample nuisance function predictions, 
    $\{\dot\chi_{m}(O_i; \widehat P)\}_{i = 1}^n = \{\dot\chi_m(O_i;\widehat P_1)\}_{i = n_1+1}^n \cup \{\dot\chi_m(O_i; \widehat P_2)\}_{i = 1}^{n_1}$.  
    \item Compute calendar time-specific treatment effects, $\widehat \chi_m = n^{-1}\sum_{i = 1}^n \dot\chi_m(O_i;\widehat P)$.
    \item For a given candidate marginal structural model, $\psi(m;\bm\beta)$, plug estimates $\{\widehat\chi_m\}_{m=1}^M$ into Equation \eqref{eqn:estimating_equation} and solve for $\widehat{\bm\beta}$.
\end{enumerate}
\end{algorithm}
\doublespacing

\subsection{Theoretical Properties of \texorpdfstring{$\widehat {\bm \beta}$}{Beta} and \texorpdfstring{$\widehat \chi_m$}{Chi}}
\label{sec:beta_theory}

The choice to estimate $\widehat\chi_m$ and $\widehat{\bm\beta}$ using an estimating equation-based approach rather than a one-step approach is a matter of practical convenience; asymptotically the two approaches are equivalent. In order to describe the asymptotic behavior of $\widehat{\bm\beta}$, we note that $\text{Var}[\dot{\bm\beta}^*(O;P)] = \bm V^{-1}\bm C\bm V^{-1}$, where $\bm V \in \mathbb{R}^{k \times k}$ is the scaling matrix of proportionality ignored in Theorem \ref{thm:beta_IF}, and $\bm C  = \text{Var}[\dot{\bm\beta}^\dagger(O;P)] \in \mathbb{R}^{k\times k}$. As we derive in Supplementary Materials Appendix S3, we have that 

\begin{equation}\label{eqn:scaleV}
\begin{aligned}
\bm V = \sum_{m = 1}^M w(m) P(E_m = 1)\Bigr\{\nabla_{\bm \beta \bm\beta}^2 \psi(m; \bm \beta)|_{\bm\beta = \bm\beta(P)}\big[\chi_m(P) - \psi\bigr(m; \bm \beta(P)\bigr)\big] \\ - \nabla_{\bm \beta} \psi(m; \bm \beta)|_{\bm\beta = \bm\beta(P)} [\nabla_{\bm \beta} \psi(m; \bm \beta)|_{\bm\beta = \bm\beta(P)}]^ T\Bigr\}.
\end{aligned}
\end{equation}

The appearance of $\chi_m(P)$ inside the second-derivative term in Equation \eqref{eqn:scaleV} arises naturally from a total-derivative argument applied to the estimating equation in Theorem \ref{thm:beta_IF}, as discussed in greater detail in Appendix S3. Finally, we summarize the asymptotic behavior of $\widehat\chi_m$ and $\widehat{\bm\beta}$ in the following theorem. Henceforth, we write $\lVert f \rVert_{(m)}$ for the $L_2(P)$ norm of a function of $(\bm L_m, A_m, Y_m)$ conditional on eligibility in the $m^\text{th}$ trial: $\lVert f \rVert_{(m)}^2 \coloneqq \mathbb{E}_P[f(\bm L_m, A_m, Y_m)^2 \mid E_m = 1]$. 

\begin{theorem}\label{thm:asymptotics}
    Suppose $\lVert \widehat{\pi}_m - \pi_m\rVert_{(m)} + \sum_{a = 0}^1\lVert \widehat{\mu}_m(a, \, \cdot\,) - \mu_m(a, \, \cdot\,)\rVert_{(m)} = o_P(1)$. Further, assume that $\epsilon \leq \widehat{\pi}_m(\bm L_m), \pi(\bm L_m) \leq 1 - \epsilon$ and $|Y_m|, |\widehat{\mu}_m(a, \bm L_m)| \leq \epsilon^{-1}$ almost surely when $E_m = 1$. Then
    $$
    \widehat\chi_{m} - \chi_{m}(P) = \mathbb{P}_n[\dot\chi^*_{m}(O;P)] + O_P\bigr(\widehat{R}_{m}\bigr) + o_P(n^{-1/2}),
    $$
    \noindent where 
    \[
    \widehat{R}_{m} = \lVert \widehat{\pi}_m - \pi_m\rVert_{(m)}\sum_{a = 0}^1\lVert \widehat{\mu}_m(a, \, \cdot\,) - \mu_m(a, \, \cdot\,)\rVert_{(m)} .
    \]
    \noindent Moreover, if $\widehat{R}_m = o_P(n^{-1/2})$ then $\sqrt{n}\bigr(\widehat \chi_m - \chi_m(P)\bigr) \overset{d}{\to} \mathcal{N}\bigr(0, \textnormal{Var}[\dot\chi^*_m(O,P)]\bigr)$, whereby $\widehat{\chi}_m$ attains the nonparametric efficiency bound.

\vspace{2mm}
    Similarly, under Assumptions S1-S3 in Appendix S3.3,
    \[\widehat{\bm \beta} - \bm \beta(P) = \mathbb{P}_n[\dot{\bm \beta}^*(O;P)] + O_P\bigg(\sum_{m=1}^M \widehat{R}_m\bigg) + o_P(n^{-1/2}).\]
    When $\sum_{m = 1}^M \widehat{R}_m = o_P(n^{-1/2})$,
    $\sqrt{n}\bigr(\widehat{\bm\beta} - \bm\beta(P)\bigr) \overset{d}{\to} \mathcal{N}\bigr(0, \textnormal{Var}[\dot{\bm\beta}^*(O;P)]\bigr)$, whereby $\widehat{\bm \beta}$ attains the nonparametric efficiency bound.
\end{theorem}

As illustrated by Theorem \ref{thm:asymptotics}, the rates of convergence for estimates of calendar time-specific treatment effects and their corresponding trend are directly tied to error rates in estimation of the underlying outcome regression and propensity score models. Importantly, convergence rates for both $\widehat\chi_m$ and $\widehat{\bm\beta}$ are faster than component nuisance functions, even when employing flexible modeling choices. When these product errors are $o_P(n^{-1/2})$---for example when each nuisance function's error rate is $o_P(n^{-1/4})$---then $\widehat\chi_m$ is $\sqrt{n}$-consistent and asymptotically normal, and attains the nonparametric efficiency bound. Furthermore, when the required rate conditions are satisfied for each component trial the same properties extend to $\widehat{\bm\beta}$. Additional commentary on Assumptions S1-S3 is provided in the Supplementary Materials.

Practically, pooled models can help ensure that the required rate conditions are met in component trials with smaller effective sample sizes. As discussed in Section \ref{sec:projection_estimation}, however, pooled parametric models for $\pi_m$ and $\mu_m$ are likely to be inconsistent in the presence of significant covariate shift, motivating the use of more flexible pooled models. Crucially, when the required rate conditions (e.g., $o_P(n^{-1/4})$ for each nuisance function) are satisfied, asymptotically valid inference for the trend in treatment effects across calendar time remains attainable, even when using such flexible pooled models. In particular, asymptotically valid pointwise Wald-style $(1-\alpha)$-level confidence intervals, for a given candidate MSM, are given by

$$
\begin{aligned}
\psi(m;\widehat{\bm\beta}) &\pm z_{1-\alpha/2}\sqrt{\nabla_{\bm \beta}  \psi(m; \bm \beta)|_{\bm {\widehat{\beta}}}^T\mathbb{P}_n[\dot{\bm\beta}^*(O;\widehat P)\dot{\bm\beta}^*(O;\widehat P)^T]\nabla_{\bm \beta}  \psi(m; \bm \beta)|_{\bm {\widehat{\beta}}} / n} \\ = \psi(m;\widehat{\bm\beta}) &\pm z_{1-\alpha/2}\sqrt{\nabla_{\bm \beta}  \psi(m; \bm \beta)|_{\bm {\widehat{\beta}}}^T\widehat{\bm V}^{-1}\widehat{\bm C}\widehat{\bm V}^{-1}\nabla_{\bm \beta}  \psi(m; \bm \beta)|_{\bm {\widehat{\beta}}} / n},
\end{aligned}
$$

\noindent where $z_\alpha$ denotes the $\alpha$ quantile of the standard normal distribution. In the above, $\widehat{\bm V}$ denotes the result of Equation \eqref{eqn:scaleV} replacing $\bm\beta(P)$, $\chi_m(P)$, and $P(E_m = 1)$ with $\widehat{\bm\beta}$, $\widehat\chi_m$, and $\mathbb{P}_n(E_m = 1)$, respectively, while $\widehat{\bm C} = \widehat{\text{Var}}[\dot{\bm\beta}^\dagger(O;\widehat P)]$.

\subsection{Selection Among Candidate Marginal Structural Models}
\label{sec:model_selection}

Unless an analyst a priori desires to summarize treatment efficacy through a single constant effect across calendar time, it is often unclear how to specify the functional form for $\psi(m;\bm\beta)$ which appropriately summarizes the nonparametric curve $\bigr\{\chi_m(P)\}_{m=1}^M$, and in practice several candidate models may be reasonable. Forcing a single specification risks either under- or over-smoothing the underlying trend in $\chi_m(P)$, obscuring clinically meaningful variation or introducing spurious structure. It is therefore preferable to consider a range of candidate MSMs, including a constant model (i.e., $\psi(m;\beta) = \beta$), a linear model (i.e., $\psi(m;\bm\beta) = \beta_0 + \beta_1m$), and more flexible MSMs including higher-order polynomials (in $m$) and splines, and select among them using a data-adaptive procedure that lets the data inform the appropriate level of complexity.

To implement such an MSM selection procedure, we adapt the ideas of Kennedy et al.\cite{kennedy2019} and van der Laan and Dudoit~\cite{vanderLaanDudoit2003cv}. Let $\{\widehat\psi_1, ..., \widehat\psi_K\}$ denote a set of $K$ candidate MSM projection models, with $\widehat\psi_k \equiv \psi_k(m;\widehat{\bm\beta}_k)$. A natural quantity by which to compare candidate MSMs is mean-squared error (relative to the true $\chi_m(P)$). To simplify comparisons, we drop terms that do not depend on the candidate MSM and proceed with the \textit{pseudorisk},

$$
L(\widehat\psi_k) = \sum_{m = 1}^M w(m)\bigr\{\psi_k(m; \widehat{\bm\beta}_k)^2 - 2\psi_k(m; \widehat{\bm\beta}_k)\chi_m(P)\bigr\}.
$$

\noindent Given the appearance of $\chi_m(P)$ in the pseudorisk, we replace $\chi_m(P)$ with estimates based on its influence function. Defining the per-observation loss

$$
\widehat\ell_k(O) \coloneqq \sum_{m = 1}^M w(m)\bigr\{\psi_k(m; \widehat{\bm\beta}_k)^2 - 2\psi_k(m; \widehat{\bm\beta}_k)\dot\chi_m(O;\widehat P)\bigr\},
$$

\noindent we proceed with the influence function-based estimator of the pseudorisk, $\widehat{L}(\widehat\psi_k) = \mathbb{P}_n[\widehat\ell_k(O)]$. This is an intuitive loss function to work with because it reflects an estimating equation-based estimator of the pseudorisk, based on its influence function, treating $\widehat\psi_k$ as fixed \citep{kennedy2019}.

In contrast to loss functions like mean squared prediction error, which can be directly evaluated on holdout data via cross validation, careful evaluation of the pseudorisk requires that $\bm\beta_k(P)$ and $\chi_m(P)$ are estimated on independent samples of data. However, we also require estimates of $\chi_m(P)$ in order to estimate the projection parameter $\bm\beta_k(P)$. Thus, we proceed with a modified version of Algorithm \ref{alg:estimation}, and introduce a second layer of sample splitting. Within each split of data at the first stage, we apply the entire Algorithm \ref{alg:estimation} to obtain our projection-based MSM on that split. Next we estimate $\dot\chi_m(O;\widehat P)$ on the opposite data split by applying steps 1-3 of Algorithm \ref{alg:estimation}, and combine results to compute $\widehat{L}(\widehat\psi)$. Finally, we swap the roles of first-stage splits and average. Consequently, each $\widehat\ell_k(O)$ is evaluated using $\widehat{\bm\beta}_k$ and $\dot\chi_m(O;\widehat P)$ constructed from opposite splits of the data.

In practice, differences between candidate projection models can be small \citep{kennedy2023semiparametric}, particularly if model classes are nested within one another. In such cases, analysts might have a preference for simpler models, such that more flexible MSMs are selected only if there are clinically meaningful trends in $\chi_m(P)$ across calendar time. To formalize this, let $\widehat\psi^* = \argmin_{\widehat \psi_k \in \{\widehat \psi_1, \dots, \widehat \psi_K\}}\widehat{L}(\widehat \psi_k)$ denote the MSM among our set of candidates which minimizes the loss function $\widehat{L}(\widehat\psi_k)$, with corresponding per-observation loss $\widehat\ell^*(O)$. For each $\widehat\psi_k \neq \widehat\psi^*$, let $\widehat\Delta_k(O) \coloneqq \widehat\ell_k(O) - \widehat\ell^*(O)$ and consider the quantity

$$
\widehat{\text{se}}_k = \sqrt{\frac{\mathbb{P}_n\bigr[\bigr(\widehat\Delta_k(O) - \mathbb{P}_n\widehat\Delta_k\bigr)^2\bigr]}{n}},
$$

\noindent that is, the standard error of the difference in loss function, relative to the minimizing MSM (treating the fitted $\widehat\psi_k$ and $\widehat\psi^*$ as fixed, as in the definition of $\widehat L$). Among $\widehat\psi^*$ and all candidates for which the standardized difference $\frac{\widehat{L}(\widehat\psi_k) - \widehat{L}(\widehat\psi^*)}{\widehat{\text{se}}_k} \leq c$, one can choose the simplest model, for some user-selected constant $c$. There is considerable debate in the literature about how large an effect size needs to be on a standard deviation scale to be clinically meaningful \citep{tsujimoto2022mic}, and the notion of clinically meaningful is very application dependent, though $c = 1/2$ or $1$ are common rules of thumb. Finally, we note that the notion of ``simpler model'' can be operationalized through any model complexity criterion so long as models can be ordered by this criterion. One reasonable such choice---which we adopt in our application---is ordering candidate MSMs by the dimension of their indexing projection parameter (i.e., the dimension of $\bm\beta$).

\section{Decomposing Sources of Variation}
\label{sec:decomposing}

While the projection parameter framework and corresponding MSM selection procedure outlined in Section \ref{sec:estimation} provide insight into \textit{how} treatment effects vary across calendar time, they fail to explain \textit{why} such changes (if any) are observed. In particular, if the candidate MSM selected is non-constant, suggesting that $\chi_m(P)$ varies over time, we would like to characterize the extent to which underlying changes in treatment efficacy, as opposed to distribution shift in baseline population characteristics, may be responsible.

\subsection{A Cross-Trial Contrast to Remove Distribution Shift}
\label{sec:ct_effect}

To isolate why treatment effects might vary over calendar time, we begin by defining new contrasts by averaging reweighted conditional average treatment effects (CATE) \cite{dahabreh2019generalizing, cole2010generalizing} over covariate distributions reflecting trial eligible populations at different points in time:
$$
\chi_{j,m}(P) \coloneqq \int_\mathcal{L} \mathbb{E}\bigr[Y_m(a_m = 1) - Y_m(a_m = 0)~|~\bm L_m = \bm \ell, g(\bm L_m) = 1\bigr]dP_{\bm L_j|E_j=1}(\bm \ell~|~E_j = 1).
$$

\noindent The estimand $\chi_{j,m}(P)$ represents the difference in mean counterfactual outcomes among patients eligible for the $m^\text{th}$ trial in a population where trial baseline covariates are standardized to the covariate distribution of the eligible population at time $j$. First, we note that when $j = m$, this contrast reduces to trial-specific ATE defined in Section \ref{sec:ctate}, that is, $\chi_{m,m}(P) = \chi_m(P)$. Given the definition of the class of estimands $\chi_{j,m}(P)$, one obtains common populations with which to standardize treatment effects across calendar time. Namely, for $m_1 \neq m_2$, differences between $\chi_{j,m_1}(P)$ and $\chi_{j,m_2}(P)$ cannot be attributable to covariate shift, as the underlying population on which the causal contrast is defined is the same. We exploit this fact in Section \ref{sec:standardization_matrix} to introduce a metric for quantifying effect modification.

In order to identify each $\chi_{j,m}(P)$, we introduce an additional assumption similar to those commonly used in transportability analyses \citep{dahabreh2019generalizing, cole2010generalizing}. 

\begin{assumption}[Cross-Trial Overlap]\label{as:transport_pos}
    For any $\bm \ell \in \mathcal{L}$,  $p(\bm \ell ~|~E_j = 1) > 0 \implies p(\bm \ell~|~E_m = 1) > 0$.
\end{assumption}

\noindent Assumption \ref{as:transport_pos} is often referred to as an overlap assumption in the transportability literature and ensures that if a patient met the study eligibility criteria at baseline of trial $j$, their covariate vector $\bm L_j$ is in the support of the covariate distribution among those eligible for trial $m$. 

\begin{lemma}\label{lemma:cross_id} Under Assumptions \ref{as:consistency}-\ref{as:transport_pos}, $\chi_{j,m}(P)$ is identified by
    $$
    \chi_{j,m}(P) = \mathbb{E}[\mu_m(1, \bm L_j) - \mu_m(0, \bm L_j)~|~E_j=1 ] = \int_{\mathcal{L}} \{\mu_m(1, \bm \ell) - \mu_m(0, \bm \ell)\} \, dP_{\bm L_j \mid E_j = 1}(\bm \ell).
    $$
\end{lemma}

\subsection{Estimation of Cross-Trial Effects}
\label{sec:cross_estimation}
Given that the cross-trial effects are just modified versions of calendar time-specific ATEs, the estimation strategy for $\chi_{j,m}(P)$ will be very similar to that of $\chi_m(P)$. We define one additional nuisance function, $\xi_{j,m}(\bm \ell) = \frac{p(\bm L_j = \ell~|~E_j = 1)}{p(\bm L_m = \ell~|~E_m = 1)}$, which represents a density ratio to transport the covariate distribution of the eligible population in trial $m$ to that of trial $j$.

\begin{theorem}\label{thm:cross_IF}
    The influence function of $\chi_{j,m}$ at $P$ in a nonparametric model is given by 
    $$
    \begin{aligned}
    \dot\chi^*_{j,m}(O;P) = &\frac{\mathds{1}(E_j = 1)}{P(E_j = 1)}\bigr\{\mu_m(1, \bm L_j) - \mu_m(0, \bm L_j) - \chi_{j,m}(P)\bigr\} \\
    & + \frac{\mathds{1}(E_m = 1)}{P(E_m = 1)}\xi_{j,m}(\bm L_m)\biggr(\frac{A_m}{\pi_m(\bm L_m)} - \frac{1-A_m}{1-\pi_m(\bm L_m)}\biggr)\Bigr(Y_m - \mu_m(A_m, \bm L_m)\Bigr).
    \end{aligned}
    $$
\end{theorem}

The influence function $\dot\chi^*_{j,m}(O;P)$ admits a very intuitive form. The first summand resembles a $g$-formula estimating function using the outcome regression for trial $m$ on the eligible covariate distribution from trial $j$. The second summand takes the inverse treatment-weighted residuals from eligible patients in trial $m$, and additionally weights them by the appropriate density ratio required to transport to trial $j$. If a subject is eligible for both trials $j$ and $m$, they will contribute in both terms, increasing efficiency, but they can still contribute if eligible for only one of the trials. 

Eventually, we will need to estimate $\chi_{j,m}(P)$ for all $M^2$ pairs of component target trials. As such, fitting a separate model for each $\xi_{j,m}$ is computationally infeasible. Moreover, conditional density estimation is a challenging statistical problem. Because $\xi_{j,m}$ is a ratio of conditional densities, however, we can directly apply Bayes' rule \citep{sugiyama2012density, benz2025} and leverage pooled models, as with $\widetilde \mu, \widetilde\pi$:

\begin{equation}\label{eqn:xi_bayes}
\xi_{j,m}(\bm \ell) = \frac{p(\bm L_j = \bm\ell~|~E_j = 1)}{p(\bm L_m = \bm\ell~|~E_m = 1)} = \frac{P(E_m = 1)P(T = j~|~\bm L =\bm \ell, E = 1)}{P(E_j = 1)P(T = m~|~\bm L =\bm \ell, E = 1)},
\end{equation}

\noindent where $T$ denotes the trial membership indicator in the full pooled data set, $\mathcal{D}$. Thus, we have turned the task of conditional density estimation into modeling the trial membership indicator conditional on $\bm L$ across all eligible subject-trials. Crucially, this is a classification task well suited for machine learning methods, and we can apply any flexible pooled model $\widetilde\xi(m~|~\bm \ell)$ to estimate $P(T = m~|~\bm L = \ell, E = 1)$.

With additional bookkeeping, we can estimate $\chi_{j,m}(P)$ for all pairs $(j,m)$ within the same routine described by Algorithm \ref{alg:estimation}. First, $\widetilde\xi$ is trained during step 2 along with the other pooled models. Following step 3, for each row in $\mathcal{D}$, we save predicted transport ratios to all other trial populations, $\{\widehat\xi_{j,m}(\bm L_m)\}_{j = 1}^M$, and fitted values for the outcome regression at all possible treatment times, $\{(\widehat\mu_m(1, \bm L_j), \widehat\mu_m(0, \bm L_j)\}_{m = 1}^M$. Extracting $\widehat\mu_m(a_m, \bm L_j) = \widetilde\mu(a_m, \bm L_j, m)$ simply requires (artificially) varying the trial index in the pooled model. Upon repeating these steps for both splits of $(O_1, \dots, O_n)$, we estimate each cross-trial effect via the following influence function-based estimator: 
\begin{align*}
    \widehat\chi_{j,m} 
    &= \mathbb{P}_n\bigg[\frac{\mathds{1}(E_j = 1)}{\mathbb{P}_n[E_j]}\bigr\{\widehat{\mu}_m(1, \bm L_j) - \widehat{\mu}_m(0, \bm L_j)\bigr\} \\
    & \quad \quad \quad + \frac{\mathds{1}(E_m = 1)}{\mathbb{P}_n[E_m]}\widehat{\xi}_{j,m}(\bm L_m)\biggr(\frac{A_m}{\widehat{\pi}_m(\bm L_m)} - \frac{1-A_m}{1-\widehat{\pi}_m(\bm L_m)}\biggr)\Bigr(Y_m - \widehat{\mu}_m(A_m, \bm L_m)\Bigr)\bigg],
\end{align*}
where, recall, hatted quantities are estimated in the training split, and $\mathbb{P}_n$ represents the empirical mean over the test split.

\subsection{Summarizing Changes in Effect Estimates}
\label{sec:standardization_matrix}

Ultimately, our interest in $\chi_{j,m}(P)$ is as a stepping stone towards determining the degree to which changes in calendar time-specific treatment effects are explained by changes in trial composition (including, for example, shifts in the distribution of effect modifiers) compared to something more fundamental in terms of changes in underlying treatment efficacy. We let $\bm S \in \mathbb{R}^{M\times M}$ denote the standardization matrix of all $M^2$ cross-trial effects, with $S_{jm} = \chi_{j,m}(P)$, and let $\widehat{\bm S}$ denote the estimated standardization matrix with entries $\widehat\chi_{j,m}$. To quantify sources of variation in cross-trial effects, we define the following variance-like quantities:

$$
\begin{aligned}
   \sigma_j^2 &\coloneqq \frac{1}{M-1}\sum_{m = 1}^M \Bigr(\chi_{j}(P) - \chi_{j,m}(P)\Bigr)^2, \\
    \gamma_m^2 &\coloneqq \frac{1}{M-1}\sum_{j = 1}^M \Bigr(\chi_{m}(P) - \chi_{j,m}(P)\Bigr)^2.
\end{aligned}
$$

Note that these quantities correspond to the row-wise and column-wise mean squared deviations of $\bm S$ about its diagonal, respectively. The quantity $\sigma_j^2$ characterizes the variability around the trial-specific ATE that would arise if the eligible population at time $j$ had been treated at alternative time points. In contrast, $\gamma_m^2$ characterizes the variability around $\chi_m(P)$ across all trial-eligible populations, had each been treated at time $m$. To quantify the degree to which trends across calendar time might be attributable to covariate shift, we examine the ratio

$$
\theta = \frac{1}{M}\sum_{m = 1}^M \theta_m = \frac{1}{M}\sum_{m = 1}^M\frac{\sigma_m^2}{\sigma_m^2 + \gamma_m^2}.
$$

\noindent Using our estimated cross-trial effects, $\widehat\chi_{j,m}$, we can readily obtain plug-in estimates $\widehat\sigma_m^2$ and $\widehat\gamma_m^2$, and subsequently $\widehat\theta$, which offers a useful way to summarize variation across estimated cross-trial effects: $\widehat\theta$ can be interpreted as the average proportion of variation in estimated cross-trial effects not attributable to covariate shift, which may reflect changes in treatment efficacy but could also reflect residual confounding or model misspecification, as discussed in Section \ref{sec:guidance_violated}.

When $\theta = 0$, all $\sigma_m^2 = 0$, meaning that within fixed populations, there is no variation in treatment effects across calendar time. In the event that a non-constant candidate MSM is selected, $\theta = 0$ indicates that the trend across treatment initiation calendar time is entirely induced by changes in underlying populations. On the other hand, when $\theta = 1$, all $\gamma_m^2 = 0$, and thus variation in cross-trial effects is fully explained by changing the calendar time index ($m$). Importantly, $\theta = 1$ implies that trends across calendar time are not the result of covariate shift. Finally, when $\theta\in (0,1)$, variation in estimated cross-trial effects is partially, but not entirely explained by distribution shift in baseline trial characteristics. Formal hypothesis testing using $\theta$ is discussed in Section \ref{sec:hypothesis_testing}. 

$\theta$ is inherently a summary measure that averages over many moving parts to characterize the extent to which observed effect changes may be explained by population composition. In particular, $\theta = M^{-1}\sum_{m=1}^M \theta_m$ averages trial-specific contributions across all $M$ target trials, collapsing potentially heterogeneous time-specific signals into a single scalar. Moreover, within any given target trial, $\theta_m$ aggregates over all factors that vary across calendar time for a fixed population, including changes in treatment efficacy, evolving standard of care, and other factors less readily characterized, such as residual confounding or model misspecification. This is an inherent limitation of any scalar summary of a complex object. However, we argue that this aggregation is a feature as much as it is a limitation --- much in the same way that a standard ATE implicitly averages conditional effects over a population-level covariate distribution, yet remains a useful and widely reported summary of treatment efficacy. In the same spirit, $\theta$ provides a tractable, interpretable summary of the aggregate role of covariate shift in explaining temporal variation in treatment effects, without requiring analysts to parse the full $M \times M$ standardization matrix $\widehat{\bm S}$. 

Complete discussion of using $\theta$, along with $\psi(m;\bm\beta)$ from Section \ref{sec:estimation}, to explain both \textit{how} and \textit{why} causal effects may vary over calendar time is presented in Section \ref{sec:toolkit}, with further discussion of alternative explanations for observed temporal variation.

\subsection{Hypothesis Testing for \texorpdfstring{$\theta$}{Variability Ratio}}
\label{sec:hypothesis_testing}

To formally assess the degree to which observed variation in cross-trial effects is attributable to changes in population composition across time, we consider the hypothesis test $H_0: \theta \in \{0, 1\}$ versus $H_1: \theta \in (0,1)$. We frame the null as simultaneously testing two point nulls because $\theta = 0$ and $\theta = 1$ correspond to the extreme cases where variation in cross-trial effects is entirely explained by covariate shift or entirely by the calendar time index ($m$), respectively, and drivers of treatment effect heterogeneity differ meaningfully between these cases. 

Given that the boundary nature of $H_0$ makes standard asymptotic characterization of the null distribution challenging, and that the computational burden of the full nonparametric bootstrap is prohibitive given the size of $\mathcal{D}$, we instead use an asymptotic parametric bootstrap \citep{van2000asymptotic} applied to $\widehat{\bm S}$, exploiting the asymptotic normality of $\sqrt{n}(\widehat{\bm S} - \bm S)$ established in Appendices S3.6 and S3.7. To handle the boundary null, we follow \cite{CavaliereNielsenPedersenRahbek2022} and adopt the modified test $H_0^{(\delta)}: \theta \in [0,\delta] \cup [1-\delta, 1]$ versus $H_1^{(\delta)}: \theta \in (\delta, 1-\delta)$ for a small user-specified $\delta > 0$, rejecting $H_0^{(\delta)}$ when a 95\% percentile-based confidence interval for $\theta$ lies fully within $(\delta, 1-\delta)$. Full details of this procedure, including derivation of the asymptotic covariance matrix $\bm\Sigma$ and discussion of the choice of $\delta$, are provided in Appendix S4.2.

\section{Practical Guidance for How and Why Effects Vary}
\label{sec:toolkit}

Thus far, we have introduced statistical tools to characterize \textit{how} treatment effects change across the study entry period, along with summary measures to quantify the role of covariate shift in explaining \textit{why} effects might change. In our experience, each set of tools in isolation reflects an incomplete picture, and both are necessary to inform analysts how best to understand and report findings. Although the goals of each study differ, and we do not advocate for a ``one size fits all'' approach, Section \ref{sec:guidance_assume} and Table \ref{tab:action_summary} summarize how analysts might use tools from this work to help interpret, characterize, and report calendar time-varying effects in their study. 

Under Assumptions \ref{as:consistency}-\ref{as:transport_pos} and the conditions of Theorems \ref{thm:asymptotics} and S1, the only possible explanations for changes in treatment effects over calendar time are due to covariate shift in effect modifiers or changes in underlying treatment efficacy. Section \ref{sec:guidance_violated} comments on alternative explanations under violations of these assumptions. 

\subsection{Guidance Under No Violations of Assumptions \ref{as:consistency}-\ref{as:transport_pos}}
\label{sec:guidance_assume}

To illustrate how MSM selection and hypothesis testing for $\theta$ can be used together to understand reporting of calendar time-varying effects, we consider two scenarios, one where a constant MSM is selected and a second where a non-constant MSM is selected. First, when the selected MSM is a constant, it is reasonable to report a common effect for the study population, $\psi(m;\widehat{\bm\beta}) = \widehat\beta$, regardless of the value of $\theta$. In some settings, $H_0^{(0)}: \theta = 0$ might be rejected despite a constant MSM having been selected. Such cases typically arise when differences in estimated treatment effects across calendar time are very small and manifest as random deviations centered around a common effect. Altogether, selecting a constant MSM indicates minimal heterogeneity in the underlying effect over calendar time, suggesting that any observed variation is unlikely to be clinically meaningful.

If the selected MSM is non-constant across calendar time but $\theta \approx 0$, analysts should at minimum acknowledge that changes across time are driven by shifting underlying populations. This is particularly important when reporting a common effect, which implicitly reflects an averaged effect over these different populations across time. Alternatively, if $\theta \approx 1$, covariate shift is unlikely to be the primary driver of observed changes, and it may be appropriate to report $\psi(m; \widehat{\bm\beta})$ as the best estimate of the calendar time-varying treatment effect. While this is still reasonable if $\theta \in (0,1)$, additional clarification should be reported to qualify that some amount of observed variation is due to changes in baseline characteristics. If there is particular interest in reporting on changes in treatment efficacy, one might consider standardization to a fixed population and report on the trend in that population (e.g., with an analogous projection-based curve). There are numerous reasonable choices for this fixed population, including both the most recent target trial population(s) (e.g., if they best represent subjects currently receiving treatment), or a subgroup defined by levels of key effect modifiers (e.g., patients with diabetes).

\begin{table}[htb]
    \centering
    \begin{tabular}{|M{3cm}|M{4cm}|M{4cm}|M{4cm}|}
        \hline
        \multicolumn{4}{|c|}{\textbf{Projection Model: Constant} $\bigr(\psi(m,\bm\beta) = \beta \bigr)$} \\
        \hline
         &
        \scriptsize \textbf{Fail to Reject $H_0^{(\delta)}$ }~~~~~~~~~~($\theta \leq \delta$) &
         \scriptsize \textbf{Fail to Reject $H_0^{(\delta)}$ }($\theta  \geq 1-\delta$) &
         \scriptsize \textbf{Reject $H_0^{(\delta)}$}($\theta \in (~\delta,1-\delta~)$) \\
        \hline
        \scriptsize \textbf{Calendar Time Varying Effect (Study Population)} & 
        \ding{55} & \ding{55} & \ding{55} \\
        \hline
        \scriptsize\textbf{Calendar Time Varying Effect (Fixed Population)} & \ding{55} & $\checkmark$ & $\checkmark$ \\
        \hline
        \scriptsize \textbf{Reason(s) for Variation} & \scriptsize No Variation  & \scriptsize Variation not clinically meaningful  & \scriptsize Variation not clinically meaningful \\
        \hline
        \scriptsize{\textbf{Action}} & \scriptsize Report common effect & \scriptsize Report common effect &  \scriptsize Report common effect \\
        \hline
        \multicolumn{4}{|c|}{\textbf{Projection Model: Not Constant} $\bigr(\psi(m,\bm\beta) \neq \beta \bigr)$} \\
       \hline
         &
        \scriptsize \textbf{Fail to Reject $H_0^{(\delta)}$ }~~~~~~~~~~~($\theta \leq \delta$) &
         \scriptsize \textbf{Fail to Reject $H_0^{(\delta)}$ }($\theta  \geq 1-\delta$) &
         \scriptsize \textbf{Reject $H_0^{(\delta)}$} ($\theta \in (\delta,1-\delta)$) \\
        \hline
        \scriptsize \textbf{Calendar Time Varying Effect (Study Population)} & 
        \checkmark & \checkmark & \checkmark \\
        \hline
        \scriptsize\textbf{Calendar Time Varying Effect (Fixed Population)} & \ding{55} & $\checkmark$ & $\checkmark$ \\
        \hline
        \scriptsize \textbf{Reason(s) for Variation} & \scriptsize Covariate shift in effect modifiers  & \scriptsize Possible changes in treatment efficacy  & \scriptsize Covariate shift in effect modifiers, possible changes in treatment efficacy  \\
        \hline
        \scriptsize{\textbf{Action}} & 
        \scriptsize Acknowledge changes across time driven by changes in underlying populations \par~\newline
        Consider standardization to fixed population and reporting common effect in that population & 
        \scriptsize Report calendar time-varying effect &  
        \scriptsize Report calendar time-varying effect
        \par~\newline
        Consider standardization to fixed population to further study changes in treatment efficacy \\
        \hline
    \end{tabular}
    \caption{Summary of possible reporting decisions based on joint results from projection marginal structural model and $\theta$.}
    \label{tab:action_summary}
\end{table}

\subsection{Guidance under Unmeasured Confounding or Model Misspecification}
\label{sec:guidance_violated}

Including $m$ as a covariate in nonparametric pooled models can make it difficult to distinguish differential forms of bias across trials from changes in the treatment effect over calendar time, i.e., it can be that $\widehat \theta$ is greater than 0 even if the underlying treatment efficacy itself is not changing. Under violations of Assumptions \ref{as:NUC}, for example, the difference between $\widehat\chi_{j,m_1}$ and $\widehat\chi_{j,m_2}$ would reflect both the difference in treatment efficacy between times $m_1$ and $m_2$ but also the difference in bias introduced by unmeasured confounding. A similar phenomenon holds under nuisance model misspecification such that trial index covariate captures residual confounding. It is exactly for these reasons why reliance on $\theta$ on its own is inadequate to characterize changing treatment efficacy. Though biases may introduce small differences between $\widehat\chi_{j,m_1}$ and $\widehat\chi_{j,m_2}$ even when $\chi_{j,m_1}(P) = \chi_{j,m_2}(P)$ (i.e. producing $\widehat \theta > 0$), a constant projection MSM will still be selected if differences between treatment effect estimates across calendar time are insubstantial. Ultimately if there is a strong concern of wrongful attribution of credit towards changes in underlying treatment efficacy, one can choose a larger value of $c$ in the projection MSM selection procedure, thereby enforcing that more complex MSMs must reduce the estimated loss by more standard errors in order to be selected.

\subsection{Simulations}
\label{sec:simulations}
Extensive simulations, tied closely to the motivating study of bariatric surgery, which validate the utility of the tools outlined in Sections \ref{sec:estimation} and \ref{sec:decomposing}, are provided in Appendix S5. Simulations highlight how MSM selection varies by sample size and choice of threshold $c$, along with the ability of the inferential procedure described in Section \ref{sec:hypothesis_testing} to distinguish between effect modification from underlying variation in treatment efficacy.

\section{Evaluating Temporal Trends in Causal Effects of Bariatric Surgery}
\label{sec:application}

\subsection{Methodological Details}
\label{sec:application_methods}

Finally, we return to our motivating study of bariatric surgery and leverage the proposed methods to examine trends over calendar time from 2005 to 2011. Analyses were conducted across 16 settings, defined by four post-surgical follow-up times (6 months, 1 year, 2 years, and 3 years) and four treatment comparisons (no surgery versus surgery, no surgery versus RYGB or SG separately, and RYGB versus SG). No SG patients met study eligibility criteria until month $m = 30$, so analysis for comparisons involving SG was restricted to trials $m\in \{30, \dots, 84\}$.

Random forests were used to estimate all pooled nuisance models, separately for each setting, adjusting for age, BMI, sex, race, KP study site, self-reported smoking status, hypertension, dyslipidemia, type II diabetes, and prescriptions for antilipemic or antihypertensive medications, or insulin. This set of potential confounders directly followed from those used by Arterburn et al.~\cite{arterburn2020}. Candidate MSMs for projection included constant, linear, and cubic models, and splines with 2 and 3 internal knots. To select among candidate MSMs, we employed the procedure described in Section \ref{sec:model_selection} with $c = 0.5$ as our primary specification, though we also report selected MSMs for $c = 1$ (i.e., a 1-SE rule). $B = 10,000$ bootstrapped replicates were used for constructing confidence intervals for $\theta$. Additional details on precise covariate and outcome definitions, as well as the loss function values underlying MSM selection, are provided in the Supplementary Materials.

\subsection{Results}
\label{sec:application_results}
\begin{figure}[!tb]
    \centering
        \includegraphics[width=\textwidth]{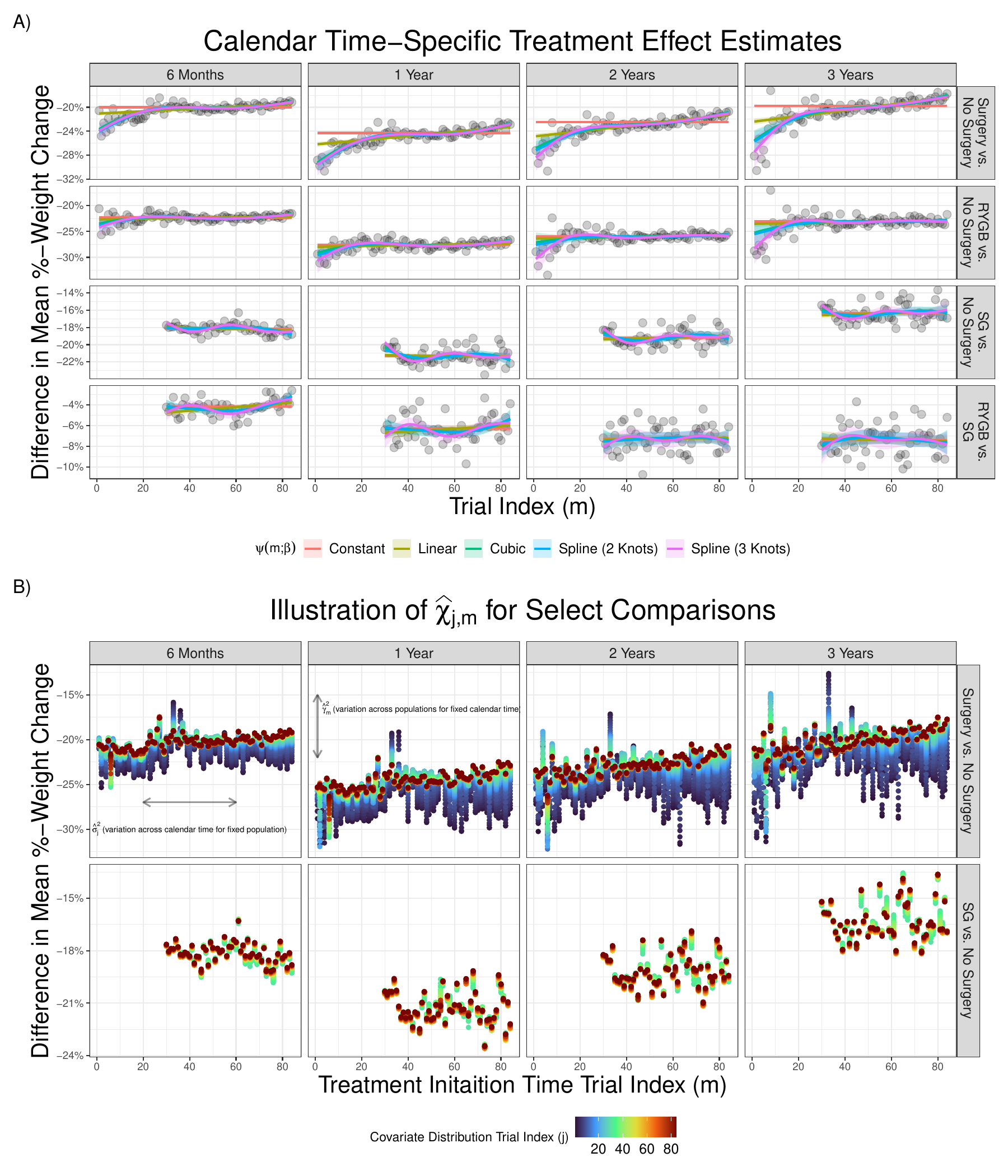}
    \caption{\textbf{(A)} Estimates of calendar time-specific treatment effects $\widehat\chi_m$ (dots) and trends based on five candidate marginal structural models, $\psi(m;\widehat{\bm\beta})$. \textbf{(B)} Estimates of cross-trial effects $\widehat\chi_{j,m}$ for select comparisons.}
    \label{fig:bmi_results}
\end{figure}

Figure \ref{fig:bmi_results}A presents estimated differences in mean relative weight change across calendar time. Gray points denote trial-specific ATEs, $\widehat\chi_m$, and overlaid lines denote estimated trends, $\psi(m;\widehat{\bm\beta})$, from each candidate MSM. As summarized in Table \ref{tab:msm_summary}, under our primary specification ($c = 0.5$), non-constant MSMs were selected for all outcomes involving comparisons between surgery versus no surgery, and RYGB versus no surgery, as well as for the 6-month weight change outcome in comparisons involving SG. Results were similar under $c = 1$, with the exception of 6-month weight change for RYGB versus no surgery, for which a constant MSM was selected.

Based on selected MSMs, surgical patients experienced excess 3-year weight change (relative to non-surgical patients) of -27.4\% (-30.1\%, -24.8\%) at $m =1$ and -18.3\% (-18.8\%, -17.8\%) at $m = 84$, reflecting a change of 9.1\% in the estimated treatment effect between 2005 and 2011. This temporal shift indicates that the common effect of –19.8\% reported in Section \ref{sec:bariatric_surgery}, obtained by projection onto a constant MSM, obscures an important aspect of understanding the efficacy of bariatric surgery.

Similar changes were observed for RYGB patients, with estimated 3-year weight change differences ranging from -28.3\% (-31.3\%, -25.3\%) at $m = 1$ to -23.0\% (-23.5\%, -22.5\%) at $m = 84$. As reflected in Figure \ref{fig:bmi_results}, there is little change in the 3-year RYGB effect estimate after $m = 20$. Unsurprisingly then, the common effect estimate of -23.1\% reported in Section \ref{sec:bariatric_surgery} comprises a reasonable summary of the treatment effect for a much greater portion of the study period.

A constant MSM was selected for the 3-year weight change outcome comparing SG patients relative to non-surgical patients. Thus, the reported estimate of -16.3\% from Section \ref{sec:bariatric_surgery} already reflects what one might reasonably report, after considering the possibility of variation over calendar time. Non-constant MSMs were selected in other comparisons involving SG patients, however, as 6-month weight change treatment effects for SG patients between $m = 30$ and $m = 84$ improved by 1.0\% [-17.5\% (-17.7\%, -17.2\%) to -18.5\% (-19.0\%, -17.9\%)] compared to non-surgical patients, and by 1.3\% [4.7 (4.0, 5.5)\% to 3.4\% (2.8\%, 4.0\%)] compared to patients undergoing RYGB.

Figure \ref{fig:bmi_results}B presents estimates of $\widehat\chi_{j,m}$ for select comparisons. For a fixed $m$, variation in the vertical direction reflects changes in cross-trial effects when varying the underlying population (captured by $\widehat\gamma_m^2$). Conversely, for a fixed color, variation in the height of points reflects changes in cross-trial effects in a particular population across calendar time (captured by $\widehat\sigma_m^2$). Notably, the greater degree of variation in the vertical direction at fixed points in time (i.e., when varying populations) in the top row of Figure \ref{fig:bmi_results}B suggests that covariate shift plays a larger role in explaining trends across calendar time for comparisons of no surgery to all surgical procedures than to SG alone. This is further reflected by 95\% confidence intervals for $\theta$ in Table \ref{tab:msm_summary}, which reject $H_0^{(\delta)}$ in all comparisons except those between SG and no surgery (with $\delta = 0.05$).

\begin{table}[tb]
    \centering\footnotesize
    \begin{tabular}{|c|c|c|c|c|c|}
        \hline
        \multirow{2}{*}{\textbf{Comparison}} & \multirow{2}{*}{\textbf{Outcome}} & \multirow{2}{*}{\textbf{Minimizing MSM}} & \multicolumn{2}{c|}{\textbf{Selected MSM}} & \multirow{2}{*}{\textbf{$\theta$ 95\% Interval}}\\
        \cline{4-5}
         & & & $c = 0.5$ & $c = 1$ & \\
        \hline
         & 6 Months & Spline (3 Knots) & Cubic & Cubic & (0.684, 0.809)\\
        \cline{2-6}
         & 1 Year & Spline (3 Knots) & Spline (3 Knots) & Cubic & (0.648, 0.767)\\
        \cline{2-6}
         & 2 Years & Spline (2 Knots) & Spline (2 Knots) & Cubic & (0.644, 0.765)\\
        \cline{2-6}
        \multirow{-4}{*}{\centering\arraybackslash Surgery vs. No Surgery} & 3 Years & Spline (3 Knots) & Spline (3 Knots) & Spline (3 Knots) & (0.682, 0.777)\\
        \cline{1-6}
         & 6 Months & Cubic & Cubic & Constant & (0.668, 0.821)\\
        \cline{2-6}
         & 1 Year & Spline (3 Knots) & Cubic & Linear & (0.658, 0.795)\\
        \cline{2-6}
         & 2 Years & Spline (3 Knots) & Spline (3 Knots) & Spline (3 Knots) & (0.647, 0.785)\\
        \cline{2-6}
        \multirow{-4}{*}{\centering\arraybackslash RYGB vs. No Surgery} & 3 Years & Spline (3 Knots) & Spline (3 Knots) & Cubic & (0.670, 0.787)\\
        \cline{1-6}
         & 6 Months & Spline (3 Knots) & Spline (3 Knots) & Spline (3 Knots) & (0.971, 0.989)\\
        \cline{2-6}
         & 1 Year & Constant & Constant & Constant & (0.974, 0.991)\\
        \cline{2-6}
         & 2 Years & Linear & Constant & Constant & (0.980, 0.992)\\
        \cline{2-6}
        \multirow{-4}{*}{\centering\arraybackslash SG vs. No Surgery} & 3 Years & Constant & Constant & Constant & (0.979, 0.992)\\
        \cline{1-6}
         & 6 Months & Spline (3 Knots) & Cubic & Linear & (0.718, 0.807)\\
        \cline{2-6}
         & 1 Year & Constant & Constant & Constant & (0.670, 0.784)\\
        \cline{2-6}
         & 2 Years & Constant & Constant & Constant & (0.666, 0.784)\\
        \cline{2-6}
        \multirow{-4}{*}{\centering\arraybackslash RYGB vs. SG} & 3 Years & Constant & Constant & Constant & (0.688, 0.793)\\
        \hline
    \end{tabular}
    \caption{Summary of MSM selection results and 95\% bootstrapped intervals for $\theta$ ($B = 10,000$ replicates). The minimizing MSM denotes $\widehat\psi^*$ minimizing $\widehat{L}(\widehat\psi)$, while the selected MSMs denote the simplest model whose loss is within $c = 0.5$ and $c = 1$ standard errors ($\widehat{\text{se}}_k$) of the minimizer, respectively.}
    \label{tab:msm_summary}
\end{table}

It is not overly surprising that underlying treatment efficacy for bariatric surgery relative to no surgery appears to have ``declined'' over time, given that the composition of bariatric procedures, specifically the proportion of RYGB and SG, has itself changed over time. Nevertheless, that our method is able to discover this well-explained trend offers evidence of its utility. When making comparisons to non-surgical patients, much of the clinical literature aggregates all surgical procedures under a single label ``bariatric surgery'' \citep{obrien2018microvascular, Harrington2024, benz2024, madenci2024, haneuse2025, fisher2018association}. As such, recognizing that both the meaning and the underlying effectiveness of this aggregated treatment label have changed is particularly relevant. Importantly, while the $m = 84$ estimate of -18.3\% is more reflective of current real-world efficacy of bariatric surgery than the common effect of -19.8\%, differences are attributable to both changing distribution of procedures and patient populations being treated. As such additional standardization efforts may be required to get a truly contemporary estimate, as the DURABLE database only follows patients through 2015.

Finally, results also hint at possible improvements in SG technique and improved weight loss outcomes 6 months post-surgery, though such improvements are not observed for longer-term outcomes. One possible explanation is that any underlying changes in technique have helped patients reach their weight-loss nadir more quickly, but have not influenced peak weight-loss or rebound effects. 

\subsection{Robustness of Clinical Conclusions}
\label{sec:application_robustness}

Several practical considerations bear on the appropriate level of confidence to place in the clinical conclusions drawn from Section \ref{sec:application_results}. We briefly discuss three such considerations here: sensitivity to the choice of threshold $c$ in the MSM selection procedure, the potential influence of missing outcome data, and the possibility of unmeasured/residual confounding.

As seen in Table \ref{tab:msm_summary} and Table S2, clinical conclusions for longer-term outcomes (2- and 3-year weight change) are stable across $c \in \{0.5, 1\}$, with non-constant MSMs selected for comparisons between surgery versus no-surgery and RYGB versus no-surgery; only the complexity of the selected curve changes. Conclusions for 6-month outcomes are somewhat more sensitive to $c$. Under $c = 1$ (i.e., a 1-SE rule), the 6-month RYGB versus no-surgery effect is summarized by a constant MSM, and the 6-month RYGB versus SG effect simplifies from a cubic to a linear trend. By contrast, the non-constant trend for 6-month SG versus no-surgery persists under both choices of $c$ (the constant model is 1.70 standard errors from the minimizer), while SG versus no-surgery comparisons at 1, 2, and 3 years are constant under both.

Missingness in outcome data is most likely to influence clinical conclusions regarding the evolution of treatment effect estimates across calendar time when ascertainment rates vary considerably over time, particularly when this variation is differential by treatment arm. Figure S3 highlights that outcome ascertainment rates for longer-term outcomes are stable over time and nearly identical across treatment arms, suggesting that observed variation in treatment effects for 2- and 3-year weight change outcomes is unlikely to be driven by differential missingness. By contrast, differential missingness by surgical status is more prominent for 6-month weight loss outcomes, with ascertainment rates for surgical patients increasing across earlier trials, potentially contributing to some of the observed treatment effect variation.

Previous studies \citep{Harrington2024, fisher2018association, coleman2022bariatric, obrien2018microvascular} using DURABLE data have examined E-Values \citep{haneuse2019evalue} as a way to quantify the strength of unmeasured confounding that would be required to overturn observed results. Each of these studies has found the strength of association that would be required by unmeasured factors (beyond the dozen plus already adjusted for) to overturn results would be far greater than any known across the literature. Given that confounders chosen for this study mirror those used in prior work, the strength of any remaining unmeasured confounding is unlikely to bias results in a clinically significant manner.

Altogether, we recommend that greater stock be placed in longer-term results (2- and 3-year weight change) while 6-month weight loss results should be interpreted with greater caution, given their greater relative sensitivity to the choice of $c$ and the potential influence of differential outcome missingness in earlier trials.

\if1\blind{\vspace{-0.5cm}}\fi
\section{Discussion}
\label{sec:discussion}

Observational studies leveraging EHR data are inherently subject to many factors beyond the control of investigators, particularly with respect to how treatments are administered and to whom. When a study reports a single causal effect (e.g., ``the'' ATE), that estimate implicitly reflects an average of calendar time-specific effects, which may not themselves be constant over time. Although the appropriate structure of reported effects depends largely on the goals of a given study, the methods in this work are designed to clarify both \textit{how} and \textit{why} causal effects may vary across calendar time. In doing so, the intent is that they will help guide decisions about what to report, particularly when potential changes to underlying treatment efficacy over time are of interest.

One key benefit of the sequential target trial approach considered in this work is that the design directly establishes initiator cohorts (i.e., each with a common time zero), through which calendar time-specific treatment effects can be defined. The proposed methods generalize to other study designs, however,  provided that the study entry period can be discretized into sufficiently narrow windows through which one can define calendar time-specific effects. For example, head-to-head comparisons of RYGB and SG do not strictly require a sequence of target trials because the start of follow-up is defined by the date of surgery, and a (matched) cohort design may be a reasonable alternative \citep{McTigue2020, benz2025}. Under such designs, one might partition the study entry period into small intervals and define contrasts analogous to $\chi_m(P)$ and $\chi_{j,m}(P)$, before applying the ideas introduced in this work. The choice of interval length will be application specific, and depend on both how quickly clinical practice can change as well as the frequency of treatment.

The decomposition framework introduced in Section \ref{sec:decomposing} shares conceptual motivation with Gilbert et al.~\citep{gilbert2025}, who propose a related decomposition to explain heterogeneity in treatment effects across distinct studies or sites. However, the two frameworks differ in important ways. Gilbert et al. decompose between-study heterogeneity into contributions from case-mix differences, mediating variables, and pure effect modification across externally defined, unordered study indices. By contrast our framework is specifically designed for the temporally ordered, calendar time setting, where the goal is to characterize the relative contributions of changes in treatment efficacy and covariate shift occurring within a single EHR-based study. 


In studies where there is no clinical rationale for changes in underlying treatment efficacy (e.g., when treatments are fixed-dose pharmaceutical medications), summarizing variation in cross-trial effects can serve as a useful diagnostic for detecting effect modification over time. For instance, $\theta$ may be particularly informative in pharmacoepidemiologic studies, helping to quantify how treatment effects vary as the characteristics of medication users evolve with broader adoption.

We note that there are natural connections between our work and prior work on violations of the consistency assumption in causal inference \citep{cole2009consistency, VanderWeele2013}. While RYGB and SG can be framed as multiple versions of bariatric surgery, it is perhaps less natural to view each procedure over time as multiple versions of treatment in the usual sense, in part because it is rarely known when treatments differ, and thus which ``version'' of treatment was received. In a sense, our setup implicitly defines trial-specific versions of treatment, i.e., through the symbols $A_m$ and $a_m$. However, we choose to view counterfactual outcomes $Y_m(a_m)$ similarly to those defined and framed in Cole and Frangakis~\cite{cole2009consistency}, $Y(a,k)$, where the additional treatment details $(k)$ required to maintain consistency are entirely captured by trial index ($m$) in our notation. In particular, we adopt the viewpoint that if $a_m = a_j$ but $Y_m(a_m) \neq Y_j(a_j)$, then this reflects changes in the efficacy of a particular treatment $a$ between two time points $m, j$, rather than always calling this phenomenon multiple versions of treatment. This choice is in line with Cole and Frangakis, who note that taking versions of treatment to the extreme precludes generalizability and hampers the utility of a study. For instance, we would find it counterproductive to define separate versions of bariatric surgery for each operating surgeon, or of non-surgical weight management based on prevailing pharmacological or diet and exercise standards.

It's worth noting that outcomes $Y_m$ are ascertained some time after baseline of trial $m$, the counterfactual $Y_m(a_m = 0)$ implicitly averages over versions of non-surgical standard of care between baseline of trial $m$ and the time of outcome ascertainment. This concern does not apply to the surgical counterfactual $Y_m(a_m = 1)$, since surgery is a discrete procedure whose effect on subsequent weight loss is not meaningfully influenced by changes in surgical technique at later time points. The degree to which this affects interpretation of non-surgical counterfactuals depends on how rapidly standard of care evolves. In our motivating study, $\chi_m(P)$ is perhaps most accurately interpreted as the average difference in weight change between undergoing surgery at time $m$ and not undergoing surgery at time $m$, but instead generally following the prevailing standard of care thereafter. Though the implication of $\theta > 0$ would not materially change, quicker changes to standard of care may make it more difficult to tease apart which treatment strategy may be driving changes in treatment effects. We note that non-surgical obesity treatment saw little improvement during our 2005-2015 study period, as poor insurance coverage and high out-of-pocket costs limited anti-obesity medication uptake \citep{Baum2015, Saxon2019}. By contrast, the subsequent rapid rise of GLP-1 receptor agonists represents a context where conflating standard of care across time points may be far more consequential.

Finally, we have assumed throughout that during the time period where a patient is enrolled in Kaiser Permanente insurance, and thus is accessible in the EHR, there is no missing data in components of $O$. In practice, missing covariates may preclude ascertainment of eligibility, or outcomes may be missing even when eligibility is ascertainable. For the purposes of this work, we treat patients with incomplete information at a particular trial as ineligible $(E_m = 0)$. However, we see combining ideas from our framework with recent work on causal inference under missing eligibility \citep{benz2024, benz2025} as a key area of future research. Simultaneously handling possibly missing treatment status and/or outcomes across trials represents another important line of further methodological development.

\singlespacing

\if0\blind{
\section*{Code Availability}
All code for analysis and simulations is made available on GitHub at \url{https://github.com/lbenz730/calendar_time}.

\section*{Acknowledgments}
This work was supported by NIH Grants R01 DK128150-01, R01 HL166324, and F31 DK141237-01
}\fi

\setlength{\bibsep}{0.8pt}
\bibliographystyle{unsrt}
\bibliography{ref}

\end{document}


\title{Supplementary Material for ``A Statistical Framework for Understanding Causal Effects that Vary by Treatment Initiation Time in EHR-based Studies''}

\if0\blind {
\author[1]{Luke Benz}
\author[2]{Rajarshi Mukherjee}
\author[2,3,4]{Rui Wang}
\author[5]{David Arterburn}
\author[6]{Heidi Fischer}
\author[7]{Catherine Lee}
\author[8,9]{Susan M. Shortreed}
\author[10]{Alexander W. Levis*}
\author[2]{Sebastien Haneuse*}
\affil[1]{Global Statistical Sciences,
Eli Lilly \& Company, Indianapolis, IN, USA}
\affil[2]{Department of Biostatistics,
Harvard T.H. Chan School of Public Health, Boston, MA, USA}
\affil[3]{Department of Population Medicine, Harvard Pilgrim Health Care Institute, Boston, MA, USA}
\affil[4]{Department of Population Medicine, Harvard Medical School, Boston, MA, USA}
\affil[5]{Kaiser Permanente Washington Health Research Institute, Seattle, WA, USA}
\affil[6]{Department of Research \& Evaluation, Kaiser Permanente Southern California, Pasadena, CA, USA}
\affil[7]{Department of Epidemiology and Biostatistics, University of California San Francisco, San Francisco, CA, USA}
\affil[8]{Biostatistics Division, Kaiser Permanente Washington Health Research Institute, Seattle, WA, USA}
\affil[9]{Department of Biostatistics, University of Washington School of Public Health, Seattle, WA, USA}
\affil[10]{Department of Biostatistics, Epidemiology and Informatics, University of Pennsylvania, Philadelphia, PA, USA }

\begingroup
\renewcommand{\thefootnote}{}%
\renewcommand{\footnoterule}{}%
\footnotetext{\noindent $^*$ denotes co-last author (AWL and SH)}
\footnotetext{Corresponding Author: Luke Benz (\url{lukesbenz@gmail.com})}
\endgroup

\date{
    \today 
}

\maketitle

}\fi

\if1\blind{

\author{}

\date{
    \today 
}

\vspace{-3in}

\maketitle

}\fi

\maketitle

\tableofcontents

\newpage{}

\section{Additional Data Application Details}
\subsection{Variable Definitions}
\underline{\textbf{Confounders}}:

\begin{itemize}
    \item Baseline Age: age at start of target trial
    \item Sex: \{Male, Female\}
    \item Study Site: \{KP Washington, KP Northern California, KP Southern California\}
    \item Race: \{White, African-American, Other\}
    \item Baseline BMI: BMI value closest to the start of each target trial, up to 1 year prior to the start of target trial
    \item Self-reported smoking status: Most recent self-reported smoking status prior to the start of the target trial
    
    \item Hypertension: Presence of ICD-9 diagnosis code 401.x-405.x up to 1 year prior to start of target trial
    \item Dyslipidemia: Presence of ICD-9 diagnosis code 272-272.2 or 272.3-272.5, up to 1 year prior to start of target trial
    \item Type II Diabetes Mellitus: Any of the following:
    \begin{itemize}
        \item Most recent hemoglobin A1c\% $\geq 6.5\%$, up to 2 years prior to the start of target trial
        \item Most recent fasting blood glucose $\geq 126$ mg/dL, up to 2 years prior to the start of target trial
        \item Active prescription for insulin or oral antiglycemic medication (other than Metformin) at the start of the target trial
        \item Presence of ICD-9 diagnosis code 250.x within 1 year prior to start of target trial and active prescription for Metformin at start of target trial.
    \end{itemize}
    \item Medications
        \begin{itemize}
            \item Insulin: active prescription for insulin at the start of the target trial
            \item Antihypertensives: active prescription for antihypertensive medication at the start of the target trial
            \item Antilipemic: active prescription for antilipemic medication at the start of the target trial  
        \end{itemize}
\end{itemize}

\noindent\underline{\textbf{Outcomes}}:\\
\noindent Let BMI$_m$ denote baseline BMI at trial $m$. Relative weight changes since baseline ($t = 0$) and at some follow-up time $t$ (e.g., one of 6 months, 1 year, 2 years, 3 years) were computed as 

$$
Y_m = \frac{\text{BMI}_{m+t} -\text{BMI}_m}{\text{BMI}_m}
$$

\noindent where BMI$_{m+t}$ was ascertained using the window method of \cite{thaweethai2021robust}. To compute BMI at 3-years, for example, the following procedure was used:

\begin{enumerate}
    \item Let $\text{Date}_{m+36}$ denote the date exactly 3 years after the start of target trial $m$.
    \item Consider a window $\pm$ 6 months around $\text{Date}_{m+36}$, $[\text{Date}_{m+30}, \text{Date}_{m+42}]$
    \item If a patient has BMI measures in both $[\text{Date}_{m+30}, \text{Date}_{m+36})$ and $(\text{Date}_{m+36}, \text{Date}_{m+42}]$, fit a linear regression to all BMI measures $[\text{Date}_{m+30}, \text{Date}_{m+42}]$ and take $\text{BMI}_{m+36}$ to be the interpolated (e.g., fitted) value at $t = 36$ months.
    \item If a patient only has BMI measures in $[\text{Date}_{m+30}, \text{Date}_{m+36})$ or $(\text{Date}_{m+36}, \text{Date}_{m+42}]$ but not both, take $\text{BMI}_{m+36}$ to be the measure recorded closest to $\text{Date}_{m+36}$.
    \item If a patient had no BMI measures in $[\text{Date}_{m+30}, \text{Date}_{m+42}]$, take $\text{BMI}_{m+36}$ to be missing. 
\end{enumerate}

\noindent Practically, we exclude patients with missing outcomes by setting $E_m = 0$ if $Y_m = \texttt{NA}$. We leave more comprehensive treatment of missing outcomes within our framework as an extension for future work. 

Outcomes for additional time points were computed in a similar manner using the following window lengths:

\begin{enumerate}
    \item BMI at 6 Months ($\pm$ 2 month window)
    \item BMI at 1 Year ($\pm$ 3 month window)
    \item BMI at 2 Years ($\pm$ 6 month window)
\end{enumerate}

\subsection{Analytic Methods}
All pooled models were estimated using random forests (\texttt{ranger}) \citep{ranger}. We opted for \texttt{ranger} rather than \texttt{SuperLearner} \citep{superlearner} (as in simulations) due to computational speed in light of the size of pooled data $\mathcal{D}$ (several million eligible patient trials). In contrast to other tree-based learners like XGBoost or LightGBM, random forests do not fit trees sequentially and thus can be heavily parallelized to reduce computational burden. Other component learners may not be optimized for datasets that large. Perhaps more importantly, employing Super Learner would require a 3rd layer of cross-fitting in Algorithm 1, again computationally burdensome given $|\mathcal{D}|$. We ultimately feel like the \texttt{\{ranger\}} implementation of random forest \citep{ranger} strikes the appropriate balance between model flexibility and computational efficiency during Stage 1 of estimation. In particular, the following configurations were used.

\begin{itemize}
     \item Outcome Model $\tilde \mu$ (stratified by treatment status): 
     \begin{itemize}
         \item \texttt{num.trees} = 500
         \item \texttt{max.depth} = 10
     \end{itemize}
     \item Treatment Model $\tilde \pi$: 
     \begin{itemize}
         \item \texttt{num.trees} = 500
         \item \texttt{max.depth} = 2
     \end{itemize}
     \item Transport Model $\tilde \xi$: 
     \begin{itemize}
         \item \texttt{num.trees} = 500
         \item \texttt{max.depth} = 10
     \end{itemize}
\end{itemize}

\noindent Inverse probability of treatment weights were truncated \citep{hernan2024} at the 99\% quantile within each treatment arm. Similarly, ratios $\frac{P(T = j~|~\bm L, E = 1)}{P(T = m~|~\bm L, E = 1)}$ were also truncated at the 99\% quantile across all $(j,m)$ pairs.

Candidate marginal structural models for projection models were as follows:
\begin{itemize}
    \item Constant $\psi_m(m;\bm \beta) = \beta$
    \item Linear $\psi_m(m;\bm \beta) = \beta_0 + \beta_1m$
    \item Cubic $\psi_m(m;\bm \beta) = \beta_0 + \beta_1m + \beta_2m^2 + \beta_3m^3$
    \item Spline (2 knots)
    \begin{itemize}
        \item Knots at $m \in \{24, 48\}$ for Surgery versus No Surgery and RYGB versus No Surgery
        \item Knots at $m \in \{48, 72\}$ for SG versus No Surgery and RYGB versus SG
    \end{itemize}
    \item Spline (3 knots)
    \begin{itemize}
        \item Knots at $m \in \{12, 36, 60\}$ for Surgery versus No Surgery and RYGB versus No Surgery
        \item Knots at $m \in \{40, 56, 72\}$ for SG versus No Surgery and RYGB versus SG
    \end{itemize}
\end{itemize}

\section{Additional Application Figures and Tables}
\subsection{Population Summary Table}

\begin{table}[H]
\centering \footnotesize
\begin{tabular}{|l|c|c|}
\hline

 & \textbf{Bariatric Surgery} & \textbf{No Bariatric Surgery} \\
 & (N Subjects=17,596) & (N Subjects=780,292$^{a}$) \\
 & (N Subject Trials=17,596) & (N Subject Trials=18,646,825) \\
\hline
\hline
\textbf{Baseline Age} & & \\
Mean (SD) & 46.4 (10.9) & 52.9 (13.3) \\
\hline
\textbf{Baseline BMI} & & \\
Mean (SD) & 44.5 (6.99) & 40.5 (5.35) \\
\hline
\textbf{Sex} & & \\
Female & 14,349 (81.5\%) & 11,561,114 (62.0\%) \\
Male & 3,247 (18.5\%) & 7,085,711 (38.0\%) \\
\hline
\textbf{Race} & & \\
White & 11,579 (65.8\%) & 11,465,962 (61.5\%) \\
African-American & 2,902 (16.5\%) & 2,750,676 (14.8\%) \\
Other & 3,115 (17.7\%) & 4,430,187 (23.8\%) \\
\hline
\textbf{Kaiser Permanente Site} & & \\
KP Washington & 1,046 (5.9\%) & 1,257,354 (6.7\%) \\
KP Northern California & 5,560 (31.6\%) & 8,669,519 (46.5\%) \\
KP Southern California & 10,990 (62.5\%) & 8,719,952 (46.8\%) \\
\hline
\textbf{Hypertension} & & \\
Yes & 10,433 (59.3\%) & 10,692,639 (57.3\%) \\
No & 7,163 (40.7\%) & 7,954,186 (42.7\%) \\
\hline
\textbf{Antihypertensive Medication Usage} & & \\
Yes & 8,581 (48.8\%) & 10,626,016 (57.0\%) \\
No & 9,015 (51.2\%) & 8,020,809 (43.0\%) \\
\hline
\textbf{Dyslipidemia} & & \\
Yes & 8,203 (46.6\%) & 8,680,596 (46.6\%) \\
No & 9,393 (53.4\%) & 9,966,229 (53.4\%) \\
\hline
\textbf{Antilipemic Medication Usage} & & \\
Yes & 4,787 (27.2\%) & 6,830,635 (36.6\%) \\
No & 12,809 (72.8\%) & 11,816,190 (63.4\%) \\
\hline
\textbf{Type II Diabetes Mellitus} & & \\
Yes & 5,914 (33.6\%) & 6,714,248 (36.0\%) \\
No & 11,682 (66.4\%) & 11,932,577 (64.0\%) \\
\hline
\textbf{Insulin Use} & & \\
Yes & 1,419 (8.1\%) & 1,574,807 (8.4\%) \\
No & 16,177 (91.9\%) & 17,072,018 (91.6\%) \\
\hline
\textbf{Self-Reported Smoking Status} & & \\
Current & 255 (1.4\%) & 1,523,825 (8.2\%) \\
Former & 6,017 (34.2\%) & 5,138,944 (27.6\%) \\
Never & 10,689 (60.7\%) & 11,462,483 (61.5\%) \\
No Self Report & 635 (3.6\%) & 521,573 (2.8\%) \\
\hline
\textbf{Bariatric Surgery Type} & & \\
RYGB & 11,659 (66.3\%) & --- \\
SG & 4,943 (28.1\%) & --- \\
Other & 994 (5.6\%) & --- \\
\hline
\end{tabular}

\vspace{2mm}
\par\noindent\footnotesize $^{a}$ Subjects contributing at least one non-surgical subject-trial, which may include patients who later underwent bariatric surgery. The 763,016 non-surgical patients reported in Section 2.2 of the main text are those who never underwent surgery.
\caption{Baseline Characteristics by Bariatric Surgery Status}
\end{table}

\subsection{Non-Surgical Patient Characteristics in DURABLE}
\begin{figure}[H]
    \centering
    \includegraphics[width=\textwidth]{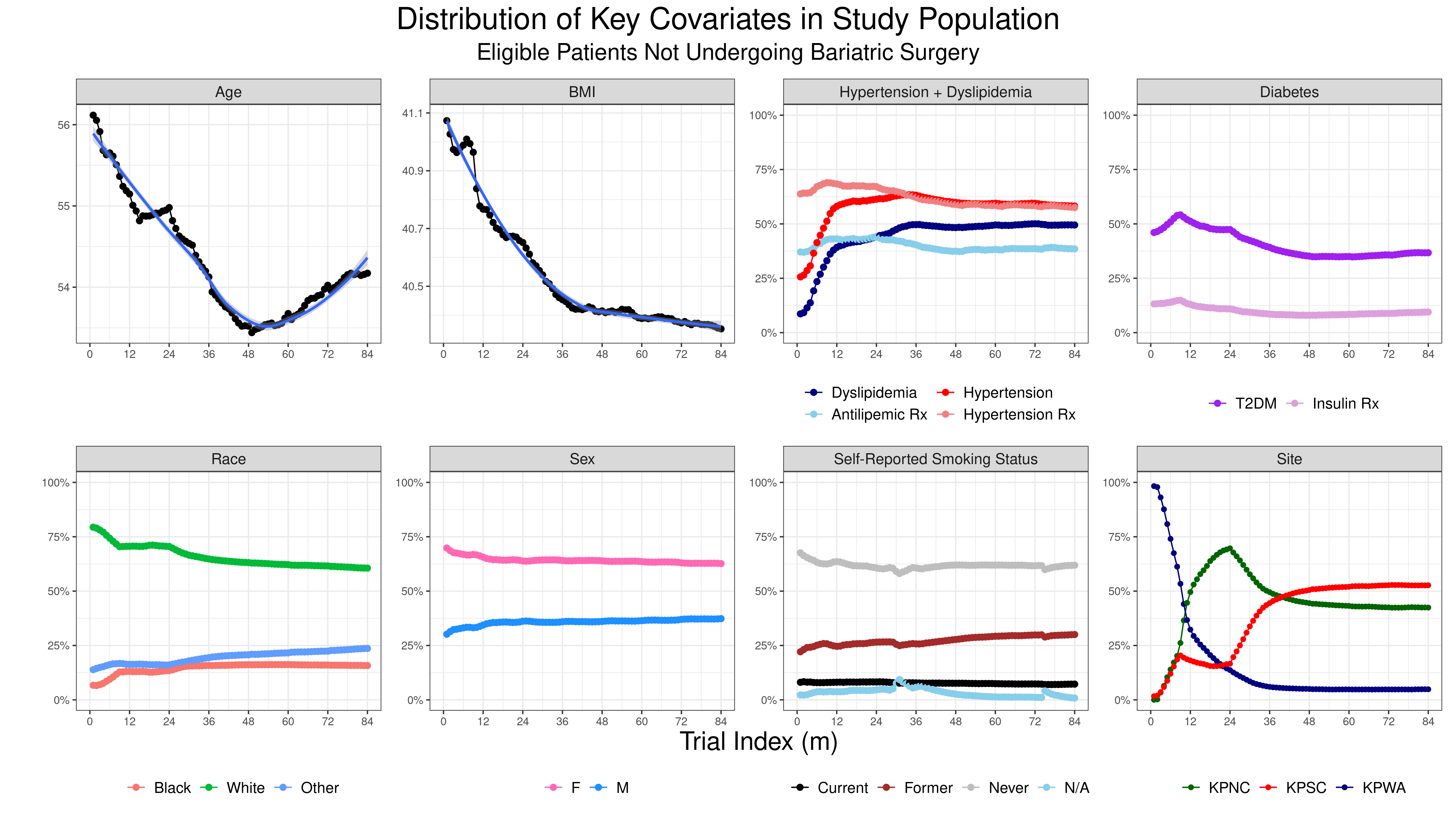}
    \caption{Distribution of patient characteristics in DURABLE electronic health record database over calendar time for eligible patients not undergoing bariatric surgery between 2005-2011. Covariate mean values (continuous covariates) or frequency (binary/categorical covariates) are plotted by month of surgery. $m = 1$ corresponds to January 2005 and $m = 84$ corresponds to December 2011.}
    \label{fig:nosurg_dist}
\end{figure}

\subsection{Number of Patients Meeting Trial Eligibility Over Time}
\begin{figure}[H]
    \centering
    \includegraphics[width=\textwidth]{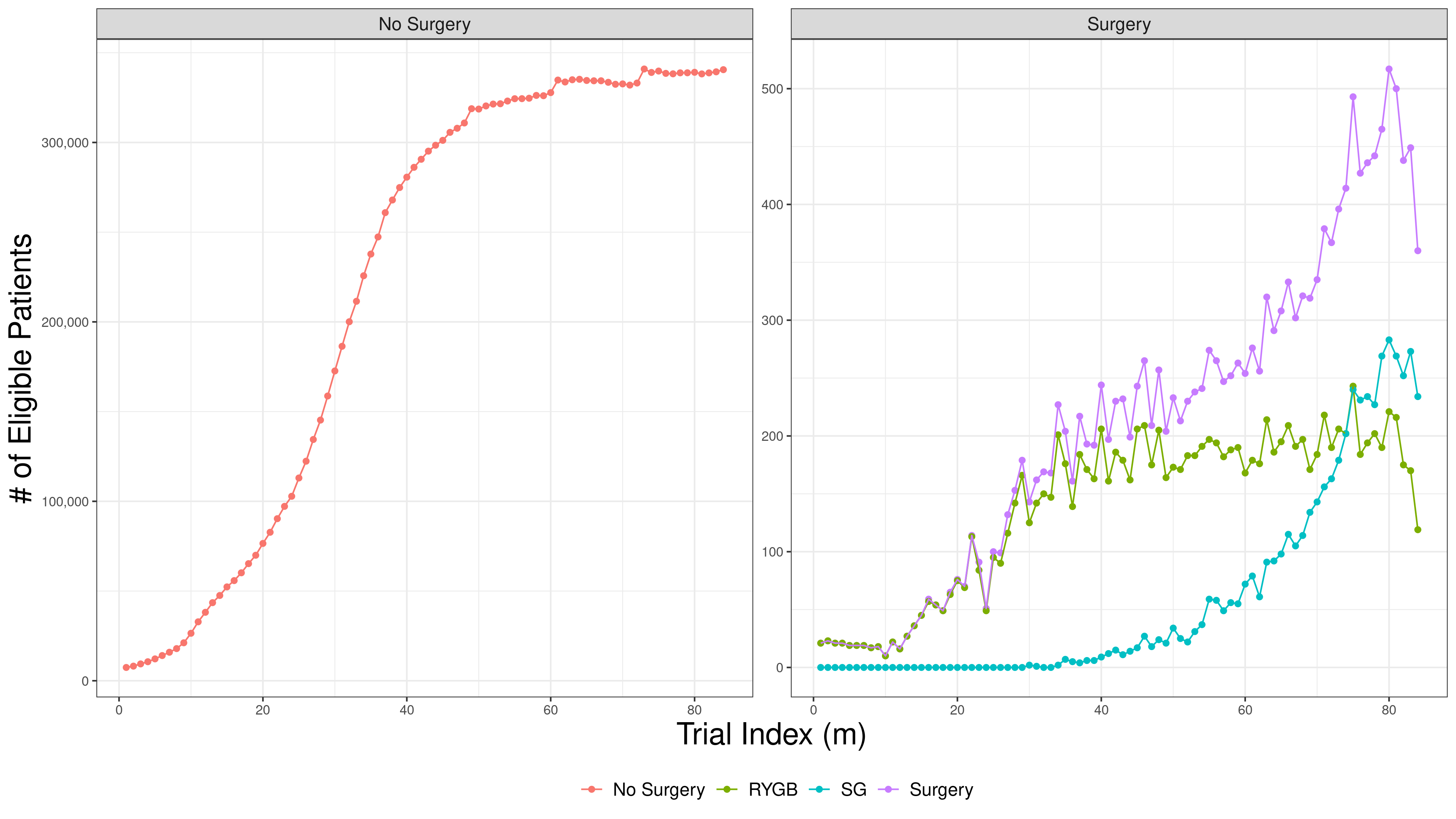}
    \caption{Number of patients meeting eligibility criteria in each target trial, by treatment status. Note that in addition to RYGB and SG procedures, the surgery group includes a small number of adjustable gastric banding procedures.}
    \label{fig:population_size}
\end{figure}

\subsection{Availability of Outcome Data}

\begin{figure}
    \centering
    \includegraphics[width=\textwidth]{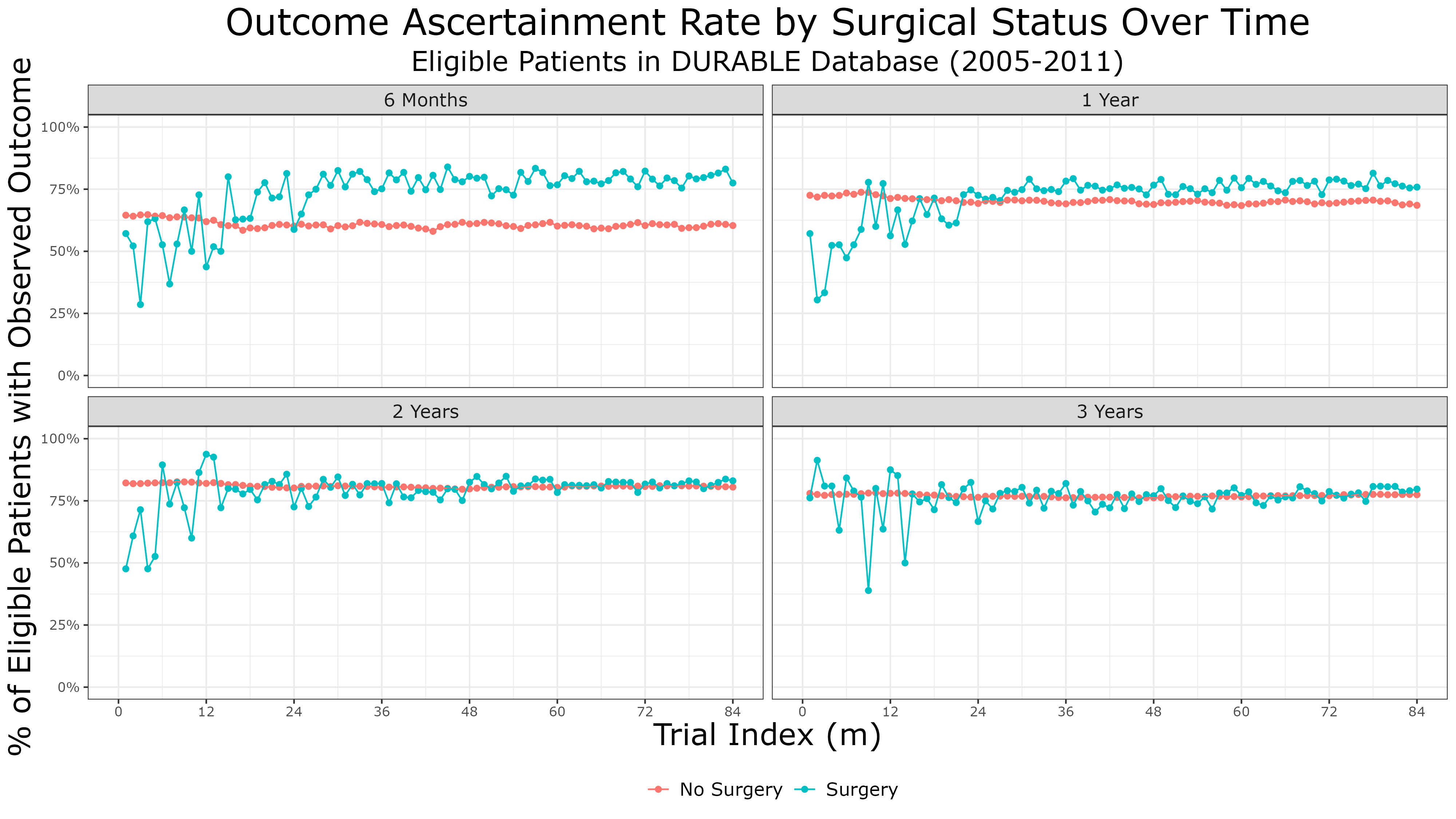}
    \caption{Outcome ascertainment rate over time among eligible patients, by surgical status.}
    \label{fig:outcome_missing}
\end{figure}

Figure \ref{fig:outcome_missing} displays the outcome ascertainment rate by surgical status over calendar time among eligible patients in the DURABLE database. Ascertainment rates for non-surgical patients are stable across all four follow-up timepoints, while ascertainment rates for surgical patients are slightly more volatile in earlier trials before stabilizing. Missingness in outcome data is most likely to influence clinical conclusions regarding the evolution of treatment effect estimates across calendar time when ascertainment rates vary considerably over time, particularly when this variation is differential by treatment arm. Longer-term outcomes (2- and 3-year weight change) do not display meaningful differential missingness by surgical status, while differential missingness is more prominent for 6-month weight loss, with ascertainment rates in the surgical arm increasing over trials $m \in \{1, \dots, 24\}$. Altogether, this suggests that observed variation in treatment effects for longer-term outcomes is unlikely to be driven by missing outcome data, while differential missingness in 6-month weight loss outcomes could explain some of the treatment effect variation observed across calendar time for that outcome. Nevertheless, analysts facing more severe or systematically differential outcome missingness than observed here are encouraged to consider methods specifically designed for missing outcome data in EHR-based studies \citep{levis2024double, sun2025quantile}.

\subsection{MSM Summary Table}
\begin{table}[H]
    \centering \scriptsize
   \begin{tabular}[t]{|Sc|Sc|Sc|Sc|Sc|Sc|Sc|}
        \hline
        \textbf{Comparison} & \textbf{Outcome} & \textbf{Constant} & \textbf{Linear} & \textbf{Cubic} & \textbf{Spline (2 Knots)} & \textbf{Spline (3 Knots)}\\
        \hline
          & 6 Months & -3.518 (2.61) & -3.525 (1.10) & -3.527 (0.08) & -3.527 (0.79) & \textbf{-3.527 (0.00)}\\
        \cline{2-7}
          & 1 Year & -5.251 (3.39) & -5.267 (1.68) & -5.273 (0.50) & -5.273 (1.01) & \textbf{-5.274 (0.00)}\\
        \cline{2-7}
          & 2 Years & -4.552 (3.43) & -4.573 (1.46) & -4.577 (0.54) & \textbf{-4.577 (0.00)} & -4.576 (0.67)\\
        \cline{2-7}
        \multirow{-4}{*}{\centering\arraybackslash Surgery vs. No Surgery} & 3 Years & -3.625 (3.60) & -3.659 (2.11) & -3.675 (1.24) & -3.675 (1.34) & \textbf{-3.678 (0.00)}\\
        \cline{1-7}
          & 6 Months & -4.222 (0.88) & -4.223 (0.66) & \textbf{-4.223 (0.00)} & -4.223 (0.19) & -4.223 (0.10)\\
        \cline{2-7}
          & 1 Year & -6.474 (1.19) & -6.476 (0.95) & -6.478 (0.39) & -6.479 (0.23) & \textbf{-6.479 (0.00)}\\
        \cline{2-7}
          & 2 Years & -5.791 (1.49) & -5.792 (1.45) & -5.794 (1.26) & -5.796 (1.14) & \textbf{-5.800 (0.00)}\\
        \cline{2-7}
        \multirow{-4}{*}{\centering\arraybackslash RYGB vs. No Surgery} & 3 Years & -4.645 (1.48) & -4.647 (1.38) & -4.655 (0.67) & -4.656 (0.57) & \textbf{-4.659 (0.00)}\\
        \cline{1-7}
          & 6 Months & -1.813 (1.70) & -1.813 (2.62) & -1.814 (1.89) & -1.814 (1.96) & \textbf{-1.814 (0.00)}\\
        \cline{2-7}
          & 1 Year & \textbf{-2.490 (0.00)} & -2.486 (9.12) & -2.486 (8.25) & -2.486 (8.59) & -2.486 (7.27)\\
        \cline{2-7}
          & 2 Years & -2.020 (0.12) & \textbf{-2.020 (0.00)} & -2.020 (1.61) & -2.020 (1.48) & -2.020 (0.21)\\
        \cline{2-7}
        \multirow{-4}{*}{\centering\arraybackslash SG vs. No Surgery} & 3 Years & \textbf{-1.466 (0.00)} & -1.463 (4.64) & -1.464 (2.79) & -1.464 (2.99) & -1.465 (2.13)\\
        \cline{1-7}
          & 6 Months & -0.101 (1.27) & -0.101 (0.74) & -0.101 (0.36) & -0.101 (0.40) & \textbf{-0.102 (0.00)}\\
        \cline{2-7}
          & 1 Year & \textbf{-0.220 (0.00)} & -0.217 (3.04) & -0.217 (2.52) & -0.217 (2.36) & -0.217 (1.93)\\
        \cline{2-7}
          & 2 Years & \textbf{-0.288 (0.00)} & -0.283 (4.74) & -0.281 (5.14) & -0.281 (5.17) & -0.281 (4.83)\\
        \cline{2-7}
        \multirow{-4}{*}{\centering\arraybackslash RYGB vs. SG} & 3 Years & \textbf{-0.298 (0.00)} & -0.297 (2.74) & -0.296 (2.38) & -0.296 (2.13) & -0.297 (1.24)\\
        \hline
\end{tabular}
    \caption{Values of the loss function $\widehat L(\widehat\psi_k)$ for each candidate MSM. Cells in bold denote the MSM $\widehat\psi^*$ which minimizes $\widehat L(\widehat\psi_k)$ in each setting. Values in parentheses denote the number of standard errors, $\widehat{\text{se}}_k$, by which $\widehat L(\widehat\psi_k)$ exceeds $\widehat L(\widehat\psi^*)$.}
    \label{tab:loss_fx}
\end{table}

\subsection{Trial Specific Estimates of Variability in Cross-Trial Effects}

\begin{figure}[H]
    \centering
    \includegraphics[width=\textwidth]{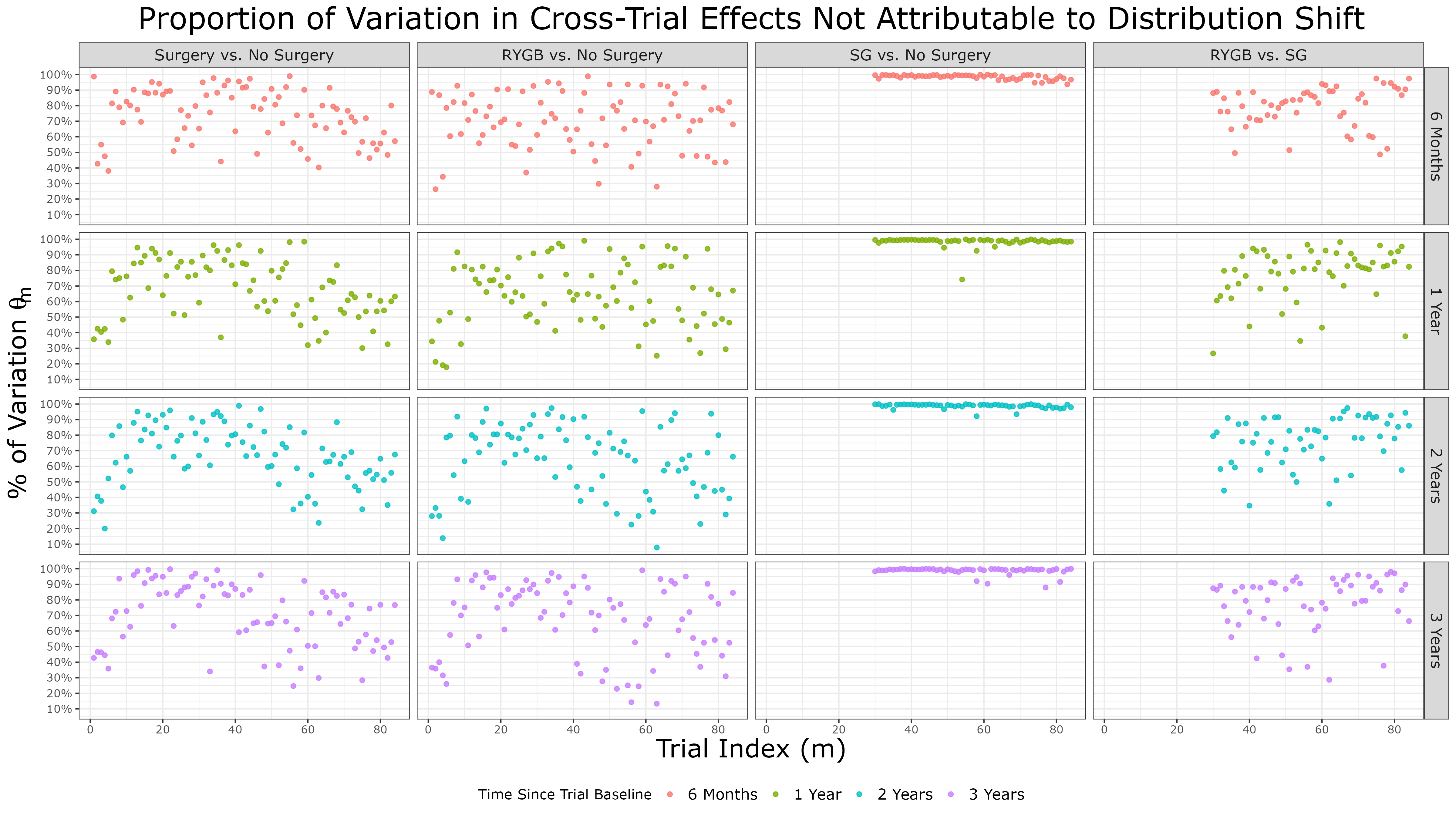}
    \caption{$\widehat\theta_m$ across calendar time}
    \label{fig:theta_m}
\end{figure}

\section{Proofs}

\subsection{Influence Function of \texorpdfstring{$\chi_m(P)$}{Chi}}

\begin{lemma}\label{lemma:VM-mm} For $m\in\{1, ..., M\}$, $\chi_m(P)$ satisfies the von Mises expansion 
$$
\chi_m(\bar P) - \chi_m(P)  = -\int \dot\chi_m^*(o; \bar P)dP(o) + R_m(\bar P, P)
$$
with remainder term, with inputs omitted for brevity, is given by
$$
\begin{aligned}
R_m(\bar P, P) &= \mathbb{E}_P\biggr[\frac{E_m}{\bar P(E_m = 1)}\biggr\{\biggr(1 - \frac{\pi_m}{\bar \pi_m}\biggr)\Bigr(\bar \mu^{(1)}_{m} - \mu_m^{(1)}\Bigr) - \biggr(1 - \frac{1-\pi_m}{1-\bar \pi_m}\biggr)\Bigr(\bar \mu^{(0)}_{m} - \mu_m^{(0)}\Bigr)\biggr\}\biggr] + \\ 
&\quad\quad\quad\Bigr(\chi_m(\bar P) - \chi_m(P)\Bigr)\biggr(1 - \frac{P(E_m = 1)}{\bar P(E_m = 1)}\biggr)
\end{aligned}
$$
\end{lemma}

\begin{proof}
$$
\begin{aligned}
R_m(\bar P, P) &= \chi_m(\bar P) - \chi_m(P) + \int\dot{\chi}^*_m(o; \bar P)dP(o) \\
&= \chi_m(\bar P) - \chi_m(P) + \mathbb{E}_P\biggr[\frac{\mathds{1}(E_m = 1)}{\bar P(E_m = 1)}\biggr\{\bar\mu_m(1,\bm L_m) - \bar\mu_m(0, \bm L_m) - \chi_m(\bar P) + \\
&~~~~~~~~~~~~~~~~~~~~~~~~~~~~~~~~~~\biggr(\frac{A_m}{\bar\pi_m(\bm L_m)} - \frac{1-A_m}{1-\bar\pi_m(\bm L_m)}\biggr)\bigr(Y_m - \bar\mu_m(A_m, \bm L_m)\bigr)\biggr\} \biggr] \\
&=\mathbb{E}_P\biggr[\frac{\mathds{1}(E_m = 1)}{\bar P(E_m = 1)}\biggr\{\bar\mu_m(1,\bm L_m) - \bar\mu_m(0, \bm L_m) - \chi_m(P) + \biggr(\frac{A_m}{\bar\pi_m(\bm L_m)} - \frac{1-A_m}{1-\bar\pi_m(\bm L_m)}\biggr)\bigr(Y_m - \bar\mu_m(A_m, \bm L_m)\bigr)\biggr\} \biggr]\\
&\quad\quad\quad+\Bigr(\chi_m(\bar P) - \chi_m(P)\Bigr)\biggr(1 - \frac{P(E_m = 1)}{\bar P(E_m = 1)}\biggr) \\
&=\mathbb{E}_P\biggr[\frac{\mathds{1}(E_m = 1)}{\bar P(E_m = 1)}\biggr\{\bar\mu_m(1,\bm L_m) - \bar\mu_m(0, \bm L_m) - \bigr(\mu_m(1,\bm L_m) - \mu_m(0, \bm L_m) \bigr) \\
&\quad\quad\quad+\frac{\pi_m(\bm L_m)}{\bar\pi_m(\bm L_m)}\bigr(\mu_m(1, \bm L_m) - \bar\mu_m(1, \bm L_m)\bigr) - \frac{1-\pi_m(\bm L_m)}{1-\bar\pi_m(\bm L_m)}\bigr(\mu_m(0, \bm L_m) - \bar\mu_m(0, \bm L_m)\bigr)\biggr\} \biggr] \\
&\quad\quad\quad+\Bigr(\chi_m(\bar P) - \chi_m(P)\Bigr)\biggr(1 - \frac{P(E_m = 1)}{\bar P(E_m = 1)}\biggr) \\
\end{aligned}
$$
\noindent The last line follows from applying the $g-$formula identification result for $\chi_m(P)$, and applying iterated expectation twice, first on $\bm L_m, A_m$, and then just on $\bm L_m$. Combining terms and omitting inputs for brevity, we are left with the desired result.

$$
\begin{aligned}
R_m(\bar P, P) &= \mathbb{E}_P\biggr[\frac{E_m}{\bar P(E_m = 1)}\biggr\{\biggr(1 - \frac{\pi_m}{\bar \pi_m}\biggr)\Bigr(\bar \mu^{(1)}_{m} - \mu_m^{(1)}\Bigr) - \biggr(1 - \frac{1-\pi_m}{1-\bar \pi_m}\biggr)\Bigr(\bar \mu^{(0)}_{m} - \mu_m^{(0)}\Bigr)\biggr\}\biggr] + \\ 
&\quad\quad\quad\Bigr(\chi_m(\bar P) - \chi_m(P)\Bigr)\biggr(1 - \frac{P(E_m = 1)}{\bar P(E_m = 1)}\biggr)
\end{aligned}
$$

By the von Mises expansion in Lemma~\ref{lemma:VM-mm} and an application of Lemma 2 of \cite{kennedy2023density}, it follows that $\dot{\chi}_{m}^*(O; P)$ is the nonparametric influence function of $\chi_{m}(P)$ at $P$.
\end{proof}

\subsection{Influence Function of \texorpdfstring{$\bm\beta(P)$}{Beta} (Proof of Theorem 1)}

Implicit in Theorem 1 is that $\bm\beta(P)$ is the unique minimizer of the projection objective, lying in the interior of the parameter space, and that $\bm V$ is nonsingular. For MSMs linear in $\bm\beta$, $\psi(m;\bm\beta)=\bm x_m^\top\bm\beta$, which includes every candidate model considered in this work, the projection objective is a weighted least squares criterion with Hessian proportional to $\sum_{m} w(m)P(E_m=1)\bm x_m\bm x_m^\top$. Full rank of this matrix, which holds trivially in our application ($M\in\{55,84\}$, $k\le 5$) and simulations ($M=36$), implies both conditions: $\bm\beta(P)$ is the unique solution of the estimating equation, and, because $\nabla^2_{\bm\beta\bm\beta}\psi \equiv \bm 0$, $\bm V = -\sum_m w(m)P(E_m=1)\bm x_m\bm x_m^\top$ is nonsingular. For MSMs nonlinear in $\bm\beta$, these conditions must be assumed directly. Finite variance of $\dot{\bm\beta}^*(O;P)$ follows from Assumption A2 and finite second moments of $Y_m$ among eligible patients.

The characterizing equation
\[0 = \sum_{m = 1}^M w(m) P(E_m = 1)\left.\nabla_{\bm\beta}\psi(m;\bm\beta)\right|_{\bm \beta = \bm \beta(P)}[\chi_m(P) - \psi(m;\bm\beta(P))]\]
is obtained by differentiating the loss function which defines $\bm \beta(P)$---this is permitted as we have assumed that $\psi(m; \bm \beta)$ is differentiable in $\bm \beta$.
We will use this equation to deduce the influence function of $\boldsymbol{\beta}(P)$. Recall from \cite{bickel1993efficient} and \cite{vandervaart2002} that the nonparametric influence function $\dot{\bm \beta}^*(O;P)$ is the unique mean-zero finite variance function satisfying pathwise differentiability:
    \[\left. \frac{d}{d\epsilon}\boldsymbol{\beta}(P_{\epsilon})\right|_{\epsilon = 0} = \mathbb{E}_P\left(\dot{\bm \beta}^*(O;P) u(O)\right),\]
    for any regular parametric submodel $\{P_{\epsilon}: \epsilon \in [0,1)\}$ such that $P_0 \equiv P$ with score function $u(O) = \left. \frac{d}{d\epsilon} \log{d P_{\epsilon}} \right|_{\epsilon = 0}$. Choosing such a regular parametric submodel and differentiating the estimating equation, we have
    \begin{align*}
        0
        &= \left. \frac{d}{d\epsilon}\sum_{m = 1}^M w(m) P_{\epsilon}(E_m = 1)\left.\nabla_{\bm\beta}\psi(m;\bm\beta)\right|_{\bm \beta = \bm \beta(P)}[\chi_m(P) - \psi(m;\bm\beta(P))]\right|_{\epsilon = 0}\\
        & \quad \quad + 
        \left. \frac{d}{d\epsilon}\sum_{m = 1}^M w(m) P(E_m = 1)\left.\nabla_{\bm\beta}\psi(m;\bm\beta)\right|_{\bm \beta = \bm \beta(P_{\epsilon})}[\chi_m(P) - \psi(m;\bm\beta(P_{\epsilon}))]\right|_{\epsilon = 0}\\
        & \quad \quad + \left. \frac{d}{d\epsilon}\sum_{m = 1}^M w(m) P(E_m = 1)\left.\nabla_{\bm\beta}\psi(m;\bm\beta)\right|_{\bm \beta = \bm \beta(P)}[\chi_m(P_{\epsilon}) - \psi(m;\bm\beta(P))]\right|_{\epsilon = 0},
    \end{align*}
    by properties of the total derivative. The first summand is simply
    \[\sum_{m = 1}^M w(m) \mathbb{E}_P(E_m u(O))\left.\nabla_{\bm\beta}\psi(m;\bm\beta)\right|_{\bm \beta = \bm \beta(P)}[\chi_m(P) - \psi(m;\bm\beta(P))],\]
    where we interchanged the derivative and integral, and used the fact that $\frac{d}{dx}\log{f(x)} = \frac{\frac{d}{dx}f(x)}{f(x)}$ for any differentiable function $f$.
    The second summand is
    $\bm V \left.\frac{d}{d\epsilon}\bm \beta(P_{\epsilon}) \right|_{\epsilon = 0}$
    by the chain rule, where
    \begin{equation*}
    \begin{aligned}
    \bm V = \sum_{m = 1}^M w(m) P(E_m = 1)\Bigr\{\nabla_{\bm \beta \bm\beta}^2 \psi(m; \bm \beta)|_{\bm\beta = \bm\beta(P)}\big[\chi_m(P) - \psi\bigr(m; \bm \beta(P)\bigr)\big] \\ - \nabla_{\bm \beta} \psi(m; \bm \beta)|_{\bm\beta = \bm\beta(P)} [\nabla_{\bm \beta} \psi(m; \bm \beta)|_{\bm\beta = \bm\beta(P)}]^ T\Bigr\}
    \end{aligned}
    \end{equation*} 
    as defined in the main text.
    The third summand is
    \[\sum_{m = 1}^M w(m) P(E_m = 1)\left.\nabla_{\bm\beta}\psi(m;\bm\beta)\right|_{\bm \beta = \bm \beta(P)}\mathbb{E}_P(\dot{\chi}_m^*(O; P)u(O))\]
    by the defining property of $\dot{\chi}_m^*(O; P)$. Combining terms and solving for
$\left.\frac{d}{d\epsilon}\bm\beta(P_\epsilon)\right|_{\epsilon=0}$
(possible since $\bm V$ is nonsingular), we find
    \begin{align*}
      \left.\frac{d}{d\epsilon}\bm \beta(P_{\epsilon}) \right|_{\epsilon = 0} &= -\bm V^{-1}  \mathbb{E}_P\bigg(\sum_{m = 1}^M w(m) \left.\nabla_{\bm\beta}\psi(m;\bm\beta)\right|_{\bm \beta = \bm \beta(P)} \big\{ E_m\big[\chi_m(P) - \psi(m; \bm \beta(P))\big] \\
      & \quad \quad \quad \quad \quad \quad \quad \quad + P[E_m = 1] \dot{\chi}_m^*(O;P) \big\} u(O) \bigg).
    \end{align*}
    Thus, after canceling terms, we have shown that
\begin{align*}
    \dot{\bm \beta}^*(O;P) 
    &= - \bm V^{-1 }\sum_{m = 1}^M w(m)\left.\nabla_{\bm \beta}  \psi(m; \bm \beta)\right|_{\bm \beta = \bm \beta(P)} E_m \Bigr\{\mu_m(1, \bm L_m) - \mu_m(0, \bm L_m) \\
    & \quad \quad \quad \quad \quad + \Bigr(\frac{A_m}{\pi_m(\bm L_m)} - \frac{1-A_m}{1-\pi_m(\bm L_m)}\Bigr)\big(Y_m - \mu_m(A_m, \bm L_m)\big) - \psi\bigr(m;\bm\beta(P)\bigr)\Bigr\}.
\end{align*}

\subsection{Asymptotic Behavior of \texorpdfstring{$\widehat \chi_{m}$ and $\widehat{\bm\beta}$}{Chi, Beta} (Proof of Theorem 2)}

The asymptotic behavior of $\widehat{\chi}_m$ is a special case of Theorem \ref{thm:ct_asymptotics}, proved in Appendix Section~\ref{app:asymp_jm}; note that $\xi_{m,m} \equiv 1$ and does not need to be estimated. We thus focus on $\widehat{\boldsymbol{\beta}}$. By definition, $\boldsymbol{\beta}(P)$ solves $\mathbb{E}_P\left[\varphi(O; \boldsymbol{\beta}, \eta(P))\right] = 0$ for $\boldsymbol{\beta}$, where for any candidate nuisance functions $\overline{\eta} = \{(\overline{\mu}_m, \overline{\pi}_m)\}_{m = 1}^M$,
\begin{align*}
    \varphi(O; \boldsymbol{\beta}, \overline{\eta}) 
    &= \sum_{m = 1}^M E_m w(m)\nabla_{\bm \beta}  \psi(m; \bm \beta)\Bigr\{\overline{\mu}_m(1, \bm L_m) - \overline{\mu}_m(0, \bm L_m) \\
    & \quad \quad \quad \quad \quad + \Bigr(\frac{A_m}{\overline{\pi}_m(\bm L_m)} - \frac{1-A_m}{1-\overline{\pi}_m(\bm L_m)}\Bigr)\big(Y_m - \overline{\mu}_m(A_m, \bm L_m)\big) - \psi\bigr(m;\bm\beta\bigr)\Bigr\}.
\end{align*}
Similarly, $\widehat{\boldsymbol{\beta}}$ is the solution to the empirical estimating equation $\mathbb{P}_n\left[\varphi(O; \boldsymbol{\beta}, \widehat{\eta})\right] = 0$ in $\boldsymbol{\beta}$. We make the following assumptions:
\begin{assumption}\label{as:S1}
For each $m \in \{1, \ldots, M\}$, the following hold:
\begin{enumerate}[(i)]
    \item $\lVert \widehat{\pi}_m - \pi_m\rVert_{(m)} + \sum_{a = 0}^1\lVert \widehat{\mu}_m(a, \, \cdot\,) - \mu_m(a, \, \cdot\,)\rVert_{(m)} = o_P(1)$,
    \item $\epsilon \leq \widehat{\pi}_m(\bm L_m), \pi(\bm L_m) \leq 1 - \epsilon$ and $|Y_m|, |\widehat{\mu}_m(a, \bm L_m)| \leq \epsilon^{-1}$ almost surely when $E_m = 1$,
\end{enumerate}
and, moreover, $\widehat{\bm \beta} \overset{P}{\to} \bm \beta(P)$.
\end{assumption}

\begin{assumption}\label{as:S2}
    The function class $\{\varphi(o; \bm \beta, \overline{\eta}): \bm \beta \in \mathbb{R}^k\}$ is Donsker in $\bm \beta$, for each fixed $\overline{\eta}$.
\end{assumption}

\begin{assumption}\label{as:S3}
    The map $\bm \beta \mapsto \mathbb{E}_P(\varphi(O; \bm \beta, \overline{\eta}))$ is differentiable at $\bm \beta(P)$ uniformly in $\overline{\eta}$, the derivative matrix $\left. \nabla_{\bm \beta} \, \mathbb{E}_P(\varphi(O; \bm \beta, \overline{\eta}))\right|_{\bm \beta = \bm \beta(P)} \eqqcolon \bm V(\bm \beta(P), \overline{\eta})$ is invertible, and $\bm V(\bm \beta(P), \widehat{\eta}) \overset{P}{\to} \bm V(\bm \beta(P), \eta(P))$.
\end{assumption}

\vspace{2mm}
\noindent Note that
{\small
\[\mathbb{E}_P(\varphi(O; \bm \beta, \overline{\eta})) = \sum_{m = 1}^M w(m) P[E_m = 1]\nabla_{\bm \beta}\psi(m; \bm \beta)\left\{\mathbb{E}_P\left[\frac{E_m}{P[E_m = 1]} B_m(\overline{\eta}, \eta(P))\right] + \chi_m(P) - \psi(m; \bm \beta)\right\},\]
}%
where
{\small
\[B_{m}(\overline{\eta}, \eta(P)) = \left\{\overline{\mu}_m(1,\bm L_m) - \mu_m(1, \bm L_m)\right\} \left(1 - \frac{\pi_m(\bm L_m)}{\overline{\pi}_m (\bm L_m)}\right) - \left\{\overline{\mu}_m(0,\bm L_m) - \mu_m(0, \bm L_m)\right\} \left(1 - \frac{1 - \pi_m(\bm L_m)}{1 - \overline{\pi}_m (\bm L_m)}\right).\]
}%
Thus, the map $\bm \beta \mapsto \mathbb{E}_P(\varphi(O; \bm \beta, \overline{\eta}))$ will be differentiable in $\bm \beta$ as we have assumed $\psi(m;\bm \beta)$ is twice differentiable in $\bm \beta$. Moreover, each candidate marginal structural model considered in the main text belongs to a finite-dimensional linear space of bounded functions, and thus satisfies the Donsker conditions in Assumption \ref{as:S2} \citep{vdvaart1996weak}.

\vspace{2mm}
With these assumptions in hand, we can directly apply Lemma 3 in \cite{kennedy2023density}, a very general result regarding sample-split $M$-estimators that involve nuisance functions. In particular, under our assumptions,
\[
    \widehat{\boldsymbol{\beta}} - \boldsymbol{\beta}(P) = -\boldsymbol{V}(\boldsymbol{\beta}(P), \eta(P))^{-1}\mathbb{P}_n\left[\varphi(O; \boldsymbol{\beta}(P), \eta(P))\right] + O_P(R_{n, \boldsymbol{\beta}}) + o_P(n^{-1/2}),
\]
where $R_{n, \boldsymbol{\beta}} = \mathbb{E}_P\{\varphi(O; \boldsymbol{\beta}(P), \widehat{\eta}) - \varphi(O; \boldsymbol{\beta}(P), \eta(P)) \}$. Note that $\varphi(O; \boldsymbol{\beta}(P), \eta(P)) = \dot{\bm \beta}^{\dagger}(O; P)$ and $- \bm V^{-1} \dot{\bm \beta}^{\dagger}(O; P) = \dot{\bm \beta}^*(O; P)$. Further, observe that
\[
    R_{n, \boldsymbol{\beta}}
    = \sum_{m = 1}^M w(m) \left.\nabla_{\bm \beta}\psi(m; \bm \beta)\right|_{\boldsymbol{\beta} = \boldsymbol{\beta}(P)}\mathbb{E}_P\big(
    E_m 
    B_{m}(\widehat{\eta}, \eta(P))
    \big).
\]
It thus follows that $\lVert R_{n, \bm \beta}\rVert \lesssim \sum_{m = 1}^M \widehat{R}_{m}$, where $\lVert\cdot\rVert$ denotes the Euclidean norm and the implied constant depends only on $\epsilon$, $w$, and $\max_m\lVert\nabla_{\bm\beta}\psi(m;\bm\beta(P))\rVert$, concluding the proof.

\subsection{Identification of \texorpdfstring{$\chi_{j,m}(P)$}{Cross-Trial Effects} (Proof of Lemma 1)}
\small
By the definition of eligibility, $g(\bm L_m) = 1$ if and only if $E_m = 1$. For $a \in \{0,1\}$ and $\bm\ell$ in the support of $\bm L_j \mid E_j = 1$, which is contained in the support of $\bm L_m \mid E_m = 1$ by A4, consistency and no unmeasured confounding apply arm by arm:
$$
\begin{aligned}
\mathbb{E}\bigr[Y_m(a)~|~\bm L_m = \bm \ell, E_m = 1\bigr] &= \mathbb{E}\bigr[Y_m(a)~|~A_m = a, \bm L_m = \bm \ell, E_m = 1\bigr] ~~~~\text{(A2, A3)}\\
&= \mathbb{E}\bigr[Y_m~|~A_m = a, \bm L_m = \bm \ell, E_m = 1\bigr] ~~~~\text{(A1)}\\
&= \mu_m(a, \bm \ell).
\end{aligned}
$$
Subtracting the $a = 0$ arm from the $a = 1$ arm and integrating over the covariate distribution of the eligible population at time $j$,
$$
\begin{aligned}
    \chi_{j,m}(P) &=  \int_\mathcal{L} \mathbb{E}\bigr[Y_m(1) - Y_m(0)~|~\bm L_m = \bm \ell, E_m = 1\bigr]dP_{\bm L_j|E_j=1}(\bm \ell~|~E_j = 1) \\
    &= \int_\mathcal{L}\Bigr(\mu_m(1, \bm \ell) - \mu_m(0, \bm \ell)\Bigr) dP_{\bm L_j|E_j=1}(\bm \ell~|~E_j = 1)\\
    &= \mathbb{E}\bigr[\mu_m(1, \bm L_j) - \mu_m(0, \bm L_j)~|~E_j = 1\bigr].
\end{aligned}
$$
\normalsize

\subsection{Influence Function of \texorpdfstring{$\chi_{j,m}(P)$}{Cross-Trial Effects} (Proof of Theorem 3)} \label{app:jm}

To prove Theorem 3, we begin by introducing the following lemma, which provides the von Mises expansion for $\chi_{j,m}(P)$.

\begin{lemma}\label{lemma:VM-jm}
    The functional $\chi_{j,m}^{(a)}(P) \coloneqq \mathbb{E}(\mu_m(a, \boldsymbol{L}_j) \mid E_j = 1)$ satisfies the following von Mises expansion:
    \[\chi_{j,m}^{(a)}(\overline{P}) - \chi_{j,m}^{(a)}(P) = - \int \dot{\chi}_{j,m}^{(a)}(o; \overline{P}) \, dP(o) + R_{j,m}^{(a)}(\overline{P}, P),\]
    where
    \begin{align*}
        \dot{\chi}_{j,m}^{(a)}(O; P) &= \frac{E_j}{P[E_j = 1]}\big\{\mu_m(a, \boldsymbol{L}_j) - \chi_{j,m}^{(a)}(P)\big\} \\
        & \quad \quad + \frac{E_m}{P[E_m = 1]}\frac{\mathds{1}(A_m = a)}{\pi_m(\boldsymbol{L}_m)^a \{1 - \pi_m(\boldsymbol{L}_m)\}^{1 - a}}\xi_{j,m}(\boldsymbol{L}_m) \big\{Y_m - \mu_m(a, \boldsymbol{L}_m)\big\},
    \end{align*}
    and
    \begin{align*}
        & R_{j,m}^{(a)}(\overline{P}, P) \\ 
        &= \left\{\chi_{j,m}^{(a)}(\overline{P}) - \chi_{j,m}^{(a)}(P)\right\}\left\{1 - \frac{P[E_j = 1]}{\overline{P}[E_j = 1]}\right\} \\
        & \quad \quad + \left\{\frac{P[E_j = 1]}{\overline{P}[E_j = 1]} - \frac{P[E_m = 1]}{\overline{P}[E_m = 1]}\right\}\mathbb{E}_P\bigg(\frac{E_j}{P[E_j = 1]}\left\{\overline{\mu}_m(a, \boldsymbol{L}_j) - \mu_m(a, \boldsymbol{L}_j)\right\}\bigg) \\
        & \quad \quad
        + \mathbb{E}_P\bigg(\frac{E_m}{\overline{P}[E_m = 1]}\left\{a\frac{\pi_m(\boldsymbol{L}_m)}{\overline{\pi}_m(\boldsymbol{L}_m)} + (1 - a)\frac{1 - \pi_m(\boldsymbol{L}_m)}{1 - \overline{\pi}_m(\boldsymbol{L}_m)}\right\}\left\{\overline{\xi}_{j,m}(\boldsymbol{L}_m) - \xi_{j,m}(\boldsymbol{L}_m)\right\} \big\{\mu_m(a, \boldsymbol{L}_m) - \overline{\mu}_m(a, \boldsymbol{L}_m)\big\}\bigg) \\
        & \quad \quad
        + \mathbb{E}_P\bigg(\frac{E_m}{\overline{P}[E_m = 1]}\left\{a\left(\frac{\pi_m(\boldsymbol{L}_m)}{\overline{\pi}_m(\boldsymbol{L}_m)} - 1\right) + (1 - a)\left(\frac{1 - \pi_m(\boldsymbol{L}_m)}{1 - \overline{\pi}_m(\boldsymbol{L}_m)} - 1\right)\right\}\xi_{j,m}(\boldsymbol{L}_m) \big\{\mu_m(a, \boldsymbol{L}_m) - \overline{\mu}_m(a, \boldsymbol{L}_m)\big\}\bigg).
    \end{align*}
\end{lemma}~\\

\begin{proof} 
Observe that
    \begin{align*}
        & \chi_{j,m}^{(a)}(\overline{P}) - \chi_{j,m}^{(a)}(P) + \mathbb{E}_P\left(\dot{\chi}_{j,m}^{(a)}(O; \overline{P})\right) \\
        &= \left\{\chi_{j,m}^{(a)}(\overline{P}) - \chi_{j,m}^{(a)}(P)\right\}\left\{1 - \frac{P[E_j = 1]}{\overline{P}[E_j = 1]}\right\} \\
        & \quad \quad + \mathbb{E}_P\bigg(\frac{E_j}{\overline{P}[E_j = 1]}\left\{\overline{\mu}_m(a, \boldsymbol{L}_j) - \mu_m(a, \boldsymbol{L}_j)\right\} \\
        & \quad \quad \quad \quad
        + \frac{E_m}{\overline{P}[E_m = 1]}\left\{a\frac{\pi_m(\boldsymbol{L}_m)}{\overline{\pi}_m(\boldsymbol{L}_m)} + (1 - a)\frac{1 - \pi_m(\boldsymbol{L}_m)}{1 - \overline{\pi}_m(\boldsymbol{L}_m)}\right\}\overline{\xi}_{j,m}(\boldsymbol{L}_m) \big\{\mu_m(a, \boldsymbol{L}_m) - \overline{\mu}_m(a, \boldsymbol{L}_m)\big\}\bigg) \\
        &= \left\{\chi_{j,m}^{(a)}(\overline{P}) - \chi_{j,m}^{(a)}(P)\right\}\left\{1 - \frac{P[E_j = 1]}{\overline{P}[E_j = 1]}\right\} \\
        & \quad \quad + \mathbb{E}_P\bigg(E_j\left\{\frac{1}{\overline{P}[E_j = 1]} - \frac{1}{P[E_j = 1]}\right\}\left\{\overline{\mu}_m(a, \boldsymbol{L}_j) - \mu_m(a, \boldsymbol{L}_j)\right\}\bigg) \\
        & \quad \quad
        + \mathbb{E}_P\bigg(\frac{E_m}{\overline{P}[E_m = 1]}\left\{a\frac{\pi_m(\boldsymbol{L}_m)}{\overline{\pi}_m(\boldsymbol{L}_m)} + (1 - a)\frac{1 - \pi_m(\boldsymbol{L}_m)}{1 - \overline{\pi}_m(\boldsymbol{L}_m)}\right\}\left\{\overline{\xi}_{j,m}(\boldsymbol{L}_m) - \xi_{j,m}(\boldsymbol{L}_m)\right\} \big\{\mu_m(a, \boldsymbol{L}_m) - \overline{\mu}_m(a, \boldsymbol{L}_m)\big\}\bigg) \\
        & \quad \quad
        + \mathbb{E}_P\bigg(\frac{E_m}{\overline{P}[E_m = 1]}\left\{a\left(\frac{\pi_m(\boldsymbol{L}_m)}{\overline{\pi}_m(\boldsymbol{L}_m)} - 1\right) + (1 - a)\left(\frac{1 - \pi_m(\boldsymbol{L}_m)}{1 - \overline{\pi}_m(\boldsymbol{L}_m)} - 1\right)\right\}\xi_{j,m}(\boldsymbol{L}_m) \big\{\mu_m(a, \boldsymbol{L}_m) - \overline{\mu}_m(a, \boldsymbol{L}_m)\big\}\bigg) \\
        & \quad \quad
        + \mathbb{E}_P\bigg(E_m\left\{\frac{1}{\overline{P}[E_m = 1]} - \frac{1}{P[E_m = 1]}\right\}\xi_{j,m}(\boldsymbol{L}_m) \big\{\mu_m(a, \boldsymbol{L}_m) - \overline{\mu}_m(a, \boldsymbol{L}_m)\big\}\bigg) \\
        & \quad \quad
        + \mathbb{E}_P\left(\frac{E_j}{P[E_j = 1]}\left\{\overline{\mu}_m(a, \boldsymbol{L}_j) - \mu_m(a, \boldsymbol{L}_j)\right\} - \frac{E_m}{P[E_m = 1]}\xi_{j,m}(\boldsymbol{L}_m)\left\{\overline{\mu}_m(a, \boldsymbol{L}_m) - \mu_m(a, \boldsymbol{L}_m)\right\}\right).
    \end{align*}
    The result follows as the last two summands simplify:
    \begin{align*}
        &\mathbb{E}_P\left(\frac{E_m}{P[E_m = 1]}\xi_{j,m}(\boldsymbol{L}_m)\left\{\overline{\mu}_m(a, \boldsymbol{L}_m) - \mu_m(a, \boldsymbol{L}_m)\right\}\right) \\
        &= \int_{\mathcal{L}} \frac{p_{\boldsymbol{L}_j \mid E_j = 1}(\boldsymbol{l})}{p_{\boldsymbol{L}_m \mid E_m = 1}(\boldsymbol{l})}\left\{\overline{\mu}_m(a, \boldsymbol{l}) - \mu_m(a, \boldsymbol{l})\right\} \, dP_{\boldsymbol{L}_m \mid E_m = 1}(\boldsymbol{l}) \\
        &= \int_{\mathcal{L}} \left\{\overline{\mu}_m(a, \boldsymbol{l}) - \mu_m(a, \boldsymbol{l})\right\} \, dP_{\boldsymbol{L}_j \mid E_j = 1}(\boldsymbol{l}) \\
        &= \mathbb{E}_P\left(\left\{\overline{\mu}_m(a, \boldsymbol{L}_j) - \mu_m(a, \boldsymbol{L}_j)\right\} \mid E_j = 1\right) \equiv \mathbb{E}_P\left(\frac{E_j}{P[E_j = 1]}\left\{\overline{\mu}_m(a, \boldsymbol{L}_j) - \mu_m(a, \boldsymbol{L}_j)\right\}\right).
    \end{align*}
\end{proof}


\begin{proof}[Proof of Theorem 3]
    By the von Mises expansion in Lemma~\ref{lemma:VM-jm} and an application of Lemma 2 of \cite{kennedy2023density}, it follows that $\dot{\chi}_{j,m}^{(a)}(O; P)$ is the nonparametric influence function of $\chi_{j,m}^{(a)}$ at $P$. Given that $\chi_{j,m}(P) = \chi_{j,m}^{(1)}(P) - \chi_{j,m}^{(0)}(P)$ and $\dot\chi_{j,m}^*(O;P) = \dot\chi_{j,m}^{(1)}(O;P) - \dot\chi_{j,m}^{(0)}(O;P)$, this concludes the proof of Theorem 3.
\end{proof}

\subsection{Theoretical Properties of \texorpdfstring{$\widehat{\chi}_{j,m}$}{Cross-Trial Effects}}

We briefly summarize the asymptotic behavior of $\widehat\chi_{j,m}$, which will be necessary for inference on $\theta$.

\begin{theorem}\label{thm:ct_asymptotics}
    Suppose $\lVert \widehat{\xi}_{j,m} - \xi_{j,m}\rVert_{(m)} + \lVert \widehat{\pi}_m - \pi_m\rVert_{(m)} + \sum_{a = 0}^1\lVert \widehat{\mu}_m(a, \, \cdot\,) - \mu_m(a, \, \cdot\,)\rVert_{(m)} = o_P(1)$. Further, assume that $\epsilon \leq \widehat{\pi}_m(\bm L_m), \pi(\bm L_m) \leq 1 - \epsilon$ and $|Y_m|, |\widehat{\mu}_m(a, \bm L_m)|, |\xi_{j,m}|, |\widehat{\xi}_{j,m}|\leq \epsilon^{-1}$ almost surely when $E_m = 1$. Then
    $$
    \widehat\chi_{j,m} - \chi_{j,m}(P) = \mathbb{P}_n[\dot\chi^*_{j,m}(O;P)] + O_P\bigr(\widehat{R}_{j,m}\bigr) + o_P(n^{-1/2}),
    $$
    \noindent where 
    \[
    \widehat{R}_{j,m} = \sum_{a = 0}^1\lVert \widehat{\mu}_m(a, \, \cdot\,) - \mu_m(a, \, \cdot\,)\rVert_{(m)} \left\{\lVert \widehat{\xi}_{j,m} - \xi_{j,m}\rVert_{(m)} + \lVert \widehat{\pi}_m - \pi_m\rVert_{(m)}\right\}.
    \]
    \noindent Moreover, if $\widehat{R}_{j,m} = o_P(n^{-1/2})$ then $\sqrt{n}\bigr(\widehat \chi_{j,m} - \chi_{j,m}(P)\bigr) \overset{d}{\to} \mathcal{N}\bigr(0, \textnormal{Var}[\dot\chi^*_{j,m}(O,P)]\bigr)$, whereby $\widehat{\chi}_{j,m}$ attains the nonparametric efficiency bound.
\end{theorem}

The asymptotic behavior of $\widehat\chi_{j,m}$ is very similar to that of $\widehat\chi_m$, with an additional requirement that the product error between transport model $\xi_{j,m}$ and outcome model $\mu_m$ is $o_P(n^{-1/2})$. This rate condition is not unreasonable in light of Equation (6) of the main text, and the choice to model trial index probabilities rather than conditional densities directly. We will leverage the asymptotic normality of $\widehat\chi_{j,m}$ in the following section when conducting inference for $\theta$.

\subsection{Asymptotic Behavior of \texorpdfstring{$\widehat \chi_{j,m}$}{Cross-Trial Effects} (Proof of Theorem S1)}\label{app:asymp_jm}
Here we directly analyze the estimator $\widehat{\chi}_{j,m}$; the calculations mirror those for the bias in Lemma~\ref{lemma:VM-jm}. Observe that
\begin{align*}
    &\widehat{\chi}_{j,m} - \chi_{j,m}(P) \\
    &= \left\{\frac{1}{\mathbb{P}_n[E_j]} - \frac{1}{\mathbb{P}[E_j = 1]}\right\}\mathbb{P}_n\bigg[E_j\bigr\{\widehat{\mu}_m(1, \bm L_j) - \widehat{\mu}_m(0, \bm L_j) - \chi_{j,m}(P)\bigr\}\bigg] \\
    & \quad \quad + \left\{\frac{1}{\mathbb{P}_n[E_m]} - \frac{1}{\mathbb{P}[E_m =1]}\right\}\mathbb{P}_n\bigg[E_m\widehat{\xi}_{j,m}(\bm L_m)\biggr(\frac{A_m}{\widehat{\pi}_m(\bm L_m)} - \frac{1-A_m}{1-\widehat{\pi}_m(\bm L_m)}\biggr)\Bigr(Y_m - \widehat{\mu}_m(A_m, \bm L_m)\Bigr)\bigg] \\
    & \quad \quad + \mathbb{P}_n\bigg[\frac{E_j}{\mathbb{P}[E_j = 1]}\bigr\{\widehat{\mu}_m(1, \bm L_j) - \widehat{\mu}_m(0, \bm L_j) - \chi_{j,m}(P)\bigr\} \\
    & \quad \quad \quad \quad \quad + \frac{E_m}{\mathbb{P}[E_m = 1]}\widehat{\xi}_{j,m}(\bm L_m)\biggr(\frac{A_m}{\widehat{\pi}_m(\bm L_m)} - \frac{1-A_m}{1-\widehat{\pi}_m(\bm L_m)}\biggr)\Bigr(Y_m - \widehat{\mu}_m(A_m, \bm L_m)\Bigr)\bigg],
\end{align*}

\noindent Under our consistency assumptions, and by the central limit theorem (and delta method), the first two summands are $O_P(n^{-1/2}) \times o_P(1) = o_P(n^{-1/2})$. The third summand can be written
\begin{align*}
& \mathbb{P}_n\left[\dot{\chi}_{j,m}^*(O;P)\right] + \mathbb{P}_n\bigg[\frac{E_j}{\mathbb{P}[E_j = 1]}\bigr\{\widehat{\mu}_m(1, \bm L_j) - \widehat{\mu}_m(0, \bm L_j) - [\mu_m(1, \bm L_j) - \mu_m(0, \bm L_j)]\bigr\} \\
& \quad \quad \quad \quad \quad \quad \quad \quad \quad + \frac{E_m}{\mathbb{P}[E_m = 1]}\bigg\{\widehat{\xi}_{j,m}(\bm L_m)\biggr(\frac{A_m}{\widehat{\pi}_m(\bm L_m)} - \frac{1-A_m}{1-\widehat{\pi}_m(\bm L_m)}\biggr)\Bigr(Y_m - \widehat{\mu}_m(A_m, \bm L_m)\Bigr) \\
& \quad \quad \quad \quad \quad \quad \quad \quad \quad \quad \quad \quad \quad \quad \quad  - \xi_{j,m}(\bm L_m)\biggr(\frac{A_m}{\pi_m(\bm L_m)} - \frac{1-A_m}{1-\pi_m(\bm L_m)}\biggr)\Bigr(Y_m - \mu_m(A_m, \bm L_m)\Bigr)\bigg\}\bigg].
\end{align*}

\noindent For the latter term, we can subtract and add the same term but replacing the outer $\mathbb{P}_n$ with $\mathbb{E}_P$; the ($\mathbb{P}_n - \mathbb{E}_P$) contribution is $o_P(n^{-1/2})$ under our assumptions by a simple application of Lemma 2 in \cite{kennedy2020sharp}, while the $\mathbb{E}_P$ contribution is given by
\begin{align*}
    & \mathbb{E}_P\bigg(\frac{E_m}{P[E_m = 1]}\frac{\pi_m(\boldsymbol{L}_m)}{\widehat{\pi}_m(\boldsymbol{L}_m)}\left\{\widehat{\xi}_{j,m}(\boldsymbol{L}_m) - \xi_{j,m}(\boldsymbol{L}_m)\right\} \big\{\mu_m(1, \boldsymbol{L}_m) - \widehat{\mu}_m(1, \boldsymbol{L}_m)\big\}\bigg) \\
    & \quad -\mathbb{E}_P\bigg(\frac{E_m}{P[E_m = 1]}\frac{1 - \pi_m(\boldsymbol{L}_m)}{1 - \widehat{\pi}_m(\boldsymbol{L}_m)}\left\{\widehat{\xi}_{j,m}(\boldsymbol{L}_m) - \xi_{j,m}(\boldsymbol{L}_m)\right\} \big\{\mu_m(0, \boldsymbol{L}_m) - \widehat{\mu}_m(0, \boldsymbol{L}_m)\big\}\bigg) \\
        & \quad
        + \mathbb{E}_P\bigg(\frac{E_m}{P[E_m = 1]}\left(\frac{\pi_m(\boldsymbol{L}_m)}{\widehat{\pi}_m(\boldsymbol{L}_m)} - 1\right)\xi_{j,m}(\boldsymbol{L}_m) \big\{\mu_m(1, \boldsymbol{L}_m) - \widehat{\mu}_m(1, \boldsymbol{L}_m)\big\}\bigg) \\
        & \quad
        - \mathbb{E}_P\bigg(\frac{E_m}{P[E_m = 1]}\left(\frac{1 - \pi_m(\boldsymbol{L}_m)}{1 - \widehat{\pi}_m(\boldsymbol{L}_m)} - 1\right)\xi_{j,m}(\boldsymbol{L}_m) \big\{\mu_m(0, \boldsymbol{L}_m) - \widehat{\mu}_m(0, \boldsymbol{L}_m)\big\}\bigg),
\end{align*}
by the exact same calculations as in the proof of Lemma~\ref{lemma:VM-jm}. Again invoking our boundedness assumptions, Theorem \ref{thm:ct_asymptotics} follows.

\section{Additional Methodological Details}
\subsection{Graphical Overview of Algorithm 1}
\begin{figure}[H]
    \centering
    \includegraphics[width=\textwidth]{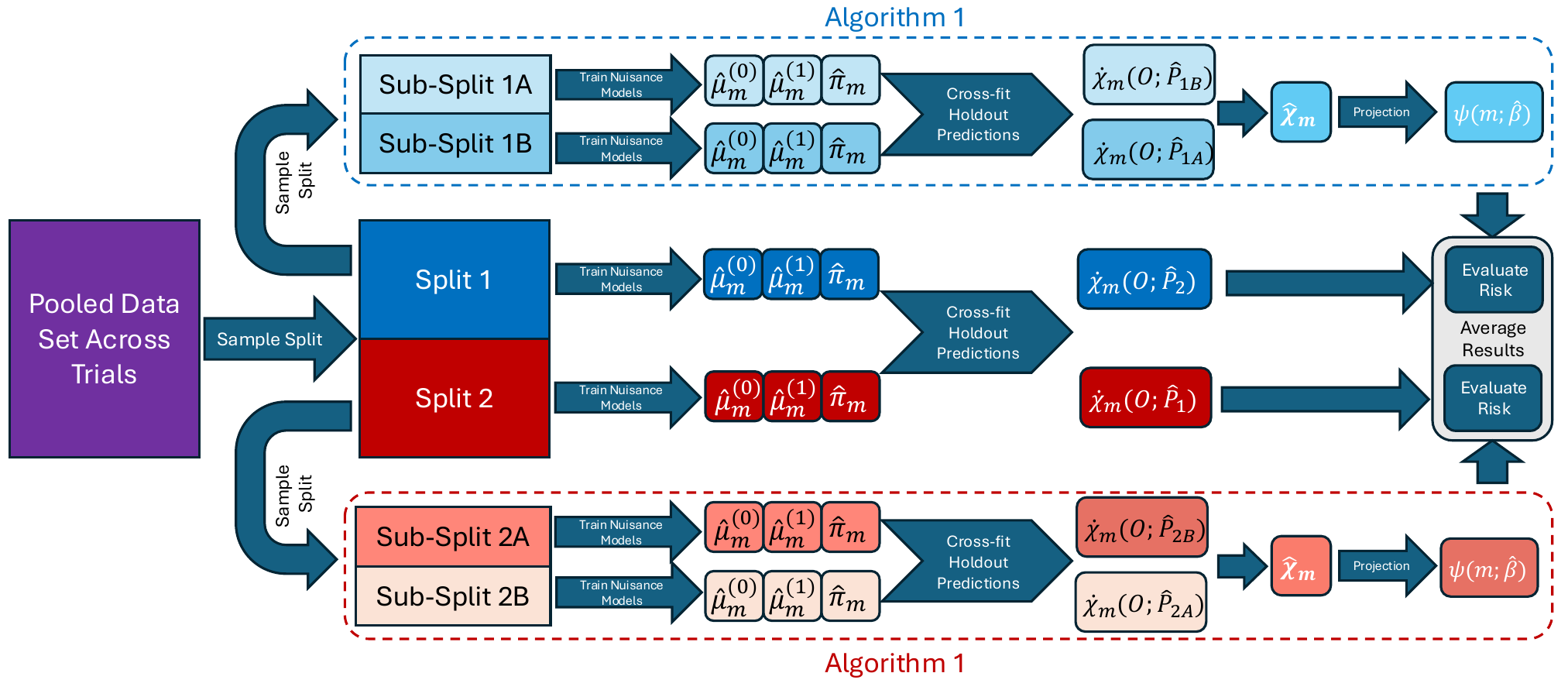}
    \caption{Overview of cross-fitting procedure for evaluating $\widehat{L}(\widehat\psi_k)$.}
    \label{fig:risk_eval}
\end{figure}
\subsection{Hypothesis Testing for \texorpdfstring{$\theta$}{Variability Ratio}}
\label{sec:sup_hypothesis_testing}

Consider the hypothesis test $H_0: \theta \in \{0, 1\}$ versus $H_1: \theta \in (0,1)$. We frame the null hypothesis as simultaneously testing two point nulls because $\theta = 0$ and $\theta = 1$ correspond to the cases where all or none of the variation in cross-trial effects, respectively, is explained by changes in underlying populations. Rejection of $H_0$ for this test is equivalent to simultaneous rejection of the two one-sided tests, $H_0^{(0)}: \theta = 0$ versus $H_1^{(0)}: \theta > 0$ and $H_0^{(1)}: \theta = 1$ versus $H_1^{(1)}: \theta < 1$. Drivers of treatment effect heterogeneity across calendar time differ based on whether $\theta = 1$ or $\theta \in (0,1)$, and thus only testing the null $H_0^{(0)}$ is insufficient. Moreover, as discussed in Section 6.1 of the main text, rejection of $H_0^{(1)}$ in favor of $\theta < 1$ may motivate different standardization strategies when the primary goal is to understand changes in treatment efficacy.

In practice, $H_0: \theta \in \{0, 1\}$ considers testing at both boundary points for $\theta$, somewhat similar to testing variance components in mixed-effects models \citep{stram1994variance} and as such characterizing the asymptotic distribution of $\widehat\theta$ under the null would likely yield a challenging ratio involving mixtures of chi-squared distributions. On the other hand, a full nonparametric bootstrap is computationally infeasible in our setting, given that the size of the pooled dataset $\mathcal{D}$ exceeds several million eligible subject-trials for comparisons involving non-surgical patients (Figure S2). Not only does re-estimation of nuisance functions across bootstrapped datasets pose a large computational burden, but so too does prediction of $\widehat\mu_m(a_m, \bm L_j)$ and $\widehat\xi_{j,m}(\bm L_m)$ given that the number of predictions required is a $M$-fold larger than the size of $\mathcal{D}$.

Instead, we use an asymptotic version of the parametric bootstrap \citep{van2000asymptotic} for $\bm S$. In particular, under the conditions of Theorem \ref{thm:ct_asymptotics}, $\sqrt{n}\,\mathrm{vec}(\widehat{\bm S} - \bm S) \overset{d}{\to} \mathcal{N}(\bm 0, \bm \Sigma)$, where $\bm\Sigma \in \mathbb{R}^{M^2\times M^2}$ has entries $\text{Cov}[\dot{\chi}^*_{j_1, m_1}(O; P), \dot{\chi}^*_{j_2, m_2}(O; P)]$. Critically, this asymptotic normality holds even when $\theta$ is at a boundary point. Letting $\widehat{\bm\Sigma}_n$ denote the empirical covariance matrix of the estimated subject-level influence function contributions, we draw $\mathrm{vec}(\widehat{\bm S}^{(b)}) \sim \mathcal{N}\big(\mathrm{vec}(\widehat{\bm S}), \widehat{\bm\Sigma}_n / n\big)$ for $b = 1, \dots, B$. Finally, to approximate the sampling distribution of $\widehat\theta$, summarize resamples $\widehat{\bm S}^{(1)}, \dots, \widehat {\bm S}^{(B)}$ by their corresponding variance ratios $\widehat\theta^{(1)}, \dots, \widehat\theta^{(B)}$.

Unfortunately, due to the boundary null, bootstrapped replicates $\widehat\theta^{(1)}, \dots, \widehat\theta^{(B)}$ will never be exactly 0 or 1 and thus 95\% percentile-based confidence intervals would always lie within $(0,1)$, rejecting $H_0$ even when it is true. We follow a similar strategy to that of Cavaliere et al.~\cite{CavaliereNielsenPedersenRahbek2022}, and modify our test to allow for a small margin away from the boundary. For some small $\delta > 0$, we adopt the test $H_0^{(\delta)}: \theta \in [0,\delta]\cup[1-\delta,1]$ versus $H_1^{(\delta)}:\theta \in (\delta, 1-\delta)$, and reject $H_0^{(\delta)}$ if a 95\% percentile-based confidence interval is fully within $(\delta, 1-\delta)$. One can interpret $\delta$ as the fraction of variability in entries in $\bm S$ for which relative changes in causal effects are deemed negligible. The choice of $\delta$ is highly context dependent, as what constitutes clinically meaningful changes is application specific; in the present context of bariatric surgery, we adopt $\delta = 0.05$.

\subsection{Inference Following MSM Selection}

The pointwise confidence intervals for $\psi(m;\widehat{\bm\beta})$ in Section 4.4 of the main text are derived for a fixed candidate MSM, and do not formally account for the data-adaptive selection among candidates described in Section 4.5. Three features limit the practical consequences of this. First, selection is among a small, nested set of $K = 5$ candidate MSMs, and each candidate's pseudorisk is evaluated out-of-sample through the nested cross-fitting procedure in Figure \ref{fig:risk_eval}, rather than from in-sample fit. Second, the quantities most central to our conclusions about \textit{why} effects vary do not depend on the selected MSM: the cross-trial effects $\widehat\chi_{j,m}$, and hence $\widehat\theta$ and its intervals, are computed independently of the selection step. Third, in our application the conclusion that the effect is non-constant for 2- and 3-year outcomes comparing surgery or RYGB to no surgery holds for both $c = 0.5$ and $c = 1$ (Table 2), although the specific functional form selected can differ. Reported intervals for $\psi(m;\widehat{\bm\beta})$ should therefore be interpreted as conditional on the selected curve. Moreover, Figure 2 of the main text presents the confidence band for each candidate MSM without selection, so that results under every candidate form are visible to the reader. Table \ref{tab:loss_fx} reports the standardized difference in loss for every candidate relative to the minimizer, so that readers can assess how clearly the selected model is separated from alternatives.

\section{Simulations}
\label{sec:simulations}

\subsection{Simulation Overview}
\label{sec:simulation_overview}

\begin{figure}[!htb]
    \centering
    \includegraphics[width=\textwidth]{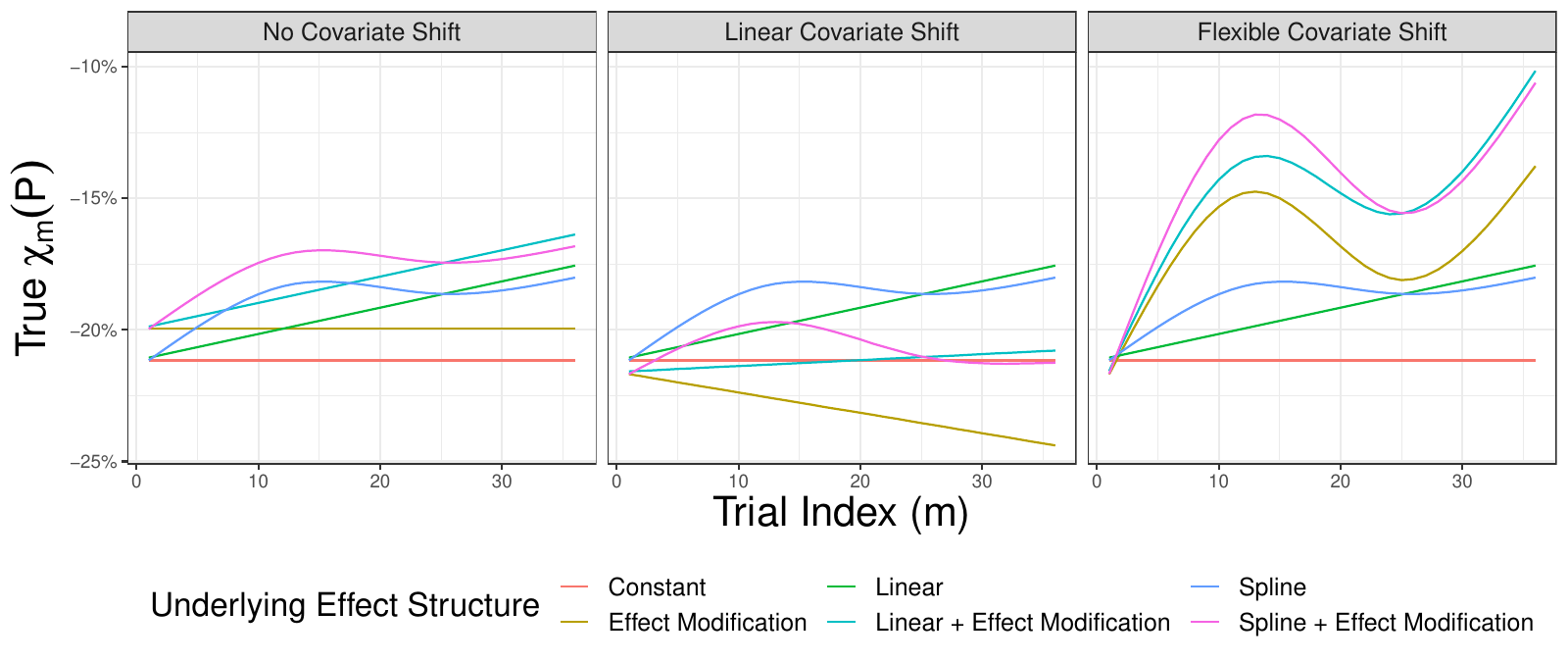}
    \caption{Overview of calendar time-specific treatment effects across simulation scenarios.}
    \label{fig:sim_overview}
\end{figure}

To evaluate the practical utility and assess the finite-sample properties of the proposed methods, we conducted an extensive plasmode simulation study \citep{Vaughan2009Plasmodes} closely linked to the EHR data underlying our bariatric surgery analysis. Our setup considered $M = 36$ target trials assessing the effect of a binary treatment $A$ on a continuous outcome $Y$, using observed covariates $\bm L$ from DURABLE (sex, age, BMI, study site, race, smoking status, hypertension, dyslipidemia, diabetes, medication usage). To maintain direct control over the structure of covariate shift across trials, we conceptualize $A$ as a treatment whose receipt does not exclude patients from future trials. In the simulated setup, exclusion of such patients would lead to a depletion of eligible patients in later trials, thus indirectly changing the distribution of covariates among the eligible population, and complicating the ability to analytically control the structure of simulated effects across calendar time. As detailed below, we used models based on $A \in \{\text{surgery}, \text{no surgery}\}$ and $Y = $ 3-year relative weight change to inform the strength of associations between $\bm L, A$ and $Y$ in our simulation study.

To generate simulated datasets, we first sampled entire covariate vectors $\bm L_1$ from the observed distribution in the DURABLE data, in order to preserve the complex correlation structure typical of EHR data. Covariate vectors for the remaining trials $m \in \{2, \ldots, M \}$ were generated under one of three settings. First, we considered no covariate shift, which we enforced by setting $\bm L_m = \bm L_1$ for $m \in \{2, \ldots, M\}$. Second, we considered a linear covariate shift by allowing both the prevalence of diabetes and mean BMI to vary linearly across trials, while age naturally increased linearly as well. Finally, we considered a form of flexible covariate shift such that the prevalence of diabetes and mean BMI followed splines on $m$ (along with linear increase in age).

Simulated treatments and outcomes were generated using covariates $\bm L_m$ and logistic/linear regression models, respectively, tied closely to our bariatric surgery example. Coefficient values for these models were informed by fits of pooled logistic regression ($\widetilde{\pi}$) and linear outcome models ($\widetilde{\mu}$) estimated from 14,353,682 subject-trials in the DURABLE database. Simulated outcomes were generated following one of six distinct outcome models, each encoding different interaction structures among $A_m$, $m$, and $\bm L_m$, and thus representing different patterns of treatment effect variation over calendar time. The six treatment effect structures were as follows: (1) Constant ($A$); (2) linear ($A\times m$); (3) spline ($A\times f(m)$ for spline basis $f(m)$), (4) effect modification ($A\times \bm L_m$); (5) linear + effect modification ($A\times m + A\times \bm L_m$); and (6) spline + effect modification ($A\times f(m) + A\times \bm L_m$). Interactions of the form $A \times \bm L_m$ were specified for age, BMI, and diabetes status, the covariates subject to controlled shifts across trials, thereby inducing change in trial-specific treatment effects independent of pattens in underlying treatment efficacy.

Together, the three forms of covariate shift and six outcome model structures comprise 18 distinct simulation scenarios. Figure \ref{fig:sim_overview} summarizes the true values of $\chi_m(P)$ over time for each scenario. Additional details on data generation in each scenario is provided in Section \ref{sec:sim_dgp_supp}, while complete tables of coefficients used to generate simulated data are provided in Section \ref{sec:sim_params}. For each scenario, 1,000 simulated datasets were generated for sample sizes $n \in \{100, 500, 1000, 5000, 10000\}$. Although these total unique patient numbers are smaller than in our motivating application (to reduce computational burden), the resulting pooled datasets still contain nearly 400,000 subject-trials when $n = 10,000$, consistent with the scale of EHR-based studies of bariatric surgery \citep{benz2024, haneuse2025}.

\subsection{Simulation Data Generating Process}
\label{sec:sim_dgp_supp}
Simulated data were according to the following steps. In the below steps, we let $\bm\alpha_f$ and $\bm X_f$ respectively denote the coefficient vector and design matrix relevant to nuisance function $f(\cdot)$. Specific values for $\bm\alpha_f$ are presented in Section \ref{sec:sim_params}.

\begin{enumerate}
    \item Draw $\bm L_1$ from observed eligible subject-trials in the DURABLE database.
    \item Update covariates $(\bm L_2, \dots, \bm L_M)$ according to one of the three rules described in more detail below.
    \item Sample treatment $A~|~\bm L_m \sim \text{Bernoulli}\bigr(\pi_m(\bm L_m)\bigr)$ where $\text{logit}[\pi_m(\bm L_m)] = \bm\alpha_\pi^T\bm X_\pi$
    \item Sample outcome $Y_m~|~\bm L_m, A_m \sim \mathcal{N}\bigr(\mu_m(A_m, \bm L_m), \sigma_y^2\bigr)$ where $\mu_m(A_m, \bm L_m) = \bm \alpha_\mu^T\bm X_\mu$. Note that $\mu_m$ takes one of 6 forms described in more details below.
\end{enumerate}

\noindent\underline{\textbf{Covariate Shift Rules}}:\\
\noindent Covariate shift across component trials followed one of three update rules.

\begin{enumerate}
    \item No covariate shift: $\bm L_m = \bm L_1$ for all $m$
    \item Linear covariate shift: 
    \begin{enumerate}
        \item BMI$_m$ = BMI$_{1} + \mathcal{N}\Bigr(\frac{m-1}{12}, 1\Bigr)$
        \item $P(\text{Diabetes}_m = 1)$ = $0.2 + \frac{0.4(m-1)}{35}$
        \item Age$_m$ = Age$_1$ + $\frac{m-1}{12}$
        \item No change in remaining covariates
    \end{enumerate}
    \item Flexible covariate shift
    \begin{enumerate}
        \item BMI$_m$ = BMI$_{1} + \mathcal{N}\Bigr(\bm x_m^T (1, -3, 1), 1\Bigr)$
        \item $P(\text{Diabetes}_m = 1)$ = $0.2 + \bm x_m^T (0.02, 0.6, -0.08)$
        \item Age$_m$ = Age$_1$ + $\frac{m-1}{12}$
        \item No change in remaining covariates 
    \end{enumerate}
\end{enumerate}

Note that the update rules for diabetes ensure that the prevalence varies between 20\% at $m = 1$ to 60\% at $m = 36$. In the above $\bm x_m^T$, denotes the $m^\text{th}$ row of the spline design matrix that is the result of a natural cubic spline on $\{1, \dots, 36\}$ with 2 internal knots, \texttt{ns(1:36, df = 3)}. \\

\noindent\underline{\textbf{Outcome Models}}:\\
\noindent Generation of outcomes $Y_m$ followed one of six different outcome models, each encoding a treatment effect structure with different reasons how and why treatment effects vary by calendar time. In the below, we use the shorthand $A_m \times \bm L_m$ to denote treatment-confounder interactions. Specific interactions are with age, BMI, and diabetes status. Effect structures across calendar time are as follows. As before, $\bm x_m$ denotes the $m^\text{th}$ row of the design matrix resulting from a natural cubic spline on $\{1, \dots, 36\}$ with 2 internal knots, \texttt{ns(1:36, df = 3)}.

\begin{enumerate}
    \item Constant: $\mathbb{E}[Y_m~|~\bm A_m, \bm L_m] = \alpha_0 + \bm\alpha_1 \bm L_m + \alpha_2 A_m$
    \item Linear: $\mathbb{E}[Y_m~|~\bm A_m, \bm L_m] = \alpha_0 + \bm\alpha_1^T \bm L_m + \alpha_2 A_m + \alpha_3 (A_m \times m)$
    \item Spline: $\mathbb{E}[Y_m~|~\bm A_m, \bm L_m] = \alpha_0 + \bm\alpha_1^T \bm L_m + \alpha_2 A_m + \bm\alpha_3^T (A_m \times \bm x_m)$
    \item Effect Modification: $\mathbb{E}[Y_m~|~\bm A_m, \bm L_m] = \alpha_0 + \bm\alpha_1^T \bm L_m + \alpha_2 A_m + \bm\alpha_3^T (A_m \times\bm L_m)$
    \item Linear + Effect Modification: $\mathbb{E}[Y_m~|~\bm A_m, \bm L_m] = \alpha_0 + \bm\alpha_1^T \bm L_m + \alpha_2 A_m + \alpha_3 (A_m \times m) + \bm\alpha_4^T (A_m \times\bm L_m)$
    \item Spline + Effect Modification: $\mathbb{E}[Y_m~|~\bm A_m, \bm L_m] = \alpha_0 + \bm\alpha_1^T \bm L_m + \alpha_2 A_m + \bm\alpha_3^T (A_m \times \bm x_m) + \bm\alpha_4^T (A_m \times\bm L_m)$
\end{enumerate}

\subsection{Analysis Methods}
\label{sec:simulation_methods}

To estimate $\widehat\chi_m$, we used \texttt{SuperLearner} \citep{superlearner} for pooled outcome and propensity models. Pooled outcomes models were estimated separately for each treatment status. Component libraries included (generalized) linear models (\texttt{SL.speedglm}), generalized additive models (\texttt{SL.gam}) and random forests (\texttt{SL.ranger}). Because the \texttt{SuperLearner} package does not provide support for categorical outcomes, we used a random forest model for $\widetilde\xi$ when estimating cross-trial effects. We considered five candidate MSMs for projection: (1) constant model, (2) linear model, (3) cubic model, (4) spline with 2 knots at $m \in \{12, 24\}$, and (5) spline with 3 knots at $m \in \{9, 18, 27\}$. \\

\noindent\underline{\textbf{Pooled Models}}
\begin{itemize}
    \item Outcome Model $\tilde \mu$ used \texttt{SuperLearner}, stratified by treatment status, with the following component libraries
    \begin{itemize}
        \item \texttt{SL.speedglm}
        \item \texttt{SL.gam} with \texttt{deg.gam = 4}
        \item \texttt{SL.ranger} with \texttt{max.depth = 10} and \texttt{num.trees = 500}
    \end{itemize}
    \item Treatment Model $\tilde \pi$ used \texttt{SuperLearner} with the following component libraries
    \begin{itemize}
        \item \texttt{SL.speedglm}
        \item \texttt{SL.gam} with \texttt{deg.gam = 4}
        \item \texttt{SL.ranger} with \texttt{max.depth = 4} and \texttt{num.trees = 500}
    \end{itemize}
    \item Transport Model $\tilde \xi$ used random forest (\texttt{ranger}) with \texttt{max.depth = 10} and \texttt{num.trees = 500}
\end{itemize}

\noindent\underline{\textbf{Thresholding near degenerate standardization matrices}}:\\
\noindent In simulation settings where the true treatment effect is constant over calendar time, even with covariate shift (i.e., there is no effect modification) differences between $\widehat\chi_{j,m_1}$ and $\widehat\chi_{j,m_2}$ can be effectively 0. However, if because there is no effect modification detected by nuisance models (if consistent), any sampling variation in outcome differences is attributed to changes in the calendar time index, thus producing estimates $\widehat\theta_m$ closer to 1. If differences between $\widehat\chi_{j,m_1}$ and $\widehat\chi_{j,m_2}$ are indeed extremely small, this undesirable behavior given that we'd like values of $\theta \approx 1$ to reflect possible shifts in treatment efficacy. Furthermore, this is somewhat of an artifact of our simulation set up given that scenarios without any covariate shift and/or effect modification are not reflective of EHR-based target trial emulations. To ``correct'' this behavior in our simulations, we apply a data-adaptive thresholding procedure, as follows.

Let $t(x,\tau)=x\,\mathds 1(|x|\ge\tau)$. In addition to $\widehat\sigma_m^2$ and $\widehat\gamma_m^2$, we compute thresholded versions

$$\widehat\sigma_m'^2=\frac{1}{M-1}\sum_{j=1}^M t\big(\widehat\chi_{m,j}-\widehat\chi_m,\tau\big)^2,\qquad \widehat\gamma_m'^2=\frac{1}{M-1}\sum_{j=1}^M t\big(\widehat\chi_{j,m}-\widehat\chi_m,\tau\big)^2,$$

with $\tau=0.005$ (on the relative weight change scale, i.e., 0.5 percentage points). Let $\widehat\theta$ and $\widehat\theta'$ denote the averages over $m$ of the unthresholded and thresholded ratios. If $|\widehat\theta-\widehat\theta'|>0.1$, all $\widehat\theta_m$ are replaced by their thresholded counterparts; otherwise the unthresholded values are used. Thresholding is triggered when a substantial share of the variability in $\widehat{\bm S}$ arises from differences smaller than $\tau$, as occurs when the true standardization matrix is (nearly) flat. In the bootstrap, the same rule is applied separately to each resampled matrix $\widehat{\bm S}^{(b)}$.

\subsection{Simulation Results}
\label{sec:simulation_results}

\begin{figure}[H]
    \centering
    \includegraphics[width=\textwidth]{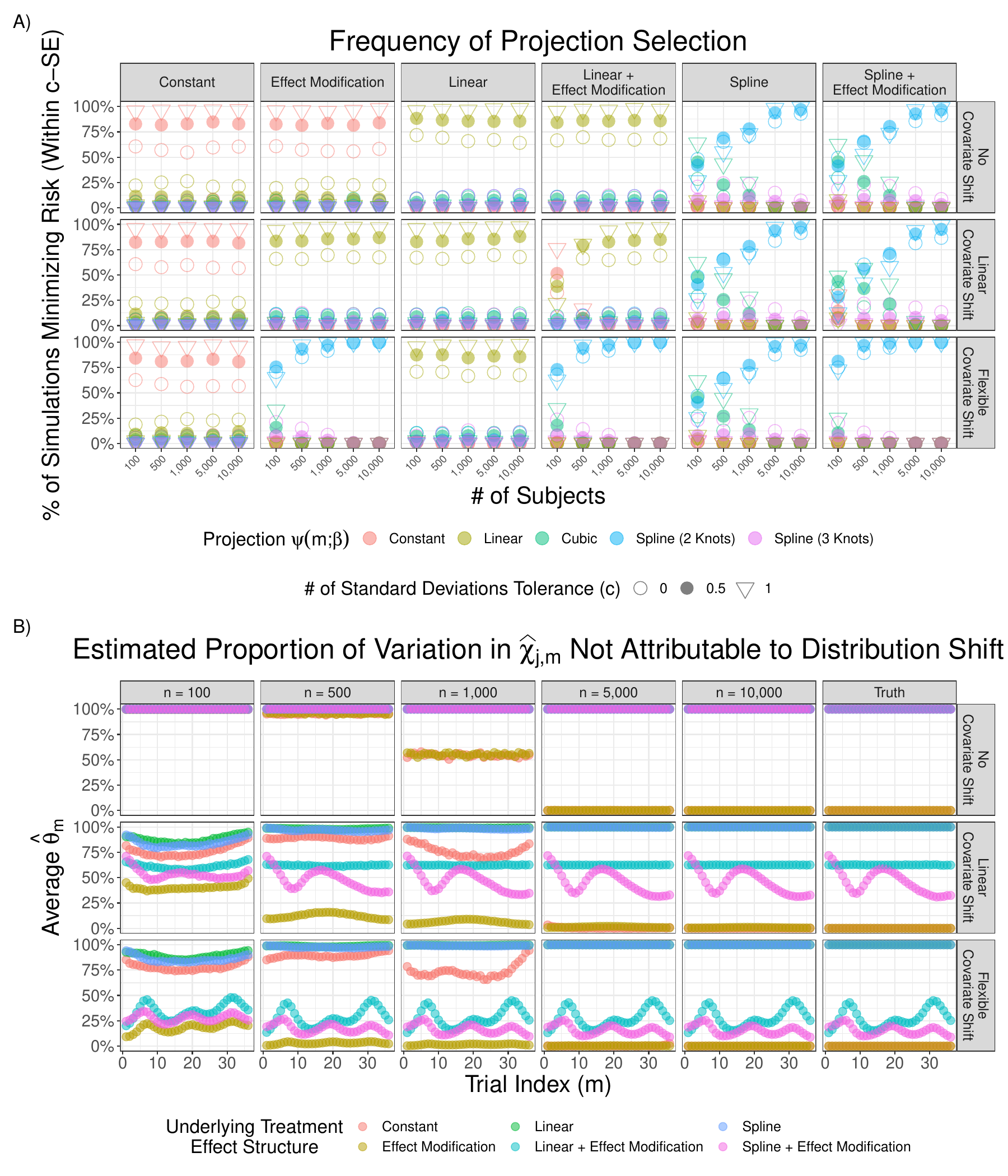}
    \caption{\textbf{(A)} Frequency with which each candidate MSM was selected across 1,000 simulations, by true effect structure (columns), covariate shift (rows), and sample size. Tolerance $c \in \{0, 0.5, 1\}$ is the number of standard errors by which $\widehat L(\widehat\psi_k)$ may exceed $\widehat L(\widehat\psi^*)$ for the simplest MSM to be selected. \textbf{(B)} Average trial-specific variability ratio $\widehat\theta_m$ (after thresholding) by sample size; the rightmost column shows true values.}
    \label{fig:sim_results}
\end{figure}

Key simulation results are presented in Figure \ref{fig:sim_results}. Figure \ref{fig:sim_results}A shows the frequency with which each candidate MSM was selected across 1,000 simulations for varying values of $c \in \{0, 0.5, 1\}$ standard error tolerance in the difference in loss functions. Simply selecting the MSM minimizing $\widehat{L}(\widehat\psi_k)$ without any tolerance ($c = 0$) frequently resulted in selection of a more flexible MSM than necessary, even when the true effect structure was constant across time. As $c$ increased, so too did the sample size required for selection of more complex MSMs, even in cases where the more complex model was the correct underlying data generating mechanism. A good example of this is illustrated by the outcome model with a linear treatment effect and effect modification under linear covariate shift. As seen in Figure \ref{fig:sim_overview}, the induced trend in calendar time is linear in $m$ with a very shallow slope, and as a result the constant MSM was still selected in a majority of simulations at $n = 100$ regardless of $c$; the linear MSM became the most frequently selected model by $n = 500$ for $c \in \{0, 0.5\}$ and by $n =$ XX for $c = 1$.

Figure \ref{fig:sim_results}A  also highlights the importance of having some tolerance $c > 0$ standard errors. For example, in scenarios with no underlying trend in $\chi_m(P)$ across calendar time, the constant MSM minimized $\widehat L(\widehat\psi_k)$ in only 58.3\% of simulations at $c = 0$, compared with 82.6\% at $c = 0.5$ and 96.1\% at $c = 1$. On the other hand, larger tolerances require larger sample sizes to select more complex MSMs. This is most visible when covariate shift in effect modifiers (linear or flexible) is present alongside a separate temporal trend in the treatment effect (e.g., Spline + Effect Modification under linear covariate shift), where $c = 1$ requires $n \geq$ 5,000 subjects before the correct MSM is selected in most simulations.

Altogether, these simulations suggest that $c \in \{0.5, 1\}$ are reasonable choices to report, as we do in our data application: $c = 0$ over-selects complex MSMs in small samples, while $c = 1$ delays selection of weak but real trends. We note that these simulations should not be construed as an exact replica of our data application, given that in EHR data treatment occurs much less frequently, subjects do not contribute to all $M$ trials, and the types of covariate shift (Figure 1 of the main text) are likely more complex than those considered here (Figure \ref{fig:sim_overview}) due to a larger number of potential effect modifiers. The simulations are therefore best read as evidence about the relative behavior of the tolerances, not as guidance on the sample size at which the choice of $c$ ceases to matter. Because the constant MSM minimizes the loss without any tolerance in several application settings (5 of 16 comparison-outcome pairs; Table \ref{tab:loss_fx}), and because $c = 1$ does not materially change the reported estimates (Figure 2A of the main text), we report final estimates under $c = 0.5$ as a compromise between parsimony and sensitivity.

True values of $\theta_m$ are plotted in the rightmost column of Figure \ref{fig:sim_results}B. The remaining columns of Figure \ref{fig:sim_results}B present mean values of $\widehat\theta_m = \frac{\widehat\sigma^2_m}{\widehat\sigma^2_m + \widehat\gamma^2_m}$ across simulations. In scenarios where the underlying treatment effect structure does not contain interactions with $m$ (e.g., constant, effect modification), $\theta_m = 0$ across time. In the constant scenario, $\sigma_m^2 = \gamma_m^2 = 0$ so that the ratio defining $\theta_m$ is undefined; when $\sigma_m^2+\gamma_m^2=0$, we set $\theta_m=0$: within every fixed population there is no variation in the effect across calendar time, so no variation is attributable to factors other than covariate shift. Conversely, in scenarios with linear or spline based effect structures, $\theta_m = 1$ for all $m$. When changes in effects are attributable to both changes in treatment efficacy and covariate shift in effect modifiers, $\theta_m \in (0,1)$ and varies across time.

Across scenarios, $\widehat\theta_m$ was unbiased for $\theta_m$ beyond $n = 5,000$ subjects. The greatest degree of bias was observed in smaller samples for scenarios where $\theta = 0$. In scenarios with a constant treatment effect, note that both $\sigma_m^2$ and $\gamma_m^2 \approx 0$, and thus extremely small differences in estimated time-specific ATEs can produce biased estimates of $\widehat\theta$ closer to 1. This issue underscores the importance of simultaneously considering the selected MSM, as in all of these cases, a constant model is most frequently selected (Figure \ref{fig:sim_results}A).

\begin{figure}[H]
    \centering
    \includegraphics[width=\textwidth]{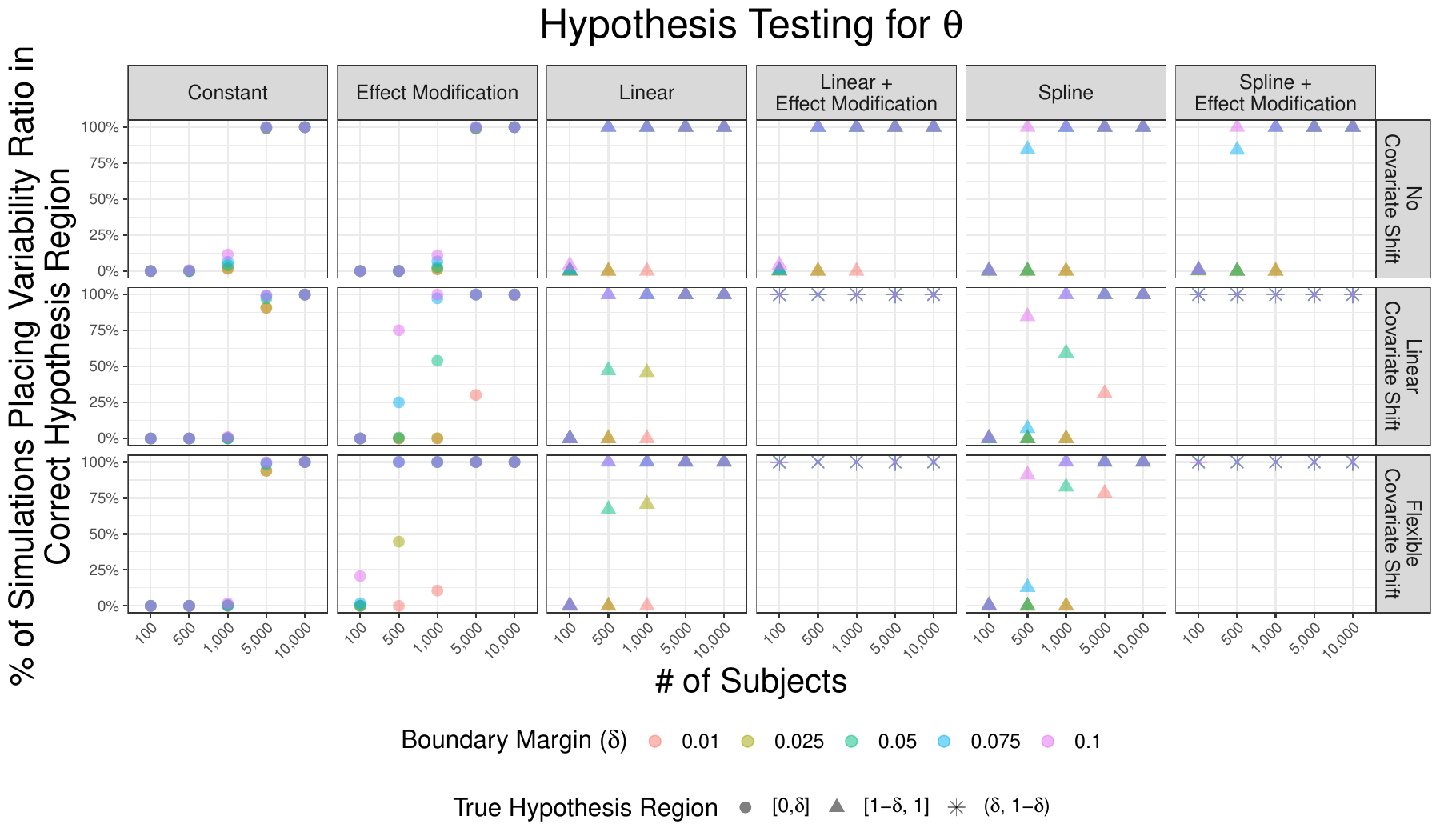}
    \caption{Results from hypothesis testing across simulations}
    \label{fig:hypothesis_testing}
\end{figure}

Figure \ref{fig:hypothesis_testing} shows hypothesis testing results from our simulation study. The y-axis of Figure \ref{fig:hypothesis_testing} displays the proportion of simulations for which the correct decision is made, that is rejecting $H_0$ when $H_1$ is true, and failing to reject $H_0$ when $H_0$ is true. For scenarios in which the true $\theta$ is not at a boundary point, this is power. For scenarios at a boundary point, this is specificity. Unfortunately, it is not directly possible to show type I error rate, as type 1 error is preserved at the nominal level only at the boundary margin, $\delta$, yet the true value $\theta \neq \delta$. In general, for larger sample sizes, there is very good power to reject $H_0$ when $\theta$ is not at a boundary point and good specificity when $\theta$ is at a boundary point.

\subsection{Simulation Parameters}
\label{sec:sim_params}

\begin{table}[H]\scriptsize
    \centering
    \begin{tabular}[t]{|Sc|Sc|Sc|Sc|Sc|Sc|Sc|Sc|}
\hline 
\multirow{2}{*}{Term} & \multirow{2}{*}{$\pi_m$} & \multicolumn{6}{c|}{$\mu_m$} \\
\cline{3-8}
& & 1 & 2 & 3 & 4 & 5 & 6 \\
\hline
\texttt{(Intercept)} & -2.49 & $9.8 \times 10^{-2}$ & $9.8 \times 10^{-2}$ & $9.8 \times 10^{-2}$ & $9.8 \times 10^{-2}$ & $9.8 \times 10^{-2}$ & $9.8 \times 10^{-2}$\\
\hline
\texttt{baseline\_bmi} & $6.7 \times 10^{-2}$ & $-1.9 \times 10^{-3}$ & $-1.9 \times 10^{-3}$ & $-1.9 \times 10^{-3}$ & $-1.9 \times 10^{-3}$ & $-1.9 \times 10^{-3}$ & $-1.9 \times 10^{-3}$\\
\hline
\texttt{gender} & -1.01 & $4.7 \times 10^{-3}$ & $4.7 \times 10^{-3}$ & $4.7 \times 10^{-3}$ & $4.7 \times 10^{-3}$ & $4.7 \times 10^{-3}$ & $4.7 \times 10^{-3}$\\
\hline
\texttt{race} & 0.41 & $-3.1 \times 10^{-3}$ & $-3.1 \times 10^{-3}$ & $-3.1 \times 10^{-3}$ & $-3.1 \times 10^{-3}$ & $-3.1 \times 10^{-3}$ & $-3.1 \times 10^{-3}$\\
\hline
\texttt{site[NC]} & -0.39 & $-2.4 \times 10^{-3}$ & $-2.4 \times 10^{-3}$ & $-2.4 \times 10^{-3}$ & $-2.4 \times 10^{-3}$ & $-2.4 \times 10^{-3}$ & $-2.4 \times 10^{-3}$\\
\hline
\texttt{site[SC]} & 0.35 & $-7.9 \times 10^{-3}$ & $-7.9 \times 10^{-3}$ & $-7.9 \times 10^{-3}$ & $-7.9 \times 10^{-3}$ & $-7.9 \times 10^{-3}$ & $-7.9 \times 10^{-3}$\\
\hline
\texttt{baseline\_age} & $-4.9 \times 10^{-2}$ & $-6.1 \times 10^{-4}$ & $-6.1 \times 10^{-4}$ & $-6.1 \times 10^{-4}$ & $-6.1 \times 10^{-4}$ & $-6.1 \times 10^{-4}$ & $-6.1 \times 10^{-4}$\\
\hline
\texttt{t2dm} & $-3.4 \times 10^{-2}$ & $-1.2 \times 10^{-2}$ & $-1.2 \times 10^{-2}$ & $-1.2 \times 10^{-2}$ & $-1.2 \times 10^{-2}$ & $-1.2 \times 10^{-2}$ & $-1.2 \times 10^{-2}$\\
\hline
\texttt{insulin} & $-2 \times 10^{-2}$ & $1.3 \times 10^{-2}$ & $1.3 \times 10^{-2}$ & $1.3 \times 10^{-2}$ & $1.3 \times 10^{-2}$ & $1.3 \times 10^{-2}$ & $1.3 \times 10^{-2}$\\
\hline
\texttt{hypertension} & 0.70 & $1.3 \times 10^{-3}$ & $1.3 \times 10^{-3}$ & $1.3 \times 10^{-3}$ & $1.3 \times 10^{-3}$ & $1.3 \times 10^{-3}$ & $1.3 \times 10^{-3}$\\
\hline
\texttt{hypertension\_rx} & -0.34 & $-8 \times 10^{-5}$ & $-8 \times 10^{-5}$ & $-8 \times 10^{-5}$ & $-8 \times 10^{-5}$ & $-8 \times 10^{-5}$ & $-8 \times 10^{-5}$\\
\hline
\texttt{dyslipidemia} & 0.42 & $-8.8 \times 10^{-4}$ & $-8.8 \times 10^{-4}$ & $-8.8 \times 10^{-4}$ & $-8.8 \times 10^{-4}$ & $-8.8 \times 10^{-4}$ & $-8.8 \times 10^{-4}$\\
\hline
\texttt{antilipemic\_rx} & -0.21 & $3.6 \times 10^{-4}$ & $3.6 \times 10^{-4}$ & $3.6 \times 10^{-4}$ & $3.6 \times 10^{-4}$ & $3.6 \times 10^{-4}$ & $3.6 \times 10^{-4}$\\
\hline
\texttt{smoking\_status[former]} & 2.16 & $-2.6 \times 10^{-3}$ & $-2.6 \times 10^{-3}$ & $-2.6 \times 10^{-3}$ & $-2.6 \times 10^{-3}$ & $-2.6 \times 10^{-3}$ & $-2.6 \times 10^{-3}$\\
\hline
\texttt{smoking\_status[never]} & 1.66 & $4.6 \times 10^{-4}$ & $4.6 \times 10^{-4}$ & $4.6 \times 10^{-4}$ & $4.6 \times 10^{-4}$ & $4.6 \times 10^{-4}$ & $4.6 \times 10^{-4}$\\
\hline
\texttt{treatment} & --- & -0.21 & -0.21 & -0.21 & 0.67 & 0.67 & 0.67\\
\hline
\texttt{trial\_id:treatment} & --- & --- & $1 \times 10^{-3}$ & --- & --- & $1 \times 10^{-3}$ & ---\\
\hline
\texttt{treatment:spline\_1} & --- & --- & --- & $1 \times 10^{-2}$ & --- & --- & $1 \times 10^{-2}$\\
\hline
\texttt{treatment:spline\_2} & --- & --- & --- & $6 \times 10^{-2}$ & --- & --- & $6 \times 10^{-2}$\\
\hline
\texttt{treatment:spline\_3} & --- & --- & --- & $1 \times 10^{-2}$ & --- & --- & $1 \times 10^{-2}$\\
\hline
\texttt{baseline\_age:treatment} & --- & --- & --- & --- & $2 \times 10^{-3}$ & $2 \times 10^{-3}$ & $2 \times 10^{-3}$\\
\hline
\texttt{t2dm:treatment} & --- & --- & --- & --- & $1 \times 10^{-1}$ & $1 \times 10^{-1}$ & $1 \times 10^{-1}$\\
\hline
\texttt{baseline\_bmi:treatment} & --- & --- & --- & --- & $-2.5 \times 10^{-2}$ & $-2.5 \times 10^{-2}$ & $-2.5 \times 10^{-2}$\\
\hline
\end{tabular}
    \caption{Summary of coefficients used for simulated treatments $A_m$ (from $\pi_m$) and outcomes $Y_m$ (from $\mu_m$). Outcome model numbering reflects the ordering of scenarios described in Section \ref{sec:sim_dgp_supp}.}
    \label{tab:sim_coefficients}
\end{table}

\newpage
\bibliographystyle{unsrt}
\bibliography{ref}